\documentclass[fleqn,usenatbib]{mnras}
\pdfoutput=1
\usepackage{newtxtext,newtxmath}
\usepackage[T1]{fontenc}
\usepackage{ae,aecompl}

\usepackage{graphicx}	% Including figure files
\usepackage{amsmath}	% Advanced maths commands
\usepackage{amssymb}	% Extra maths symbols

\usepackage{aas_macros}
\usepackage{color}
\usepackage{url}

\definecolor{orange}{rgb}{1,0.5,0}
\definecolor{violet}{rgb}{0.95,0.52,0.95}

\newcommand{\rvir}{$r_\text{vir}$}
\newcommand{\rdm}{$r_\text{DM}$}
\newcommand{\rhot}{$r_\text{hot}$}
\newcommand{\rshot}{$r_\text{s,hot}$}

\newcommand{\mcold}{$M_\text{cold}$}
\newcommand{\mstar}{$M_\star$}
\newcommand{\mbh}{$M_\text{BH}$}

\newcommand{\sag}{\textsc{sag}}
\newcommand{\sagb}{\textsc{sag$_{\beta1.3}$}}

\newcommand{\sage}{\textsc{sage}}
\newcommand{\darksage}{\textsc{dark sage}}
\title[Semi-Analytic Galaxies I ]{Semi-Analytic Galaxies - I. Synthesis of environmental and star-forming regulation mechanisms}

\author[Sof\'ia A. Cora et al.]{
Sof\'ia A. Cora,$^{1,2,3}$\thanks{E-mail: sacora@fcaglp.unlp.edu.ar}
Cristian A. Vega-Mart\'inez,$^{1,3}$
Tom\'as Hough,$^{1,2,3}$
\newauthor{ Andr\'es N. Ruiz$^{3,4,5}$
\'Alvaro Orsi,$^{6}$
Alejandra M. Mu\~noz Arancibia,$^{7,8}$ } 
\newauthor{ Ignacio D. Gargiulo,$^{1,2,3}$
Florencia Collacchioni,$^{1,2,3}$
Nelson D. Padilla,$^{8,9}$ }
 \newauthor{ Stefan Gottl\"ober$^{10}$
and Gustavo Yepes$^{11}$ }
\vspace{0.2cm}\\
$^{1}$Instituto de Astrof\'isica de La Plata (CCT La Plata, CONICET,  UNLP), 
   Observatorio Astron\'omico, Paseo del Bosque,\\ B1900FWA La  Plata, Argentina\\
$^{2}$Facultad de Ciencias Astron\'omicas y Geof\'{\i}sicas, 
   Universidad Nacional de La Plata, 
   Observatorio Astron\'omico, Paseo del Bosque,\\ B1900FWA La Plata, Argentina\\
$^{3}$Consejo Nacional de Investigaciones Cient\'ificas y T\'ecnicas
   (CONICET), Rivadavia 1917, Buenos Aires, Argentina\\
$^{4}$Instituto de Astronom\'ia Te\'orica y Experimental (CCT  C\'ordoba, CONICET, UNC),
   Laprida 854, X5000BGR C\'ordoba, Argentina\\
$^{5}$Observatorio Astron\'omico de C\'ordoba, Universidad Nacional de C\'ordoba, 
   Laprida 854, X5000BGR, C\'ordoba, Argentina\\
$^{6}$Centro de Estudios de F\'isica del Cosmos de Arag\'on, 
   Plaza de San Juan 1, Teruel, 44001, Spain\\
$^{7}$Instituto de F\'isica y Astronom\'ia, Universidad de Valpara\'iso,
   Av. Gran Breta\~na 1111, Valpara\'iso, Chile\\
$^{8}$Instituto de Astrof\'isica, Pontificia Universidad Cat\'olica de Chile,
   Av. Vicu\~na Mackenna 4860, 7820436 Macul, Santiago, Chile\\
$^{9}$Centro de Astro-Ingenier\'ia, Pontificia Universidad Cat\'olica de Chile,
   Av. Vicu\~na Mackenna 4860, 7820436 Macul, Santiago, Chile\\
$^{10}$Leibniz Institute for Astrophysics, An der Sternwarte 16, 
   14482 Potsdam, Germany\\
$^{11}$Departamento de F\'isica Te\'{o}rica and CIAFF, M\'{o}dulo 8, 
   Facultad de Ciencias, Universidad Aut\'{o}noma de Madrid, \\ 28049 Madrid, Spain
}

\date{Accepted XXX. Received YYY; in original form ZZZ}

\pubyear{2017}

\begin{document}
\label{firstpage}
\pagerange{\pageref{firstpage}--\pageref{lastpage}}
\maketitle

\begin{abstract}
We present results from the semi-analytic model of galaxy
formation \sag~applied on the \textsc{MultiDark} simulation MDPL2. 
\sag~features an updated supernova (SN) feedback scheme and 
a robust modelling of the environmental effects on satellite galaxies.
This incorporates a gradual starvation of the hot gas halo driven by the action of ram pressure stripping (RPS), that can affect the cold gas disc, and tidal stripping (TS), which can act on all baryonic components. 
Galaxy orbits of orphan satellites are integrated 
providing adequate positions and velocities for the estimation of RPS and TS. 
The star formation history and stellar mass assembly of galaxies are sensitive
to the redshift dependence implemented in the SN feedback model. 
We discuss a variant of our model that allows
to reconcile the predicted star formation rate density at $z\gtrsim 3$ with 
the observed one, at the expense of an excess in the faint end
of the stellar mass function at $z=2$.
The fractions of passive galaxies as a function of stellar mass, 
halo mass and the halo-centric distances
are consistent with observational measurements.
The model also reproduces the evolution of the
main sequence of star forming central and satellite galaxies.
The similarity between them is a result of the 
gradual starvation of the hot gas halo suffered by satellites, in which
RPS plays a dominant role.
RPS of the cold gas does not affect the fraction of quenched satellites
but it contributes to reach
the right atomic hydrogen gas content
for more massive satellites
($M_{\star} \gtrsim 10^{10}\,{\rm M}_{\odot}$).
\end{abstract}


%%%%%%%%%%%%%%%%%%%%%%%%%%%%%%%%%%%%%%%%%%%%%%%%%%

%%%%%%%%%%%%%%%%% BODY OF PAPER %%%%%%%%%%%%%%%%%%

\section{Introduction}

The properties of galaxies we observe at the present epoch are the result 
of a variety of physical mechanisms acting throughout their assembly history. 
Galaxy evolution leads to a bimodal colour distribution that features 
a red and a blue branch in the colour-stellar mass diagram 
\citep[e.g. ][]{Baldry04}. 
Interestingly, each colour branch is also associated to a different 
galaxy morphology: early type galaxies are located in the 
well defined red sequence, whereas late-type galaxies populate 
the blue cloud \citep{conselice2006}.
The fraction of galaxies in the red sequence increases as a function of 
local environmental density \citep{baldry2006},
especially at high stellar masses. 
However, the environmental
dependence of the morphology fractions is milder than that of colour
fractions for low-masses \citep{Bamford09}.
Outliers of these two broad colour branches include galaxies in a so-called
intermediate green valley. These galaxies are interpreted as experiencing a 
decrease of their star-formation rate (SFR) 
\citep{Schawinski14,Powell17,Bait17,Coenda18}.
Star formation (SF) quenching is 
higher for more massive galaxies, 
and increases at fixed stellar mass with environmental density (given either
by local density or halo mass) as becomes
evident from the strong correlations of the fractions of quiescent galaxies with stellar mass, environment and halo-centric radius,
found both locally \citep{Wetzel12}
and at high redshift 
\citep{Peng10, Muzzin12, Lin14, Jian17, Kawinwanichakij17}.
Besides, the way in which red and quiescent fractions vary with
environment and stellar mass are different for central and satellite galaxies
\citep{weinmann2006, Wetzel12, Kovac14, Darvish17, Smethurst17, Wang17}.

Central and satellite galaxies of a given stellar mass are subject 
to very different
processes. Central galaxies are believed to reside at the centres of
their host dark matter (DM) haloes. They can receive new gas via cooling flows,
and they cannibalize satellite galaxies whose orbits in the group
potential decay due to dynamical friction. 
Dry mergers (both minor and major) 
can explain the shallower slope of the massive end of the red sequence 
\citep{Jimenez11}. On the other hand, major mergers contribute 
to produce massive, passive galaxies \citep{vanderWel09}, doubling the
mass of the massive brightest cluster galaxies since $z\sim 1$ 
\citep[e.g. ][]{Shankar15}.
While orbiting in the main
group, satellite galaxies are also affected by
galaxy encounters that may result in morphological transformations
\citep{Kannan15}. However, the key processes that are believed to quench
the SF in satellite galaxies are triggered by 
gas disruption due to tidal stripping \citep[TS, ][]{Merritt83}
and ram pressure stripping (RPS, \citealt{gg72, Abadi99}).

During TS, the material is pulled
from a galaxy by the global tidal field of the host halo
or a larger neighbour galaxy, the former being 
more relevant than the latter in low-velocity close encounters taking place
in galaxy groups \citep{Villalobos14}.
Tidal shocks at the pericentres of the orbit of satellite galaxies
induce dynamical
instabilities and impulsive tidal heating of the stellar distribution
that are particularly relevant for dwarf galaxies in environments
similar to our Local Group \citep{Kazantzidis11}.
Besides, the impact of TS (from negligible
effect to complete galaxy disruption) also depends on 
the satellite morphology \citep{Chang13}. 

RPS is a result of the ram-pressure (RP) exerted by the hot diffuse gas 
of a group 
or cluster of galaxies
on satellites moving through it at velocities that could
be close to supersonic. 
This physical process may account for the gradual
removal of both the hot gas halo and the cold gas disc
\citep{BoselliGavazzi06,Jaffe13,Poggianti17}, giving place
to `jellyfish' galaxies in cases of extreme RPS \citep{Bellhouse17}
where the trailing stripped tails show signs of shock heating and gas
compression.
Hydrodynamical simulations show that, on one hand, RP contributes 
to SF quenching through gas stripping and, on the other, 
can also produce a temporary and moderate enhancement of SF
(more likely to occur in Milky Way-type disc galaxies)
as a result of gas compression that takes place 
at pericentre passage
\citep{Bekki14, YozinBekki15, Steinhauser16, Ruggiero17}.

The instantaneous complete removal
of the hot diffuse gas halo of a galaxy after its
infall into a larger halo,
that takes place 
regardless of which physical mechanism may actually be responsible
for it,
has been standard ingredient in SAMs. 
This effect is generally referred to as
strangulation
\citep*{larson80,balogh2000}.
Although this simple prescription for halo stripping can
account for the presence of passive galaxies in clusters at high
redshift, it results in fractions of red satellite galaxies which are
higher than those observed in groups of galaxies 
\citep[e.g.][]{baldry2006, weinmann2006, kimm2009};
this excess of red satellites constitutes the satellite overquenching problem.
Based on these results, there is
now general consensus that the instantaneous stripping of the hot gas
of satellites due to shock-heating 
is a crude approximation
of the process. This is additionally supported by the observational evidence 
that large fractions of near-IR-bright, early-type galaxies in groups
\citep{jeltema2008} and also in clusters \citep{sun2007, Wagner17} 
have extended
X-ray emission, indicating that they retain significant hot gas haloes
even in these dense environments.
Studies based on smoothed particle hydrodynamics (SPH) simulations 
have found that the hot gas haloes of
satellites are not stripped instantly \citep{mccarthy2008}, although
RPS of halo gas is much more effective than that of disc gas
\citep{bekki2009}.
Moreover, the effect of RPS on the hot corona seems to be
even milder than expected because of
turbulence in the interface with the intracluster medium (ICM) 
and fuelling of gas from this
component to the satellite hot gas halo \citep{Quilis17}.

This mounting evidence motivated several authors
to include prescriptions
for a gradual removal of hot halo gas in SAMs, 
with the inherent limitations of this numerical technique. 
Some of them only consider the effect of RPS
\citep[e.g. ][]{Kang08, font2008, weinmann2010, GonzalezPerez14}
while others also take into account the
influence of TS \citep{guo11, kimm2011, Henriques17, Stevens17}.
If hot gas removal proceeds
gradually by the action of RPS and/or TS, the
time-scale of SF quenching
becomes longer and the 
discrepancies in the fraction of red/quiescent galaxies 
between models and observations are alleviated
although some disagreements remain.  
The prescriptions adopted are generally based on
the analytic formulation of \citet{mccarthy2008}.
The effect of RPS on the cold gas of galaxy discs 
has been implemented only in few SAMs
\citep{on2003,lanzoni2005, bdl2008, tecce10, luo16, Stevens17}. 
The combination of suppressed cooling flows due to
strangulation and RPS of cold disc gas as in \citet{tecce10}
worsens the problem of excess of
red satellites. The presence of hot gas halo during gradual
removal acts as a shield that helps to prevent a too efficient action of
RPS on the cold gas phase. Although this is taken into account
by \citet{Stevens17}, the condition used  
seems to still produce larger amounts of removed cold gas than
expected. 

The star formation history of galaxies and, consequently,
the fractions of star forming and passive galaxies, are
strongly affected by supernova (SN) feedback.
\citet{hirschmann16} tested the impact of different
schemes for stellar feedback
and gas recycling using the semi-analytic model \textsc{Gaea}.
They find that a model characterized by
a strong ejection coupled to
a mass-loading that is explicitly dependent
on redshift
is reasonably successful
in reproducing the observed `anti-hierarchical' trends
in galaxy mass assembly, with
a delayed
metal enrichment and a realistic present-day stellar and
gaseous metallicity.
They adopt a redshift dependence of
the mass-loading factor (ratio between the gas outflow rate
from galaxies
and the star formation rate) similar to the one provided by \citet{muratov15}, who analyse the galaxy-scale gaseous outflows in the cosmological hydrodynamical
zoom FIRE (Feedback
in Realistic Environments) simulations. 

In this work, we present   
an updated version of the semi-analytic model 
\sag~\citep[acronym for Semi-Analytic Galaxies,][]{cora2006, lcp08, orsi14,
munnozarancibia2015, Gargiulo15},
which includes an improved treatment of 
environmental effects 
through the implementation of 
the gradual removal of hot gas in satellites by RPS and TS,
allowing also the action of these processes on the cold gas disc,
under certain conditions; the modelling of 
supernova (SN) feedback is also modified.
This version of \sag~was used to generate one of the galaxy catalogues of the
\textsc{MultiDark Galaxies} project \citep{knebe17},
which is based on the Planck cosmology $1\,h^{-1} \,{\rm Gpc}$
\textsc{MultiDark} simulation MDPL2
\citep{Klypin16}. This catalogue is publicly available%
\footnote{\url{http://dx.doi.org/10.17876/cosmosim/mdpl2/007}} 
in the \textsc{CosmoSim} database%
\footnote{\url{https://www.cosmosim.org}}
together with those generated
by the semi-analytic models \textsc{Sage} \citep{Croton16}
and \textsc{Galacticus} \citep{Benson12}.

This paper presents a detailed description 
of the modifications introduced,
emphasizing the advantages with respect to previous works.
Details of
the MDPL2 simulation and a brief summary of the basic
aspects of \sag~are given in Section~\ref{sec:model}.
Improvements introduced regarding
environmental effects and SN feedback are explained 
in Sections~\ref{sec:env} and~\ref{sec:snfeedFIRE}, respectively.
The method applied to tune the free parameters of the model
and the comparison of  model results
with observational constraints imposed to calibrate the
model are presented in Section~\ref{sec:calibration}.
Section~\ref{sec:SAG-predictions}
includes an analysis of general
galaxy properties predicted by \sag~\citep[complementing the discussion already presented in ][]{knebe17}, 
and of those related with SF quenching:
dependence of the fraction of quenched galaxies as a function of stellar mass,
halo mass and halo-centric distance.
We show that the relations involving the fraction of passive galaxies
are better reproduced by modifying a parameter introduced in 
the modelling of SN feedback which has impact on the SF history of galaxies.
For this variant of the model, we analyse the role of environmental
processes in Section~\ref{sec:ts-rps}, focussing on the stripped mass and atomic gas content.
In Section~\ref{sec:SAMcomp}, we compare details of our model and its main results with other SAMs.
In Section~\ref{sec:conclu}, we present the summary and conclusions of this work.
Appendix~\ref{ap:rpscalc} describes the estimation
of the hot gas stripping radius due to RP.   

\section{Galaxy formation model}
\label{sec:model}
Our semi-analytic model of galaxy formation and evolution
\sag~uses DM haloes extracted from a cosmological
DM simulation and their corresponding merger trees as the
basic inputs to construct the galaxy population.
We use the \textsc{MultiDark} simulation MDPL2,
which is part of the \textsc{CosmoSim} database.

\subsection{MDPL2 simulation}
\label{sec:mdpl2sim}

MDPL2 simulation follows the evolution of $3840^3$ particles within a box of
side-length $1\,h^{-1}\,{\rm Gpc}$,
with a mass resolution 
$m_\textrm{p} = 1.5 \times 10^{9}\, h^{-1}\, \textrm{M}_{\odot}$ per
DM~particle. This simulation is
an analogue of the MDPL simulation described in \cite{Klypin16}, 
but ran with a different realisation of the initial conditions.
It is consistent with
a flat $\Lambda$CDM model characterized by Planck cosmological parameters:
$\Omega_{\rm m}$~=~0.307, $\Omega_\Lambda$~=~0.693, 
$\Omega_{\rm B}$~=~0.048, $n_{\rm s}$~=~0.96
and $H_0$~=~100~$h^{-1}$~km~s$^{-1}$~Mpc$^{-1}$, where $h$~=~0.678 
\citep{Planck2013}.

DM haloes have been identified with the
\textsc{Rockstar} halo finder
\citep{Behroozi_rockstar}, and merger trees were constructed with 
\textsc{ConsistentTrees} \citep{Behroozi_ctrees}.
Overdensities with at least $N_\text{min}$~=~20~DM 
particles were considered in the detection by the halo finder. 
Each halo is characterized by measuring the physical properties 
defined by the particle distribution, assuming spherical overdensity 
approximation.  
The virial mass is defined as the mass enclosed by a sphere of radius
\rvir, so that the mean density reaches a constant factor $\Delta=200$ times
the critical density of the Universe $\rho_\textrm{c}$, i.e.
\begin{equation}
   M_\textrm{vir}(<r_\textrm{vir}) = \Delta \rho_\textrm{c}
            \frac{4 \pi}{3} r_\textrm{vir}^3.
   \label{eq:Mvir}
\end{equation}
Moreover, the virial velocity of each halo is defined in terms of these
properties as $V_\textrm{vir}= \sqrt{G M_\textrm{vir}/r_\textrm{vir}}$, 
where $G$ is the gravitational constant. 
The detected DM haloes can be over the background density or lie
within another DM haloes. To differentiate them, henceforth the
former will be referred to as \textsl{main host} haloes, whereas the latter as
subhaloes.
The calculation of the physical properties of subhaloes
considers only
the bound particles of the substructure identified by the halo finder.

\subsection[]{Semi-analytic model of galaxy formation SAG}
\label{sec:sag}

The version of our semi-analytic model of galaxy formation and evolution
\sag~presented here is a further
development of the model described by \citet{cora2006},
which is based on \citet{springel2001},
and later improved in \citet{lcp08}, \citet{tecce10}, 
\citet{orsi14}, \citet{munnozarancibia2015} and  
\citet{Gargiulo15}.

\sag~includes the effects of radiative cooling of hot gas,
star formation, feedback from SN explosions,
chemical enrichment with a scheme that 
tracks several chemical elements contributed by 
different sources (stellar winds and supernovae Type Ia and II)
taking into account the lifetime of progenitors
\citep{cora2006}, 
growth of supermassive black holes (BHs) in galaxy centres and the consequent
AGN feedback, and starbursts triggered by disc
instabilities or galaxy mergers \citep{lcp08}. 
These starbursts contribute to the formation of a bulge component,
which sizes are estimated as described in \citet{munnozarancibia2015};
the cold gas that has been transferred to the bulge is gradually
consumed, thus starbursts are  
characterised by a time-scale 
\citep{Gargiulo15} 
instead of being instantaneous. 
\citet{tecce10} added the effect of RPS on the cold
disc gas of satellite galaxies
implementing the criterion 
from \citet{gg72}.

Two important new features have been added to \sag~as described in the present work.
One is related to the inclusion of an explicit redshift dependence
in the estimation of the reheated and ejected mass as a result
of SN feedback,   
guided by one of the ejective feedback schemes implemented in the \textsc{Gaea} semi-analytic model \citep{hirschmann16}.
This redshift dependence becomes crucial to reproduce
observed galaxy properties at high redshift, as we will show in 
Sections~\ref{sec:SAG-const} and \ref{sec:SAG-predictions}. 
The second major improvement
replaces the strangulation scheme 
that removes the hot gas from satellites instantaneously
(standard practice in SAMs until recently)
by a gradual removal of the halo gas by the
action of TS and/or RPS (gradual starvation%
\footnote{
The process
of removal of the hot gas haloes of satellite galaxies is usually
indistinctly called `strangulation' or `starvation', regardless of which
physical mechanism may actually be responsible for it. 
In this study, we will use the term `strangulation' to refer to the 
instantaneous removal of hot gas from satellites, and we
will use `gradual starvation' to refer to the gradual removal of halo gas by any process.}).

The features that the current version of
\sag~inherits from its predecessors are described in \citet{Gargiulo15}, 
except for those related to the RPS of cold
gas \citep{tecce10}, 
which will be described in Section~\ref{sec:env} along with the other
environmental processes introduced in this work. 
The new SN feedback scheme is described in Section~\ref{sec:snfeedFIRE}. 
Furthermore, the model of radio mode AGN feedback, in which
black holes generate jets and bubbles as they accrete gas from 
hot atmospheres suppressing gas cooling, has been replaced by that
described in
\citet[][see their eq. S24 in the supplementary material]{henriques_mcmc_2015}.
This means that
the dependences with the fraction of hot gas with respect to the virial mass
and with the virial velocity of the halo (introduced by \citealt{lcp08})
are substituted by a dependence
with the hot gas mass. The aim of this change is to make AGN feedback more
efficient at late times.

The semi-analytic model assigns 
one galaxy to each new detected halo in the simulation, and it 
follows the halo merger trees to compute the evolution of the galaxy
properties. 
Each considered system of haloes is constituted by only one 
group/cluster central
galaxy, the one associated to the main host halo, 
so that the other galaxies act as satellites. 
When two haloes merge, the smaller one loses mass
to tides as it orbits within the larger structure, until  
the satellite
is no longer identified by the halo finder. 
We assume that the galaxy contained within
this disappeared subhalo
survives until it eventually merges with the central galaxy of its
host halo. During this temporary stage, these galaxies are called
\textsl{orphan} satellite galaxies.

\section{Environmental processes}
\label{sec:env}

As satellite galaxies orbit within their main host haloes,~DM and
the baryonic components in their
own subhaloes 
are affected by numerous environmental processes.
We model mass stripping produced by tidal forces and RP.
TS will act on every component of the galaxy (stars, gas and~DM) 
whereas~RPS is a hydrodynamical process which will only affect the gas.
Thus, a two-stage model for gas removal is
now included in \sag. TS and RPS both act first on the hot gas halo: whichever
effect is stronger determines the amount of gas stripped. 
Once a significant fraction of the hot halo is
removed, RPS can start affecting the cold gas disc, following
\citet{tecce10}.
The combination of the RPS of the cold gas phase 
with the gradual starvation scheme
applied to the hot gas  
constitutes an improvement
of our model
with respect to previous works that ignore the RPS of the cold gas
\citep[e.g. ][]{guo11, kimm2011, GonzalezPerez14}. Recently, this effect
has been taken into account by \citet{luo16} and \citet{Stevens17};
while the former authors do not consider the hot gas as a shield
that regulates the action of RP on the cold gas, the latter
take into account this possibility using a different criterion
than the one adopted in \sag.

\subsection{Estimation of RP}
\label{sec:RPfit}

There have been numerous studies using hydrodynamic simulations
focused on the effects of RPS on cold gas discs 
\citep[e.g.][]{Abadi99, RoedigerBrueggen07, Tonnesen07}
and hot gas haloes \citep{mccarthy2008, Quilis17}.
RP is defined as the product of the ICM density at the location
of the satellite galaxy and the square of its velocity relative to the ICM.
The importance of the local variations of the ambient gas makes
analytical calculations of the RP effect slightly inaccurate, 
particularly for the resulting
size of the gas disc. 
\citet{Tonnesen08}
show that these variations can produce
differences in the pressure of more than one order of magnitude at fixed 
cluster-centric distance.
As demonstrated by \citet{tecce10} from the analysis
of adiabatic hydrodynamic simulations of galaxy clusters,
the values of RP present a radial profile mainly determined
by the dependence on cluster-centric distance of the ICM 
\footnote{We refer to the hot gas associated with a
  main host halo using the term `intracluster medium', even for host halo
  masses smaller than that of a galaxy cluster.}
density. 
Values of RP of the order of $10^{-12}\,h^2\,{\rm dyn}\,{\rm cm}^{-2}$,
$10^{-11}\,h^2\,{\rm dyn}\,{\rm cm}^{-2}$ and
$10^{-10}\,h^2\,{\rm dyn}\,{\rm cm}^{-2}$ 
are considered weak, medium and strong, respectively 
\citep{RoedigerBrueggen06}.
At $z=0$, 
mean RP values range from weak to medium 
for clusters masses of $\approx 10^{14}\,{\rm M}_{\odot}$ and 
$\approx 10^{15}\,{\rm M}_{\odot}$, respectively.
At $z\approx 2$, most of  
satellite galaxies within progenitors of currently massive clusters are already
experiencing medium-level RP, and at $z \lesssim 0.5$
a significant fraction of galaxies ($\approx 20$ per cent) 
suffer strong levels of RP \citep[see fig. 5 of ][]{tecce10}. 
This evolution 
depends mainly on the build-up of the
ICM density over time.

We take into account all these effects through
the implementation in \sag~of fitting formulae that estimate the RP 
experienced by galaxies in haloes of different
mass as a function of halo-centric distance and redshift,
which are derived from a procedure similar
to that described by \citet{tecce11}.
The new fit uses 
alternative formulae that capture more adequately the RP behaviour,
as described in Vega-Mart\'inez et al. (in preparation)
using the self-consistent information provided by \citet{tecce10}
from the analysis of hydrodynamic simulations of galaxy clusters. 
This procedure, called the `gas-particles' method in \citet{tecce10}, 
automatically takes into account any local variation 
of the density and/or the velocity field
which may result from the dynamics of the gas in the hydrodynamic simulation. 
The fitting formulae generated from these data allow
to include RP effects into SAMs applied 
on dark matter-only simulations which lack gas physics, as the one used in this
work.
This way of estimating RP values
constitutes   
the main advantage of our treatment of RPS. 

\subsection{Orbits of orphan galaxies}
\label{sec:orbits}

For those satellites
with identified host subhaloes, the effects of dynamical
friction and TS are provided self-consistently by tracking the subhalo
evolution in the base simulation.
This approach breaks down at the subhalo resolution
limit, when the galaxy host by a subhalo that is no longer identified 
becomes an orphan satellite.
In previous versions of \sag, positions and velocities of orphans
were assigned using the current position and velocity of the
most-bound DM particle of the galaxy's host subhalo at the time
it was last identified \citep{lcp08, tecce10}, or 
assuming a circular orbit with decaying radial distance
estimated from the dynamical friction and 
initial halo-centric distance and velocity
determined by the virial radius and velocity of the host (sub)halo, 
respectively
\citep{Gargiulo15}.
In any case, the merging criterion applied to orphans compares
the time elapsed since the time of infall 
with a dynamical friction time-scale; when the former is larger than the latter,
an orphan galaxy is considered merged with a central galaxy regardless of the 
relative distance between them. 

In the current version
of \sag, the phase-space of orphans is obtained from the integration of
the orbits of subhaloes that will give place to an orphan galaxy
once \sag~is applied 
to the merger trees extracted from the underlying
simulation,
as described in detail by
Vega-Mart\'inez et al. (in preparation).
When a subhalo is not longer tracked, 
the last known position, velocity and virial mass are taken as
initial conditions to integrate its orbit numerically.
The orbital evolution 
is consistent with the potential well of the
host halo, and takes into account mass loss by TS and dynamical friction effects,
following some aspects of the works by 
\citet{gan2010} and \citet{kimm2011}.
The integration is carried out until
a merger event occurs, which is defined 
when the halo-centric distance 
becomes smaller than $10$ per cent of the virial radius of the host halo, 
taken as an estimation of the radius of a central galaxy,
since in this stage previous to the application of \sag~we do not have 
information on the properties of galaxies that will populate these haloes.
This merging criterion allows to reach general good agreement with the redshift 
evolution of the merger rate of galaxies obtained by considering 
the merger time-scale given by \citet{jiang2008}, which is inferred from 
a high-resolution cosmological hydro/N-body simulation with star formation.

\subsection{Assumptions for the action of RPS and TS}

We assume that the time-scale for RPS and TS 
of any of the baryonic components is given
by~$t_\text{dyn,orb}=2\,\pi\,{\omega}^{-1}$, where $\omega$ is the angular
velocity of the satellite, according to \citet{Zentner05}.
In each interval~$\Delta T$ between simulation snapshots,
the RPS and/or TS effect removes 
a fraction~$\Delta T/t_\text{dyn,orb}$ of gas and stars outside the
corresponding  
stripping radius,~$r_\text{s}$, 
if this fraction is less than unity; otherwise, the whole 
estimated stripped mass,  
$M_\text{strip}$, 
is removed.
In the following, we describe the way in which the stripping 
radius $r_\text{s}$ is derived for hot gas, cold gas disc, 
disc stars and bulge stars,  
referring to this quantity as $r_\text{s,hot}$, $R_\text{s,cold}$, 
$R_\text{s,disc}$ and $r_\text{s,bulge}$, respectively. 

\subsection[]{Gradual stripping of the hot gas haloes of satellites}
\label{sec:hotgasstrip}

The new version of \sag~presented in this paper allows 
galaxies to keep their hot gas haloes when they become
satellites. These haloes are then gradually stripped by the action of~RPS
and/or~TS, and as long as they
survive they can replenish the satellite's cold gas via gas cooling.
This is the
scheme we call gradual starvation.

In this new scenario, the mass of hot gas $M_\text{hot}$ available for cooling
to the central galaxy of each main host halo 
is calculated at the beginning of each 
simulation snapshot as
\begin{equation}
  \begin{split}
    M_\text{hot} = & f_{\rm b} M_\text{vir} - M_{\star,\text{cen}} -
    M_\text{cold,cen} - M_\text{BH,cen} \\
    & - \sum_{i=1}^{N_\text{sat}} \left( M_{\star,i} +
    M_{\text{cold,}i} + M_{\text{BH,}i} + M_{\text{hot,}i} \right),
  \end{split}
  \label{eq:hotgas}
\end{equation}
where $f_{\rm b}$~=~0.1569 is the universal baryon fraction \citep{Planck2013}
and \mstar, \mcold~and
\mbh~are, respectively, the total masses of stars, of cold gas and
of the central black hole
(BH) of each of the galaxies contained within a given main host halo.
Central and satellite galaxies are considered separately in order
to discount the total mass in the remaining hot gas haloes of
the $N_\text{sat}$ satellite galaxies within the same main host halo. 

Our prescription for~RPS of the hot gas halo is based on the 
model from \citet{font2008}, which
uses the criterion for~RPS for a spherical distribution of gas determined by
\citet{mccarthy2008} from the results of hydrodynamic simulations. The gas beyond a
satellite-centric radius~$r_\text{sat}$ will be removed if the
value $P_{\rm ram}$ of RP, which is given by the fitting formulae
described in Section~\ref{sec:RPfit}, meets the 
condition
\begin{equation}\label{eq:rpshot}
P_{\rm ram} > \alpha_\text{RP} \frac{G
    M_\text{sat}(r_\text{sat}) \rho_\text{hot}
    (r_\text{sat})}{r_\text{sat}},
\end{equation}
where $\rho_\text{hot}$ is the density of the satellite's hot gas halo,
and $\alpha_\text{RP}$ is a geometrical constant of order unity chosen to match the results of~hydrodynamic simulations. 
\citet{mccarthy2008} find that their simulations support $\alpha_\text{RP}=2$.
They also test the effect of values as large as $\alpha_\text{RP}=10$, which
produce milder effects. Since they do not consider gas cooling that can 
also reduce the hot gas content, 
we adopt an intermediate value of $\alpha_\text{RP}=5$ in our model.
The total mass $M_\text{sat}$ of a satellite is
\begin{equation}\label{eq:msat}
  \begin{split}
    M_\text{sat}(r_\text{sat}) = & M_\star + M_\text{cold}\\
    &+ 4\pi \int_0^{r_\text{sat}} \left[ \rho_\text{hot}(r) +
    \rho_\text{DM}(r) \right] r^2 dr,
  \end{split}
\end{equation}
assuming that $r_\text{sat}$ is large enough to contain all the stars and cold
gas. 
All density profiles are represented using
isothermal spheres, i.e.,  
$\rho_\text{hot} = {M_\text{hot}}/({4\pi\, r_\text{hot}\, r^2})$, 
where $r_\text{hot}$ is the radius that contains all of $M_\text{hot}$.
This radius 
initially adopts the value of the subhalo 
virial radius, \rvir; in the case of orphan satellites, \rvir~preserves
the value corresponding to the last time the subhalo was identified.
Combining equations~\eqref{eq:rpshot} and~\eqref{eq:msat} one can
numerically solve for the hot gas stripping radius due to~RP, 
$r_\text{s,hot}^\text{RPS}$ (see Appendix~\ref{ap:rpscalc}). 

To determine the halo gas loss of satellite galaxies via TS
we assume that the hot gas distributes
parallel to the~DM.
The bounding radius for the DM,~\rdm, 
is given by 
\rvir.
This radius gives the tidal radius determined by TS, that is
$r_\text{s,hot}^\text{TS}=r_\text{DM}$, and 
is compared with the hot gas stripping radius due to~RP, 
$r_\text{s,hot}^\text{RPS}$.
The smaller of the
two will be the stripping radius~\rshot~which will contain the remaining 
hot gas mass; all hot gas beyond that radius can be stripped.
Thus, the value of~\rhot~is updated such that 
$r_\text{hot}^\text{new}=r_\text{s,hot}$. 

Once the hot gas outside the new stripping radius is removed, 
we assume that the
remaining gas quickly redistributes its mass and restores an isothermal profile,
but truncated at~\rhot,
as in~\citet{font2008} and \citet{kimm2011}.
The pressure of the ambient~ICM could act to confine the satellite's hot
gas, preventing any outflow from going beyond the stripping radius
\citep{mj2010}. 
If in a subsequent simulation time step the value
obtained for~\rhot~is larger than the previous one, 
\rhot~is not updated and no gas is lost in that time step.

In previous versions of \sag, based on the strangulation scheme, 
all feedback processes occurring in satellite
galaxies were assumed to transfer cold gas and its associated metals from the
satellite into the hot phase of the central galaxy of their main host
halo (i.e. the ICM). In our new implementation, we assume 
that the hot gas halo of a satellite is depleted
when 
the hot gas mass 
drops below a fraction $f_\text{hot,sat}$ of the baryonic mass of the 
satellite, 
specified in
Section~\ref{sec:parameters}.
Before this situation takes place, 
all feedback processes will transport gas
and metals to the hot phase of the same galaxy, 
proportionally to its content of 
hot gas; the remaining reheated mass and associated metals are transferred 
to the corresponding central galaxy.  
If the latter is a central galaxy of a subhalo (i.e. it is also a satellite) 
and its
hot gas halo also drops below $f_\text{hot,sat}$ times the baryon fraction,
then the reheated gas and metals of the orphan satellite
are transferred to the main host halo.

In any given simulation time step,
all those galaxies, either central or satellite, 
that still have a hot gas reservoir will proceed to cool gas.
Gas cooling rates are calculated using the simple model presented in
\citet{springel2001}, but considering 
the total radiated power per
chemical element given by \citet{foster2012}. 

\subsection[]{Ram pressure and tidal stripping of cold gas disc}
\label{sec_RPS_TS_cold}

Ram pressure exerted by the intra-group/cluster medium can also affect 
the cold gas disc of satellite galaxies. Its action could
be regulated by 
the presence of hot gas halo in the gradual starvation scheme. 
Hot gas haloes in satellite galaxies will be replenished by SNe feedback as
long as there is some star formation and the ratio between
the hot gas halo and the baryonic mass of the galaxy is larger than
the fraction $f_\text{hot,sat}$
(see section~\ref{sec:hotgasstrip}). At a certain point, this ratio will
become smaller than $f_\text{hot,sat}$, and the hot gas halo will be gradually
reduced by gas cooling and/or stripping processes (RPS, TS) reaching very 
low values.
We assume that for as long as this ratio is larger than $0.1$, the
hot gas halo shields the cold gas
disc from the action of~RPS.
Once this condition is not fulfilled anymore,
the ambient~RP starts affecting the cold gas. 
This threshold has been chosen small enough to allow the role of the hot gas as a 
shield for a sufficiently long time; the action of RPS on the cold disc gas
becomes too effective 
without any restriction of this kind.

We consider the model for~RPS of cold gas disc introduced in \citet{tecce10}, 
which is
based on the simple criterion proposed by \citet{gg72}. The cold gas of the
galactic disc located at a galactocentric radius~$R$ will be stripped away if
the~RP exerted by the ambient medium on the galaxy exceeds the restoring force
per unit area due to the gravity of the disc,
\begin{equation}\label{eq:rps}
  P_{\rm ram} > 2\pi G \Sigma_\text{disc}(R)
  \Sigma_\text{cold}(R). 
\end{equation}
Here $\Sigma_\text{disc}$, $\Sigma_\text{cold}$ are the surface densities of the
galactic disc (stars plus cold gas) and of the cold gas disc, respectively.

The discs of stars and gas are modelled by 
an exponential surface density profile given by
$\Sigma(R) = \Sigma_0 \exp(-R/R_\text{d})$
where $\Sigma_0$ is the central surface density and~$R_\text{d}$ is the  scale length
of the
disc. This scale length is estimated as  
$R_\text{d}=(\lambda/\sqrt{2})R_\text{vir}$ \citep{mmw98},
where $\lambda$ is the spin parameter of the DM halo in which
the galaxy resides.
Starting from
condition~\eqref{eq:rps} it can be shown that~RPS will remove from a galaxy all
cold gas beyond a stripping radius~$R_\text{s,cold}$ given by
\begin{equation}\label{eq:rstrip}
  R_\text{s,cold}^\text{RPS} = -0.5 R_\text{d} \ln \left( \frac{P_{\rm ram}}{2\pi G \Sigma_\text{0,disc}
  \Sigma_\text{0,cold}} \right),
\end{equation}
where $\Sigma_\text{0,disc}$ and $\Sigma_\text{0,cold}$ are the central surface
densities of the stellar disc and of the cold gas disc, respectively. We assume
for simplicity that both disc components have the same scale length.

To account for the effect of tides, at each simulation
snapshot
the value for~$R_\text{s,cold}^\text{RPS}$ obtained from equation~\eqref{eq:rstrip} is compared
to~\rdm, which determines the stripping radius due to TS, 
$R_\text{s,cold}^\text{TS}$; 
if the former is smaller than the latter we 
set~$R_\text{s,cold}=R_\text{s,cold}^\text{RPS}$. 
If the resulting~$R_\text{s,cold}$
is smaller than the current value of the cold gas disc radius,~$R_\text{cold}$, all the gas beyond~$R_\text{s,cold}$
is stripped away. 
The stripped gas is added to the hot gas component of the central
galaxy (either of a main halo or a subhalo).
If the latter is 
central galaxy of a subhalo already stripped of halo gas, the stripped gas goes
to the ICM.

After a stripping event produced by RPS and/or TS, the
remaining disc gas is assumed to form an exponential disc 
truncated at a
radius~$R_\text{cold}^\text{new}=R_\text{s,cold}$ 
and with a new scale length defined as 
$R_\text{d,cold}^\text{new}=R_\text{cold}^\text{new}/7$ 
(assuming that $99$ per cent of the cold gas disc is contained within $7*R_\text{d,cold}$).

\subsection[]{Tidal stripping of stars}
\label{sec:TSstars}

Unlike~RPS,~TS may also affect the stellar components of a satellite galaxy. To
process the stellar stripping we consider the disc and bulge separately. In the
case of the disc, we compare the current value 
of~\rdm, with the size of the
stellar disc, $R_{\rm disc}$. 
If $R_{\rm disc} > r_\text{DM}$, we assume that the galaxy loses
all stars beyond~\rdm~and that the remaining disc stars still form an
exponential disc, 
truncated at $R_{\text{s},\text{disc}} = r_\text{DM}$,
but with the stellar mass redistributed with a new scale length $R_{\text{d},\text{disc}} = R_{\text{s},\text{disc}}/7$, as it is considered for the cold gas disc.
The scale length of the galaxy disc composed by cold gas and stars is
then defined from the mass-weighted scale length of their respective 
components as $R_\text{d}=(M_\text{cold}\,R_\text{s,cold}+M_{\text{disc}}\,R_{\text{s},\text{disc}})/(M_\text{cold}+M_{\text{disc}})$.

If a bulge component is present, to evaluate the TS of its stars we assume that
they are distributed according to a \citet{hernquist1990} profile,
\begin{equation}
  \rho(r) = \frac{M_\text{bulge}}{2\pi}\, \frac{a_\text{b}}{r}\, \frac{1}{(r + a_\text{b})^3}
\end{equation}
where $a_\text{b}$ is a scale length related to the bulge half-mass radius by $r_\text{b,h}=
(1 + \sqrt{2}) a_\text{b}$. To calculate~$r_\text{b,h}$ 
we follow the procedure outlined in
\citet{cole2000} 
and adapted to our model in \citet{munnozarancibia2015}.
If the radius that contains the $99$ per cent of the bulge mass, given by
$r_\text{bulge}=198.5*a_\text{b}$, 
satisfies the condition $r_\text{bulge} > r_\text{s,bulge}$ 
with $r_\text{s,bulge}=r_\text{DM}$, 
we assume that the stars in the bulge 
beyond
$r_\text{s,bulge}$
are stripped and that the remaining bulge stars are
redistributed still following a Hernquist profile truncated at
$r_\text{s,bulge}$, with a new scale length
defined as $a_\text{b}^\text{new}=r_\text{s,bulge}/198.5$.

The stars removed by~TS
are assumed to orbit freely within the main host halo.
For practical purposes they are assigned to a new `stellar halo' component of
the corresponding central galaxy. 
As these halo stars evolve and die, they inject
gas and metals directly into 
the ICM.

\section{Supernovae feedback}
\label{sec:snfeedFIRE}

In previous versions of~\sag,
the amount of reheated mass produced by the SNe arising 
in each star forming event 
is assumed to be 
\begin{equation}
\Delta M_{\rm reheated} = \frac{4}{3} \epsilon {\frac{\eta E_\text{SN}}{V_{\rm vir}^2}} \Delta M_{\star},
\label{eq:feedbackSN}
\end{equation}
where $\eta$ is the number of
SNe generated from the stellar population of mass $\Delta M_\star$ formed,
$E_\text{SN}=10^{51}\,{\rm erg}$ is the energy released
by a SN, $V_{\rm vir}$ is the virial velocity of the host (sub)halo, which
is a measure of its potential well, and $\epsilon$ is the
SNe feedback 
efficiency, i.e., 
a free parameter that controls the amount of cold gas reheated by the
energy generated by SNe. 
The number of SNe 
depends on the initial mass function (IMF) adopted and for core collapse supernovae (SNe CC) is
estimated as
\begin{equation} 
\eta = \frac{\int^{\infty}_8 \phi(m)\,\,{\rm d}m}{\int_0^{\infty} \phi(m)\, m\,\,{\rm d}m}, 
\end{equation}
\noindent where $m$ is
the stellar mass and $\phi(m)$ is the IMF. 
This quantity
is constant for an universal IMF, like the Chabrier IMF
\citep{Chabrier03}
adopted here.
The energy and metals generated by SNe and stellar winds produced
by progenitors in different mass ranges are released with different time-scales
depending on the lifetime of the progenitors, as given by \citet{pm93}.   

In order to satisfy observational constraints at high redshifts, we found it
necessary to modify this model of SNe feedback. 
\citet{hirschmann16} find that the observed trends in galaxy assembly 
are reproduced by replacing 
the `fiducial' feedback scheme implemented in \textsc{Gaea}
model 
with a parametrization of the mass-loading factor similar to
the one inferred by
\citet{muratov15} from the analysis of the FIRE simulations.
We use this information
to modify the estimation of the reheated mass (eq.~\ref{eq:feedbackSN})
by simply adding new factors that take into account the dependence on redshift and an additional modulation with virial velocity. 
Thus, the feedback scheme in the
current version of~\sag~produces a reheated gas mass given by
\begin{equation}
\Delta M_{\rm reheated} = \frac{4}{3} \epsilon {\frac{\eta E_\text{SN}}{V_{\rm vir}^2}} \,(1+z)^{\beta}\,\left(\frac{V_{\rm vir}}{60\,{\rm km}\,{\rm s}^{-1}}\right)^{\alpha_\text{F}}\Delta M_{\star},
\label{eq:feedfire}
\end{equation}
\noindent where 
the exponent $\alpha_\text{F}$ takes the values $-3.2$ and $-1.0$ 
for virial velocities
smaller and larger than $60\,{\rm km}\,{\rm s}^{-1}$, respectively.
Although the fit provided by \citet{muratov15}
indicates that the value of $\beta$ is 1.3,
we consider this exponent to be a free parameter of
the model~\sag, allowing it to absorb
aspects of physical processes that are not properly captured by the model
and helping to reproduce the observational constraints imposed,
detailed in Section~\ref{sec:parameters}.
In section~\ref{sec:SAG-const}, we show the results of
this procedure and also discuss the impact of assigning to
$\beta$
the value
found by \citet{muratov15}.

The reheated gas is transferred from the cold to the hot phase, 
subsequently returning to the cold phase through gas cooling 
taking place in both central and satellite galaxies.
However, to avoid an excess of stellar mass at high redshifts, 
some of the hot gas must be ejected out of the halo reducing
the hot gas reservoir available for gas cooling \citep{guo11, henriques13,
hirschmann16}. Hence, we also consider the energy conservation argument
presented by \citet{guo11} to calculate the ejected hot gas mass
\begin{equation}
\Delta M_{\rm ejected}= \frac{\Delta E_\text{SN} - 0.5\,\Delta M_{\rm reheated}\,V_{\rm vir}^2}{0.5\,V_{\rm vir}^2},
\label{eq:EnergyCons}
\end{equation}
\noindent where $\Delta E_\text{SN}$ is the energy injected by massive stars 
which we model in a way similar to the modified reheated mass, 
as also done by \citet{hirschmann16},
\begin{equation}
\Delta E_{\rm SN} = \frac{4}{3} \epsilon_\text{ejec} {\frac{\eta E_\text{SN}}{V_{\rm vir}^2}} \,(1+z)^{\beta}\,\left(\frac{V_{\rm vir}}{60\,{\rm km}\,{\rm s}^{-1}}\right)^{\alpha_{\rm F}}\Delta M_{\star}\,0.5\,V_\text{SN}^2. 
\label{eq:ejecfire}
\end{equation}
\noindent Here, $\epsilon_\text{ejec}$ is the corresponding efficiency,
considered as another free parameter of the model, 
and $0.5\,V_\text{SN}^2$ is 
the mean kinetic energy of SN ejecta per
unit mass of stars formed. Following the analysis of \citet{muratov15},
we adopt the fit for 95th percentile wind velocity as a function of the
virial velocity of the halo
(see their fig. 8 and eq. 10), such that  
$V_\text{SN}=1.9\,V_\text{vir}^{1.1}$.

In the gradual starvation scheme implemented in~\sag, galaxies
keep their hot gas halo when they become satellites. These
haloes are reduced by gas cooling and environmental effects (RPS, TS),
but they  also can be reconstructed by the injection of reheated gas
or the reincorporation of ejected gas. 
The ejected gas mass is assumed to
be re-incorporated back onto the (sub)halo from which it was expelled
within a time-scale that depends on the inverse of (sub)halo mass, $M_{\rm vir}$,
as assumed by \citet{henriques13} in order to reproduce the
observed evolution of the galaxy stellar mass function.
Thus, the reincorporated mass is given by
\begin{equation}
\Delta M_{\rm reinc}= \gamma\,\Delta M_\text{ejected}\,\frac{M_{\rm vir}}{10^{10}\,{\rm M}_{\odot}},
\label{eq:reinc}
\end{equation}
\noindent where the parameter $\gamma$ regulates the efficiency of the process
and is also a free parameter of \sag~model.

In all the above parametrizations, we have added the energy injected by
SNe Ia. We adopt the single degenerate
model in which a SN Ia occurs by carbon deflagration in C--O
white dwarfs in binary systems whose components have masses
between $0.8$ and $8\,{\rm M}_{\odot}$ \citep{gr83}.
We implement the formalism presented by \citet{Lia2002},  
choosing a fixed fraction of binary systems, $A_\text{bin}=0.05$.
As for SNe CC, we also consider the lifetime of the progenitors.

SNe CC and SNe Ia, together with low- and intermediate-mass stars,
contribute with metals that pollute the cold and hot gas, affecting
the cooling rates and regulating also in this way the subsequent events
of star formation.  
Details on the chemical model implemented in~\sag~are given in \citet{cora2006},
with the latest updates on chemical yields presented in \citet{Gargiulo15}.

\section{Calibration of the model SAG}
\label{sec:calibration} 

The physical processes included in \sag~model, as occurs for all SAMs, 
involve several free parameters which regulate the modelling of these 
processes.  
Proper values of these free parameters are found by imposing constraints 
given by observed galaxy properties that the SAM should satisfy. 

\subsection{Calibration process}
\label{sec:parameters} 

The calibration process of \sag~model is performed by implementing the Particle
Swarm Optimization (PSO) technique presented in \citet{ruiz2015}. 
This numerical tool allows us to tune the free parameters via the exploration of
the \sag~parameter space by random walks of a set of `particles' (the
\textit{swarm}, equivalent to the chains in Monte Carlo Markov techniques)
that share information between them, thus determining the evolution of the
exploration from both their individual and collective experience. 
By comparing the model results against a given set of observables, the PSO
method yields a set of best-fitting values for the free parameters.

For this current version of ~\sag, we consider nine free parameters for the
calibration process.
The corresponding equations that involve these free parameters and a brief
description of the related physical processes can be found in \citet{ruiz2015},
except those related to the new SN feedback and ejection/reincorporation scheme
implemented in the current version of~\sag, presented in
Section~\ref{sec:snfeedFIRE}.   
The free parameters are
the star formation efficiency ($\alpha$), the
efficiency of SN feedback from stars formed in both the disc and the bulge
($\epsilon$, equation~\ref{eq:feedfire}), the efficiency of ejection of gas from the
hot phase ($\epsilon_\text{ejec}$, equation~\ref{eq:ejecfire}) and of its
reincorporation ($\gamma$, equation~\ref{eq:reinc}), the exponent that regulates the
redshift evolution of the mass-loading factor of the reheated and ejected mass
($\beta$, equations~\ref{eq:feedfire} and~\ref{eq:ejecfire}), the growth
of super massive BHs and efficiency of AGN feedback ($f_\text{BH}$ and
$\kappa_\text{AGN}$, respectively), the factor involved in the distance scale
of perturbation to trigger disc instability events ($f_{\rm pert}$), and the
fraction that determine the destination of the reheated cold gas
($f_\text{hot,sat}$) introduced in Section~\ref{sec:hotgasstrip}. 

The set of observables used for the calibration process is the stellar mass
functions (SMF) at $z=0$ and $z=2$, the star formation rate distribution
function (SFRF), the fraction of mass in cold gas as a function of stellar mass
(CGMF) and the relation between bulge mass and the mass of the central
supermassive BH (BHB).
For the SMFs, we adopt the compilation data used by \citet{henriques_mcmc_2015},
which for $z=0$ is a combination of the SMF of the 
SDSS
from \citet{baldry08} and \citet{li_smf_2009}, and of the Galaxy And
Mass Assembly (GAMA) from \citet{baldry_smf_2012}, while for $z=2$  is a
combination of the data of the Cosmic Evolution Survey (COSMOS) from
\citet{dominguezsanchez11}, the Ultra Deep Survey (UltraVISTA) from
\citet{muzzin13} and \citet{ilbert13}, and the FourStar Galaxy Evolution Survey
(ZFOURGE) from \citet{tomczak14}.
For the SFRF, which  is the number density of galaxies in a certain SFR
interval, we use data from a flux-limited sample of galaxies observed with the
\textsl{Herschel} satellite which gives the total (IR+UV) instantaneous SFR for the
redshift interval $z \in [0.0,0.3]$, as presented in \citet{gruppioni15}.
This observable is compared with the \sag~output at $z=0.14$, which is the closest
output of the model to $z=0.15$, the mean value of the redshift range of observations.
For the CGMF, we adopt observational data from \citet{boselli14}, which is
based on a volume limited sample, within the range 
$\rm{log}(M_{\star}[{\rm M}_{\odot}]) \in [9.15,10.52]$ in stellar mass.  
Finally, for the BHB relation, we combine the datasets from
\cite{mcconnell_bhb_2013} and \cite{kormendy_bhb_2013}. 
This particular set of constraints and the observational data adopted were
defined as a common set for calibration during the Cosmic CARNage
workshop
aimed to compare different calibrated galaxy formation models based on the same
cosmological DM-only simulation \citep{Knebe18}. 
This comparison project was initiated in the
nIFTy Cosmology workshop \citep{knebe15}.

Due to the large size of the simulation, the computing time needed to run the model several times becomes quickly prohibitively long. Hence, 
the calibration is carried out by running \sag~with the
merger trees extracted from a smaller box of the whole MDPL2 simulation that
constitute a representative sample of all the merger trees contained in the
$1\,h^{-1}\,{\rm Gpc}$ side-length box.   
We divide the full volume of the MDPL2 simulation in
$9^3$ parallelepiped sub-boxes, each of them with roughly the same number density of
haloes of the whole volume.
Despite each of these samples constitutes only $\sim 0.137$ per cent of the
whole simulation, they are large enough to be considered themselves simulations
from which a statistically significant set of merger trees can be extracted for
the further application of the~\sag~model.
Therefore, we compared the halo mass function for all these subboxes 
with the one obtained for the whole simulation, choosing the most similar
one simply by visual inspection, selecting for this case a subbox with an
effective length side of $111.19\,h^{-1}\,{\rm Mpc}$.

\begin{table}
  \centering
  \begin{tabular}{l r r}
    \hline
    Parameter &  Best-fitting value\\
    \hline
    $\alpha$           & 0.04                    \\
    $\epsilon$         & 0.33                    \\
    $\epsilon_{\rm ejec}$  & 0.022               \\
    $f_{\rm BH}$       & 0.06                    \\
    $\kappa_{\rm AGN}$ & 3.02 $\times$ 10$^{-5}$ \\
    $f_{\rm pert}$     & 14.56                    \\
    $\gamma$      & 0.055                    \\
    $f_{\rm hot,sat}$      & 0.277                    \\
    $\beta$      & 1.99                    \\
    \hline                                                                
  \end{tabular}
  \caption{Best-fitting values of the free parameters of \sag~model obtained with the~PSO technique. This set of values is
    obtained from the application of \sag~to the merger trees 
    of the subbox selected from the MDPL2 simulation. 
     }
      \label{table1}
\end{table}

The values of the parameters obtained from the calibration of the model to
match the aforementioned set of observables are shown in Table~\ref{table1}.

\subsection[]{Functions and relations used as constraints}
\label{sec:SAG-const}

The new features included in~\sag~allow the model 
to reach a good match with the observational constraints
imposed to calibrate the model, i.e.,  
the SMF at $z=0$
and $z=2$, the SFRF at $z=0.14$, and the CGMF and BHB and relations at $z=0$,
as 
we detail below.
In addition to the results of the calibrated model, 
we also evaluate the impact on these functions and relations of fixing
the value of the parameter $\beta$
involved in the redshift dependence of the reheated and ejected mass. 
For this alternative model, we adopt the value given 
by the fit of \citet{muratov15}, i.e. $\beta=1.3$,
leaving the values of the remaining free parameters of \sag~unchanged.
We refer to this slightly modified version of the model as \sagb.

It is worth noting that the small value of the parameter $\gamma$ that regulates the amount of ejected gas that is reincorporated makes the results of the model insensitive to the fate of the reincorporated gas, i.e. subhalo from which it was ejected or main host halo where the satellite resides. We tested the latter possibility and found no change in the galaxy properties shown in this work.

\subsubsection[]{Stellar mass functions}

\begin{figure}
  \centering
  \includegraphics{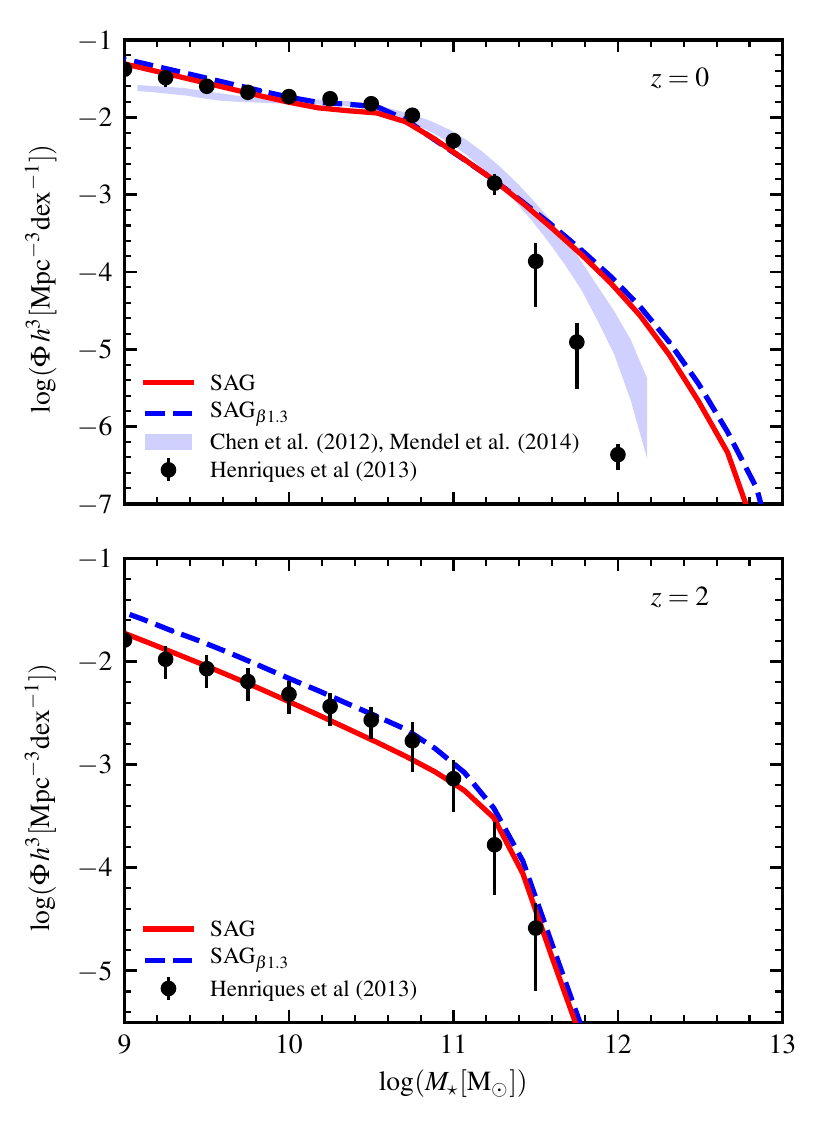}
  \caption{
Stellar mass functions used as constraints to calibrate the \sag~model.
Results obtained from the calibrated model \sag~with the best-fitting
values given in Table~\ref{table1} are represented by red solid lines.
{\it Top panel}:
Stellar mass function at 
$z=0$. 
{\it Bottom panel}:
Stellar mass function at $z=2$. 
Observational data
compiled by \citet{henriques_mcmc_2015} is represented by filled 
circles with error bars.
The agreement between model and observations is rather good, except
for the excess at the high-mass end at $z=0$.
This excess becomes less evident when comparing with 
estimations compiled and analysed 
by \citet[][grey shaded area]{Bernardi17}, which include data from 
\citet{Chen12} and \citet{Mendel14}.
In both panels, results obtained from model \sagb~are shown by a blue dashed line. 
The SMF at $z=0$ is not significantly affected by the change in the value of the 
parameter $\beta$,
whereas this gives rise to an excess in the low-mass end of the SMF at $z=2$.
}
  \label{fig:SAGconstraints_SMF}
\end{figure}

In Fig.~\ref{fig:SAGconstraints_SMF}
both the local SMF (top panel) and the SMF at $z=2$ (bottom panel)
generated by the model are shown, where  
the break and low-mass end given by observations
are reproduced. 
This is a due to the new implementation of SN feedback.
The classical estimation of the reheated mass
considered in previous versions of \sag~(equation~\ref{eq:feedbackSN}) 
allows a good match of the SMF only at $z=0$.
The addition of ejection acting on the hot phase
and the redshift dependence of both reheated and ejected mass
(equations~\ref{eq:feedfire} and \ref{eq:ejecfire})
become key ingredients to avoid an excess of star formation 
overproducing stellar mass at high redshifts.
This good agreement is obtained by a stronger redshift dependence than
the one found by \citet{muratov15}, whose parametrization is characterized
by a parameter $\beta=1.3$.
From Table~\ref{table1}, which shows the best-fitting values 
of the free parameters of \sag,
we can see that the exponent that regulates this  
redshift dependence 
takes a value 
$\beta=1.99$.
On the other hand, model \sagb~produces an
excess in the low-mass end of the
SMF at $z = 2$, as shown by the dashed line
in Fig.~\ref{fig:SAGconstraints_SMF}. This is
due to the higher levels of
SFR this model features at higher redshifts, as shown and
discussed in Section~\ref{sec:SAG-pred-sfr}. However,
the SMF at $z = 0$ is not significantly affected by
the change in the value of the parameter $\beta$. 

None of the improvements included in \sag~leads to a better fit of the massive
end of the SMF at $z=0$, which shows an excess of galaxies with stellar mass
$M_{\star} \gtrsim 2\times 10^{11}\,{\rm M}_\odot$. 
This excess 
is not avoided by the implementation
of the more locally efficient radio mode AGN feedback 
\citep{henriques_mcmc_2015}.
In any case, AGN feedback allows to recover the break of the SMF
which characteristic stellar mass highly depends
on the feedback efficiency regulated by the two free parameters 
involved in the AGN feedback ($\kappa_{\rm AGN}$ and $f_{\rm BH}$).
However, the slope 
of the high-mass end of the SMF 
is insensitive to the AGN feedback efficiency, 
as we have checked by manually exploring values of the parameters
related with this process around the best-fitting values.

It is important to take into account the uncertainties in the
observational data used as constraints, 
considered in the comparison plot of the SMF at $z=0$. 
As specified in Section~\ref{sec:parameters}, the compilation made by 
\citet{henriques_mcmc_2015} 
includes data from \citet{li_smf_2009} for $z=0$.
As discussed by \citet{Bernardi17}, this particular set of data gives
the lowest values of the comoving number density of galaxies for
stellar masses $\gtrsim 10^{11}\,{\rm M}_\odot$ in comparison with
other results presented in the literature because of 
the use of inappropriate algorithms for estimating the observed flux
of the most luminous galaxies and 
inadequate assumptions about the stellar population modelling.
While systematic effects on the SMF
related with photometry account for differences
of only 0.1 dex, systematics arising from different treatments of the
stellar population can be as large as $0.5$ dex.
The estimations compiled and analysed by \citet{Bernardi17},
which include data from \citet{Chen12} and \citet{Mendel14}, 
are represented by a grey shaded area.
When comparing with these sets of data,
the excess in the comoving number density of massive galaxies 
predicted by \sag~becomes less pronounced.

\subsubsection[]{Star formation rate function}

\begin{figure}
  \centering
  \includegraphics{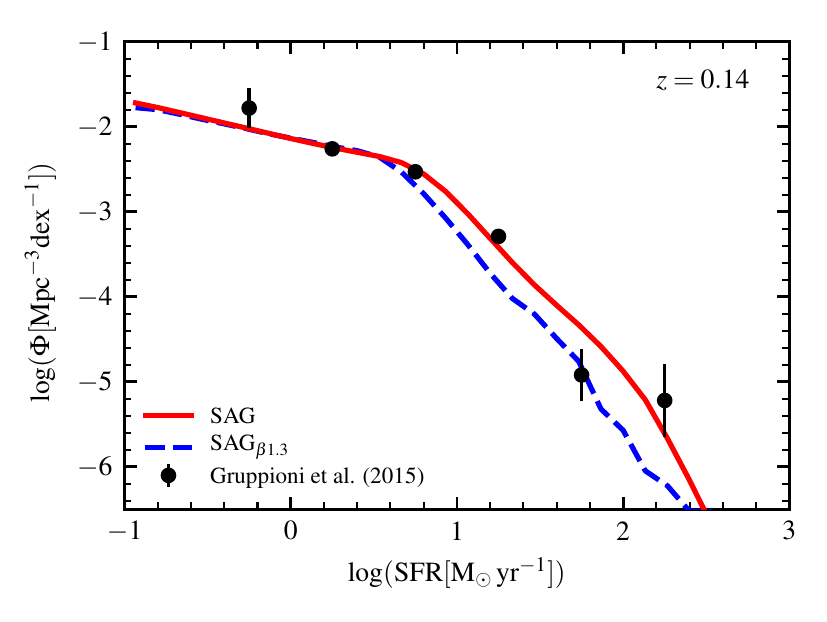}
  \caption{
Star formation rate distribution function at $z=0.14$
used as constraint to calibrate the \sag~model.
Results obtained from the calibrated model \sag~with the best-fitting values
given in Table~\ref{table1}, and from the \sagb~model are shown by red solid
and blue dashed lines, respectively.
The latter produces a decrease in the number density of high SFR values.
Model results are compared with observations by \citet{gruppioni15} 
represented by filled circles with error bars.
}
  \label{fig:SAGconstraints_SFRF}
\end{figure}

In addition to the SMF at different redshifts, 
the SFRF at $z=0.14$ 
helps to recover the observed  
evolution of SFR and mass growth in
galaxies. 
Fig.~\ref{fig:SAGconstraints_SFRF}
presents 
the SFRF at $z = 0.14$ for the galaxy population
generated by the calibrated model \sag~(solid line).
The agreement with the 
observations by \citet{gruppioni15} 
is very good for intermediate
values of the SFR, in the range 
$1\,{\rm M}_{\odot}\,\text{yr}^{-1} \lesssim {\rm SFR} \lesssim 20\,{\rm M}_{\odot}\,\text{yr}^{-1}$,
being slightly underpredicted (still within the dispersion of observed data)
for lower values. \sag~gives higher number density
of galaxies for $\text{SFR} \approx 60\,{\rm M}_{\odot}\,\text{yr}^{-1}$, 
reaching again a good agreement for the highest SFR bin 
($\approx 200\,{\rm M}_{\odot}\,\text{yr}^{-1}$).
This particular behaviour of our model is quite similar to the predictions
at low redshifts provided by other SAMs considered 
by \citet{gruppioni15} in a direct comparison with their data.
The agreement shown here for SFR above $100\,{\rm M}_{\odot}\,\text{yr}^{-1}$ 
is better
than the one reached by the `FIRE feedback model' analysed 
by \citet{hirschmann16}
from which the current feedback scheme implemented in \sag~is inspired.
This better agreement could have been achieved simply because the
SFRF was used to calibrate the model. 
However, it is not always possible
to recover the complete behaviour of an observational constraint,
as it is the case of the massive end of the SMF at $z=0$. 
This tension that emerges during the calibration procedure denotes
the need of improvement in the physical processes modelled and/or
inconsistencies between different sets of observed galaxy properties. 
On the other hand, 
model \sagb~produces a decrease in the number density of high SFR values,
as shown by the dashed line.

\subsubsection{Cold gas mass fraction}

\begin{figure}
  \centering
  \includegraphics{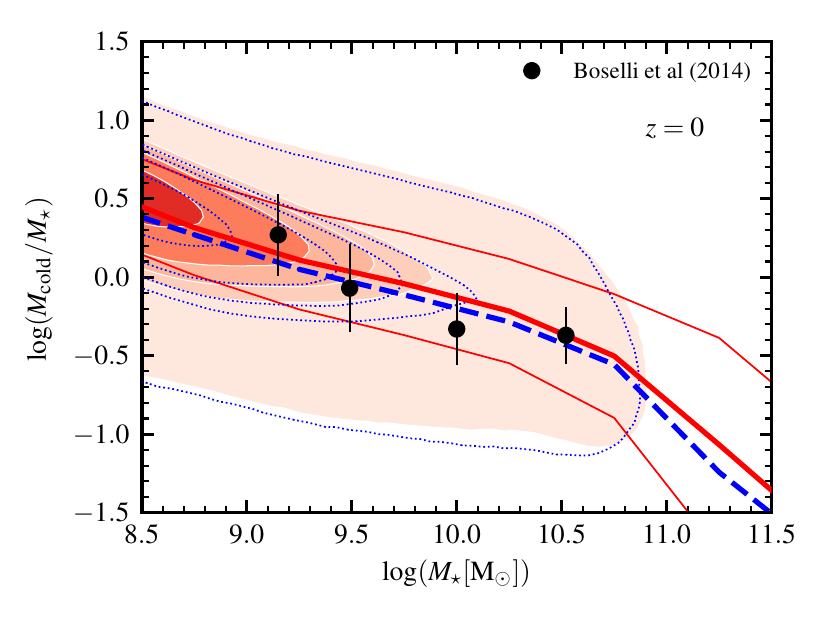}
  \caption{
Fraction of mass in cold gas as a function of stellar mass at $z=0$ 
used as constraints to calibrate the \sag~model.
Results obtained from the calibrated model \sag~with the best-fitting
values given in Table~\ref{table1} are represented by red solid lines
and reddish filled contours. 
Red thin solid lines denote the standard deviation around the mean
(thick red solid line).
In both cases, 
the contour levels are:
$[0.01, 0.19, 0.26, 0.38, 0.68, 0.95, 0.997]$ in terms of the maximum 
number density of points of each sample.
CGMF is almost unaffected by adopting $\beta=1.3$,
as shown by results from the \sagb~model represented by a blue dashed line
and empty dotted contours.
This relation is compared
with observations by \citet[][filled circles with error bars]{boselli14}.
}
  \label{fig:SAGconstraints_CG}
\end{figure}

The fact that galaxy growth and star formation quenching
are well captured by \sag~becomes evident in the cold gas content
of model galaxies (Fig.~\ref{fig:SAGconstraints_CG}).
The CGMF obtained from \sag~also behaves as expected within
the stellar mass range considered for the calibration of the model
(${\rm log}(M_{\star}[{\rm M}_{\odot}])\in [9.15,10.52]$).
Their mean values (thick solid line)
are in good agreement with observational data from
\citet[][filled circles with error bars]{boselli14};
thin solid lines represent the standard deviation around the mean and
the shaded contours show the distribution of model values.
For high stellar masses (${\rm log}(M_{\star}[{\rm M}_{\odot}]) > 10.52$),
\sag~predicts lower fractions of cold gas with respect to those
expected from a simple extrapolation of the 
trend found at lower masses.
However, these values are still contained within the range 
of cold gas fractions inferred
from the atomic and molecular gas fractions presented by 
\citet[][see their fig.~3]{Saintonge16}, obtained from data of
the Arecibo Legacy Fast ALFA (ALFALFA), 
the GALEX Arecibo SDSS Survey (GASS) and the CO Legacy Database for GASS 
(COLD GASS) surveys.
The same trend is obtained with model \sagb.

\subsubsection[]{Black hole-bulge mass relation}

\begin{figure}
  \centering
  \includegraphics{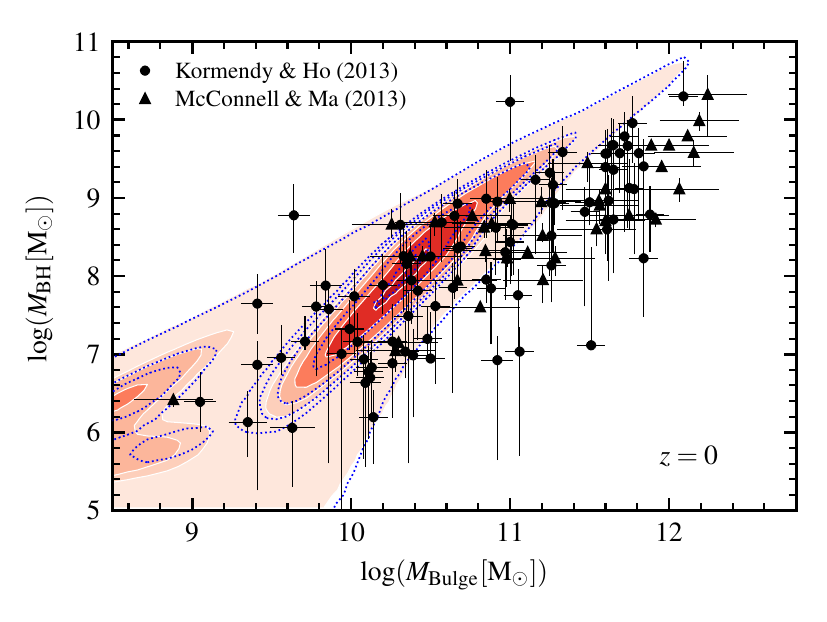}
  \caption{
Relation between black hole and bulge mass at $z=0$
used as constraints to calibrate the \sag~model.
Results obtained from the calibrated model \sag~with the best-fitting
values given in Table~\ref{table1} are represented by reddish filled contours, while
results obtained from the \sagb~model are shown by empty dotted contours. 
In both cases, the
contour levels are the same as those of Fig.~\ref{fig:SAGconstraints_CG}.
The BHB relation is almost no affected when
adopting $\beta=1.3$.
Observational data are taken from 
\citet[][black circles]{kormendy_bhb_2013}
and 
\citet[][black triangles]{mcconnell_bhb_2013}.
}
  \label{fig:SAGconstraints_BHB}
\end{figure}

The BHB relation displayed by the calibrated \sag~model
(Fig.~\ref{fig:SAGconstraints_BHB}, filled contours) is in good general agreement
with the observational trend denoted by data from 
\citet{mcconnell_bhb_2013} and \citet{kormendy_bhb_2013}.
However, black holes 
seem to have higher masses
than expected for large bulge masses,
but still within
the allowed ranges defined by 
the dispersions of the observational data.
This trend is preserved in
model \sagb~(empty dotted contours).
This is a result of the restriction imposed by both the high mass end of the 
SMF at $z=0$ and the BHB relation.
During the calibration process, 
the parameter that controls the growth of super massive BHs, $f_{\rm BH}$,
takes a high value in order to make the AGN feedback as effective as possible.
Due to the existing degeneration between the $f_{\rm BH}$ and $\kappa_{\rm AGN}$
parameters, a similar behaviour of the high-mass end of the SMF at $z=0$ can be achieved by 
taking lower values of $f_{\rm BH}$ at the expense of increasing the feedback 
efficiency, $\kappa_{\rm AGN}$. 
Nevertheless, large values of $f_{\rm BH}$ are preferred 
as a direct consequence of the statistical test used in the calibration process.
The chi-square calculated for each constraint 
\citep[][see their eq. 27]{ruiz2015} gives heavier 
weights to the mass bulge ranges where the bulk of the galaxies are located, 
which decreases the statistical significance of the high-mass galaxies in 
the selection of the best-fitting parameters.

\subsubsection[]{Origin of the excess in the high-mass end of the SMF at $z=0$}
\label{sec:originSMF}

The AGN feedback together with
additional modifications
introduced in the model, that is,
the redshift dependence of the reheated and ejected material
and the scalings adopted for the reincorporation of the latter
(Equations~\ref{eq:feedfire},~\ref{eq:ejecfire} and~\ref{eq:reinc}),
allow to recover downsizing
in the star formation rates, as shown later in Fig.~\ref{fig:SFRDvsz}.
Hence, there is not an excess of SF in high-mass galaxies
that could be responsible for the excess of the high-mass end of the
SMF at $z=0$. 

A test made with the semi-analytic model 
\sage~discussed in \citet{knebe17}
shows that
stars added to an intra-cluster component as a result of tidal disruption
of satellite galaxies would enhance the massive end of the SMF if
they would have merged with the central galaxy. 
In \sag, satellite galaxies are not tidally disrupted when becoming orphans as in \sage~but suffer TS. 
The stripped mass is added to the stellar halo of the corresponding central galaxy (see Section~\ref{sec:TSstars}) and
represents the intra-cluster stars. The stellar mass density of this component is
$\sim 2$ and $\sim3$ orders of magnitude lower than 
the one characteristic of the stellar mass of the whole galaxy population, for redshifts $z=0$ and $2$, respectively.
This means that the excess of the high-mass end of the SMF at $z=0$ could be attributed to inefficient stripping or disruption of satellites by effects of tides in the model.

In order to evaluate the efficiency of TS in our model,
we estimate the fraction of mass in stars contained in the stellar
halo of central spiral galaxies and compare them with results
from \citet{Merritt16}.
They estimate the stellar mass in excess of a
disc+bulge beyond 5 half-mass radius of eight spiral galaxies in the Dragonfly Nearby Galaxies Survey (DNGS)
and obtain an average halo fraction of $0.009 \pm 0.005$. 
In our model, spiral galaxies are those characterized by a ratio between stellar bulge and total stellar mass
${\rm B/T} < 0.85$; this cut allows to reproduce the observed morphological distribution \citep{conselice2006}. 
Median values of the halo fractions (estimated with respect to the total stellar mass of central galaxies, including the stellar halo) increase with stellar mass and are comprised within the range 
$\approx 3\times 10^{-5} - 7\times 10^{-4}$
for stellar masses $M_{\star} \approx 10^{10} - 10^{11} \,{\rm M}_{\odot}$. 
These median values are at least 
one order of magnitude smaller than the average halo fraction obtained
by \citet{Merritt16}. 
Although 
the difference between the methods used to obtain the mass of the stellar halo in the observations and in the model might contribute to such a discrepancy, it seems quite plausible that TS is not efficient enough in our model, as hinted by the excess in the high-mass end
of the SMF at $z=0$.
In other words, the mass of merging satellites might be higher than expected because they do not suffer enough TS. 

\begin{figure}
  \centering
  \includegraphics{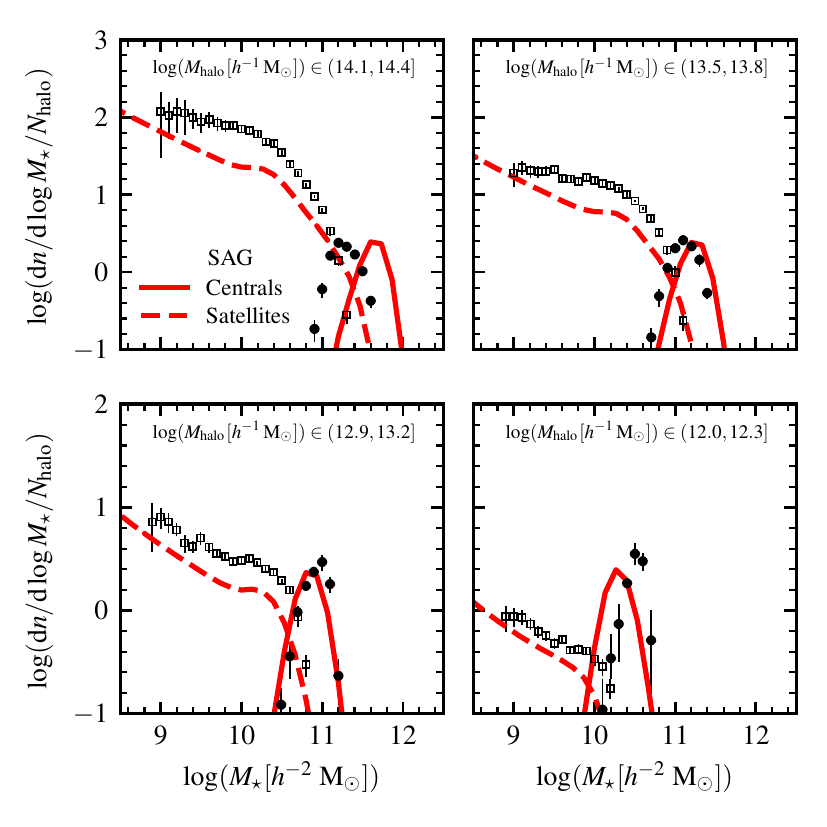}
  \caption{
Conditional stellar mass function for central (solid line) and satellite (dashed line) galaxies in model \sag. Galaxies are selected according to the mass of the main host halo they reside in as indicated in the legend of the different panels. Model results are compared with data from the group catalogues constructed by 
\citep{Yang09b} from SDSS DR4; central galaxies are identified by filled circles and satellites are represented by open squares.
}
  \label{fig:CSMFz0}
\end{figure}

Another possible cause of the excess in the high-mass end of the SMF at $z=0$ might be 
related to 
dry mergers at $z\lesssim 2$ of low-mass galaxies with massive ones.
In this context, it is worth mentioning a caveat regarding the 
merger condition adopted for orphan galaxies. As we briefly explain
in Section~\ref{sec:orbits}, it is based on a criterion that demand
an assumption about the radius of the central galaxy (assumed to be
$10$ per cent
of the virial radius of the host halo). 
The drawback of this approximate estimation 
is that it 
can be very large for cluster-size haloes. Therefore, the
integrated orbit can easily take the orphan satellite 
within the sphere determined by those radii, thus merging with its central.
In order to test this possibility, we estimate the conditional stellar mass functions for central and satellite galaxies within different ranges of halo mass in \sag, which are shown in Fig.~\ref{fig:CSMFz0}. Following \citet{Kang14}, they are compared with data from SDSS DR4 \citep{Yang09b}.
There are less model satellites than observed 
in all the halo mass ranges analysed.
This deficiency is more pronounced in more massive haloes
($\log (M_{\rm halo} [{\rm M}_{\odot}]) \in [13.5, 13.8], [14.1, 14.4]$), and
it is accompanied by a shift to higher stellar masses of the population of central galaxies.

Hence, both an inefficient TS and over-merging of satellite galaxies are responsible for the excess
of high-mass galaxies found in the SMF at $z=0$.
These drawbacks in the model are mainly related to aspects of the integration of orphan galaxies, like the treatment of TS and the merging criterion, which will be revisited (Vega-Mart\'inez et al. in preparation).

\section{Model predictions}
\label{sec:SAG-predictions}

We present galaxy properties predicted by the model,
that are not used as constraints to calibrate it.
We first focus on the evolution of the 
cosmic star formation rate density (SFRD) and 
of the specific star formation rate (sSFR).
The former is given by  
the volume-averaged sum of SF of all galaxies at any
given time, and the
latter is
defined as the ratio between the SFR 
and the stellar mass of the galaxy.
We connect the trends of the sSFR of satellites and centrals
with the stellar mass content of galaxies populating
DM haloes of different mass through
the stellar-to-halo mass relation (SHM).
According to the value of the sSFR, we classify galaxies as active and passive. 
We analyse the dependences of the fraction of passive galaxies
with stellar mass, halo mass and halo-centric distance.

\subsection[]{Star formation rate density}
\label{sec:SAG-pred-sfr}

\begin{figure}
  \centering
  \includegraphics{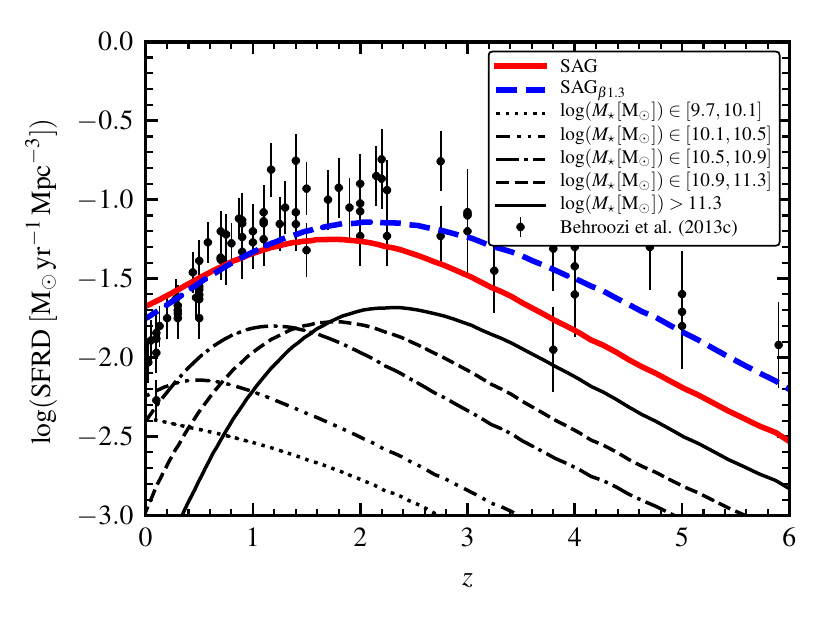}
  \caption{
Evolution of the cosmic star formation rate density for all galaxies 
(red thick solid line)
compared with observational data compiled by \citet{behroozi13c}. 
Contribution from galaxies lying in different mass ranges at $z=0$ are 
represented by thin black lines of different style, 
as indicated in the legend. 
Downsizing in stellar mass assembly becomes evident. 
Intermediate-mass galaxies  
$\log(M_{\star} [{\rm M}_{\odot}]) \in [10.1,10.5]$
are the main responsible
of the excess in the SFRD at $z=0$.
The global SFRD obtained from the calibrated model \sag~but 
adopting $\beta=1.3$ (model \sagb) is 
added for comparison, being represented by a blue thick dashed line.
Model \sagb~is characterized by higher (lower) levels of SFR at high (low)
redshifts with respect to the calibrated model \sag, in better agreement
with observations.
}
  \label{fig:SFRDvsz}
\end{figure}

The evolution of the cosmic SFRD predicted by \sag~is presented in 
Fig.~\ref{fig:SFRDvsz} (thick solid line) and is
compared with data compiled by 
\citet{behroozi13c}.
Although model results show consistency with observations for $z<4$,
they under-predict the SFRD at higher redshifts. Besides, 
the decline in the SFRD towards low redshifts is less pronounced in the model than in observations, thus resulting in an excess of SFRD at $z=0$. The peak and normalization of the SFRD vary significantly among different galaxy formation models  
\citep[e.g. ][]{guo16}. 

The low SFRD predicted at high redshifts is 
a consequence of the redshift dependence of the reheated and ejected gas fractions in 
the new feedback scheme implemented in \sag~(Equations~\ref{eq:feedfire} and~\ref{eq:ejecfire}), 
an effect that 
becomes even stronger in our model because of the preferred high value
of $\beta$ in the calibration process, as discussed
in Section~\ref{sec:SAG-const}. The lower value 
suggested by the fit found by \citet{muratov15}, e.g. $\beta=1.3$,
allows to reconcile model predictions with observed SFRD at high redshifts,
as shown in Fig.~\ref{fig:SFRDvsz}
by model \sagb~(thick dashed line).
This better match takes place at
the expense of increasing 
the number
density of low-mass galaxies 
at $z=2$ (dashed line in bottom panel of Fig.~\ref{fig:SAGconstraints_SMF}).
Therefore, it is not possible to satisfy both observational constraints
simultaneously.
This tension between the observed 
redshift
evolution of the SFRD and 
$z=2$ galactic SMF has been also 
noted by \citet{hirschmann16}
from their analysis of the evolution of the correlation of sSFR with
the stellar mass (main sequence of star-forming galaxies, see their fig.~9);
they find that ejective models that
successfully reproduce the measured
evolution of SMF
tend to under-estimate the sSFR at
high redshifts. 
In the same line, the recent work of \citet{Rodrigues17}, based on the exploration of the
parameter space of a version of the \textsc{GALFORM} semi-analytic model, 
shows that the evolution of the SMF
is recovered by a particular set of parameters 
that favours larger SN feedback efficiency at higher redshifts, although 
the rise of the SFRD with redshift in the range $0.5 <z < 1$ is not as steep as observed.

\begin{figure*}
 \centering
  \includegraphics{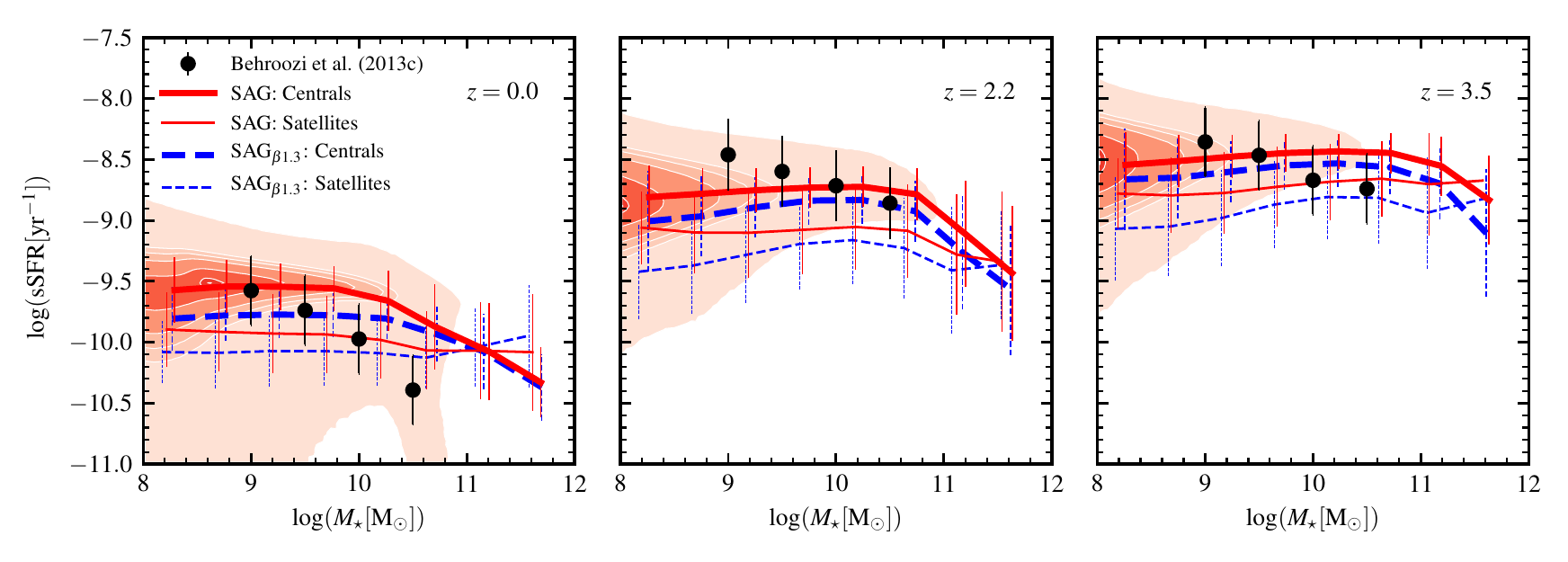}
  \caption{Specific star formation rate as a function of stellar mass
for all galaxies in the calibrated model \sag~(filled contours, considering the same
levels as the Fig.~\ref{fig:SAGconstraints_CG})
at different redshifts ($z=0$, $2.2$ and $3.5$).
Mean values of star-forming galaxies 
(${\rm sSFR}>10^{-10.7}\,{\rm yr}^{-1}$) 
are estimated for central and satellite galaxies 
(red solid thick and thin lines, 
respectively); error bars represent $1\sigma$-standard deviation around the 
mean. 
They are compared with redshift dependent fits to
observational data compiled by
\citet{behroozi13c}
(filled circles with error bars). 
The gradual removal of hot gas in satellites through RPS and TS allows 
satellite galaxies
to have a behaviour similar as central ones. 
Results of model \sagb~for both central and satellite galaxies
are represented by blue dashed thick and thin lines, respectively.
}
  \label{fig:sSFRmstarz}
\end{figure*}

The difficulty 
in satisfying simultaneously the observed evolution of the SFRD and the SMF at high redshift
also emerges in the analysis of observational results.
It is well known that
the star formation history inferred from the observed evolution of the stellar mass differs from instantaneous indicators of star formation, being the former $\approx 0.6$ dex smaller than the latter at $z=3$ \citep{Wilkins2008}.
Thus, the integrated star formation history implies a local
stellar mass density in excess of that measured.
According to recent studies \citep{Madau2014}, this discrepancy in the
stellar mass density can be $\approx 0.2$ dex for $z\lesssim 3$ depending on the data considered,
being smaller than previously estimated.
In any case, 
these
discrepancies can arise for 
several reasons such as inaccurate dust extinction corrections, underestimation
of stellar masses due to the outshining of old stellar populations in star
forming galaxies, or the evolution of the integrated stellar IMF in galaxies.
Both the star formation history implied by instantaneous indicators, which are typically dominated by very massive stars, and the star formation history inferred from the evolution of
the average stellar mass density are affected by the IMF assumed. The discrepancies between them might be mitigated when considering an evolving IMF that is
top heavy at high redshifts \citep{Wilkins2008}.

The stronger suppression of SF at
high redshifts is evident in galaxies of all masses 
being more pronounced in low-mass galaxies, thus 
allowing to reach the antihierarchical assembly of stellar mass.
This becomes evident from the evolution of the
galactic SMF presented by
\citet[][see their fig.~1]{hirschmann16};
for ejective feedback models,
high-mass galaxies 
($M_{\star} \gtrsim 10^{10}\,{\rm M_{\odot}}$) are already 
in place
at $z\approx 2-3$, while the population of low-mass galaxies keeps
increasing towards lower redshifts, in agreement with the trend denoted by
observational measurements.
Downsizing in galaxy assembly is also produced by \sag, as
can be appreciated from the contribution
to the global SFRD of galaxies in different mass ranges selected at $z=0$,
as indicated by thin black lines of different style in Fig.~\ref{fig:SFRDvsz}.
The peak of the SFRD of galaxies with stellar mass 
$M_{\star}> 10^{11}\,{\rm M}_{\odot}$
is located at $z \approx 2$, while intermediate-mass galaxies
$(\log(M_\star [{\rm M}_\odot]) \in [10.5,10.9])$ reach the
peak at $z \approx 1$. Lower-mass galaxies are still in the regime
of increasing SFRD while approaching to $z=0$. 
Downsizing is also 
present in model \sagb~but with a change in the SF history of galaxies
which  start to form stars earlier.
As expected, this general
trend is consistent with the predictions of ejective feedback models in
\citet[][see bottom row of their fig.~4]{hirschmann16}.

Galaxies with stellar masses $M_{\star}\lesssim 3\times 10^{10}\,{\rm M}_{\odot}$
are responsible for the excess of the local SFRD with respect to observations,
especially those in the mass range
$\log(M_\star [{\rm M}_\odot]) \in [10.1,10.5]$ that dominate 
the contribution to the SFRD at $z \lesssim 0.3$. 
\citet{hirschmann16} note this problem for the most massive galaxies
($M_{\star} \sim 10^{12}\,{\rm M}_{\odot}$) and attribute this failure
in reproducing the SFRD at $z \lesssim 0.5$ to a 
radio-mode AGN feedback not efficient enough in suppressing SF in these
galaxies. Although this could be a possible
explanation for galaxies within the stellar mass range
$\log(M_\star [{\rm M}_\odot]) \in [10.1,10.5]$ in our model,
the restrictions imposed to calibrate \sag~with the PSO method
could also favour this excess.
Both the high-mass end of the SMF at $z=0$ and the local BHB relation
help finding appropriate values of the free parameters involved
in the AGN feedback scheme.
However, the SFRF at $z=0.14$ also plays a role.
We have shown that the tuned parameters allow to achieve a
highly satisfactory agreement with the data presented by \citet{gruppioni15} (see Fig.~\ref{fig:SAGconstraints_SFRF}).
However, 
the IR+UV SFRD estimated from these data 
through the integration of
the best-fitting modified Schechter function
to the IR+UV SFRFs down to $\log{\rm (SFR)}=-1.5$
is higher than the optical SFRD presented by \citet{behroozi13c}
at low redshift ($z<0.5$), which is reflected in 
the excess shown by the model in Fig.~\ref{fig:SFRDvsz}.
We evaluate the impact of the constraint used in the calibration process
by imposing the
evolution of the SFRD upto $z=2$ as an observational restriction 
in place of 
the SFRF at $z=0.14$.
We consider the values for the SFR at several redshifts within 
that range taken from the
fit to observational data presented in \citet{behroozi13c}.
We find that the excess of the SFRD at $z=0$ is not avoided even with 
this restriction.    
This aspect deserves more investigation, in order to disentangle 
what particular combination of physical processes is causing this effect, 
or what process 
is still not well captured by the modelling. The analysis done in 
Section~\ref{sec:originSMF} 
points to TS, 
tidal disruption and over-merging as processes
that deserve special attention.

\begin{figure}
  \centering
  \includegraphics{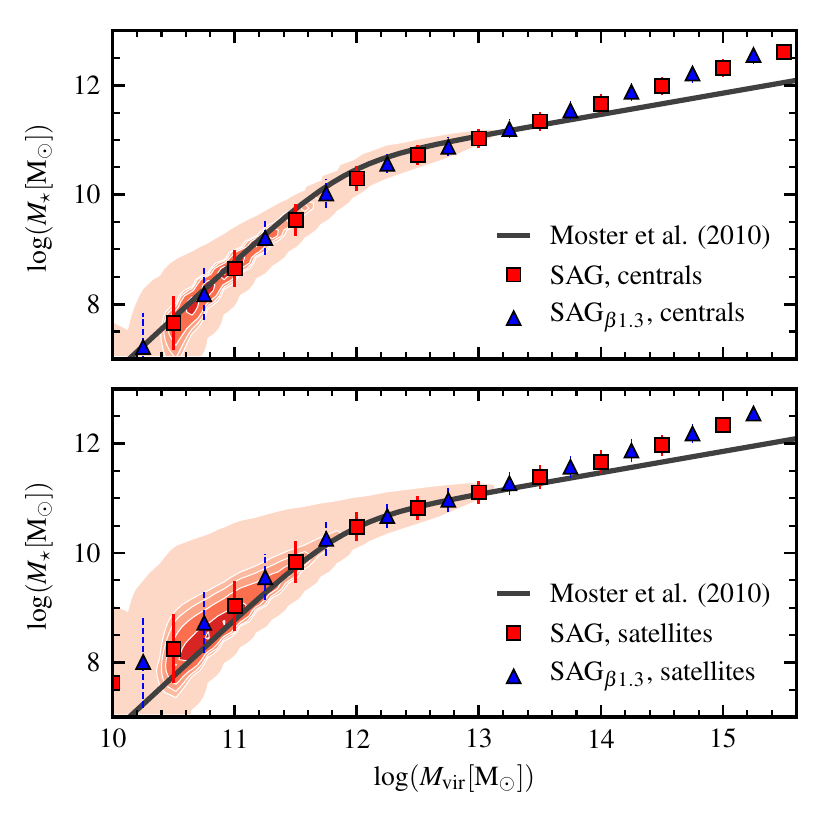}
  \caption{
Stellar mass as a function of DM halo mass for galaxies generated by both the 
calibrated model \sag~and model \sagb~at $z=0$
compared with the parametrization given by \citet{Moster10}.
{\it Top panel}: Relation for central galaxies of main host haloes
represented by 
a coloured contour map, 
considering the same contour levels as those of Fig.~\ref{fig:SAGconstraints_BHB}.
Mean values of the stellar mass for each bin of halo mass 
for models \sag~and \sagb~are depicted by red squares and blue triangles, respectively;
vertical error bars are $1\sigma$-standard deviation around the mean.
{\it Bottom panel}: Same as top panel but for 
satellite galaxies within DM subhaloes.
Differences between this relation for central and satellite galaxies translate
into the different normalization and shape of their respective main sequences 
(see Fig.~\ref{fig:sSFRmstarz}).
}
  \label{fig:SHMratio}
\end{figure}

\subsection[]{Specific star formation rate: main sequence}
\label{sec:SAG-pred-ssfr-mainseq}

Fig.~\ref{fig:sSFRmstarz} shows the relation
between the sSFR and stellar mass of
galaxies generated by
the calibrated model \sag~at different
redshifts ($z=0$, $0.7$, $2.2$ and $3.5$)
by the density map.
At $z=0$ (top-left panel), 
this distribution resembles very much the one presented
by \citet{Salim07}, both in shape and normalization.  
They estimate the SFR of 
50,000 optically selected galaxies in the local Universe
from gas-rich dwarfs to massive ellipticals
using ultraviolet and optical data from the 
{\it Galaxy Evolution Explorer} (GALEX) and 
SDSS Data Release 4, respectively.
Comparing with their fig. 15, it is evident that the dependence of
sSFR with stellar mass for model galaxies is shallower than observed.
This can be better quantified by considering the mean values of 
sSFR for different stellar mass bins for star-forming galaxies.
The tight correlation followed by them is known as the main sequence of galaxies.
Following \citet{Brown17}, who investigate environment driven gas depletion 
in satellite galaxies using a sample from SDSS, we classify galaxies as 
star-forming when they have 
${\rm sSFR}>10^{-10.7}\,{\rm yr^{-1}} $.
This limit allows a better separation between
active and passive galaxies in our model than 
the cut 
${\rm sSFR}=10^{-11}\,{\rm yr^{-1}}$,
commonly used in the         
literature \citep[e.g.][W12 hereafter]{Wetzel12}.
Although this is
inferred from the bimodality
that emerges for massive galaxies within groups and clusters
(see Section~\ref{sec:SAG-pred-ssfr-distr}),
we apply this cut to galaxies in all mass ranges, following W12. This criterion also comprises the star-forming
sample of galaxies selected by \citet[][see their fig. 17]{Salim07} 
through optical emission lines
and the BPT diagram. 

Mean values of sSFR of star-forming galaxies are shown for 
central and satellite
galaxies separately.
These main sequences are compared with
redshift dependent fits 
to observational data 
collected by  
\citet{behroozi13c}, 
all corrected to a Chabrier IMF
(see their table $8$ in appendix F). The data set, 
specified in their table 5, includes the 
sSRF measured by \citet{Salim07} described above. 
The general agreement is rather good at all redshifts, especially 
regarding the normalization.
Reproducing the right normalization of this correlation has been 
challenging;
galaxy formation models have failed in reproducing this feature
underpredicting it, specially at high redshifts \citep{Daddi07, Weinmann12, Xie17}. 

The similarity between 
the main sequence of central and satellite galaxies
for any of the redshifts considered highlights the importance
of modelling environmental effects through RPS and TS, with the consequent
gradual removal of hot halo gas.
The strangulation scheme considered in previous versions of \sag~quenches 
star formation in satellite
galaxies too early, leading to values of SFR one order of magnitude lower than
those characterising central galaxies, thus generating a 
complete separate main sequence for satellites.
Satellite galaxies in the current model lie on a main sequence
that is systematically lower than the one traced by centrals at all redshifts,
but differences are within the $1\sigma$-standard deviation around the mean.
Although the normalization is quite good for both central and
satellite galaxies, 
the model predicts flatter sequences than those inferred from observations.
Mean values of sSFR of central galaxies
decrease for stellar masses $M_{\star}\gtrsim 10^{10}\,{\rm M}_{\odot}$,
being this trend more pronounced for centrals than for satellite galaxies;
the main sequence for satellites remains almost flat at $z=3.5$.
This leads to 
more similar values of mean sSFR for these two galaxy populations
at high stellar masses. 
For low-mass galaxies ($M_{\star} \lesssim 10^{10}\,{\rm M}_{\odot}$),
the agreement between \sag~predictions and observational fits
is better at $z=0$ than at higher redshifts, 
while the opposite situation occurs for high-mass galaxies. The
higher values of sSFR at $z=0$ for galaxies with stellar masses
within the range
$\log(M_\star [{\rm M}_{\odot}]) \in [10.1,10.5]$
are consistent 
with the fact that these galaxies are the main responsible of
the excess of the cosmic SFRD at $z=0$
(see Fig.~\ref{fig:SFRDvsz}).

Both the normalization and shape of the main sequence of central
and satellite galaxies are affected when considering
model \sagb, as it is shown in Fig.~\ref{fig:sSFRmstarz}
by dashed thick and thin lines, respectively.
Central and satellite galaxies from model \sagb~achieve smaller sSFR 
values than those characterizing the same galaxy types generated by model 
\sag~in all the redshifts considered,
being these differences larger for smaller stellar masses.
They originate in the large amount of stellar mass acquired by low
mass galaxies at high redshift in model \sagb~(see
bottom panel of Fig.~\ref{fig:SAGconstraints_SMF})
because the SFRD in this model is higher than in \sag~for $z\gtrsim 1$ (see Fig.~\ref{fig:SFRDvsz}). Besides, the SFRD in \sagb~is slightly lower at $z=0$, 
a pattern that is common to galaxies
within different $z=0$ stellar mass ranges,
contributing to reduce the sSFR. 
Moreover, the decrease of sSFR with decreasing stellar mass 
for $M_{\star} \lesssim 10^{10}\,{\rm M}_{\odot}$, which
is a trend opposite to the one typical of observational data, becomes more evident in model \sagb.
This particular behaviour is similar to the one obtained
by \citet[][see their fig.~11]{Xie17} 
who use an updated version of the
semi-analytic model GAEA \citep{hirschmann16}
considering the same ejective
model of SN feedback
in which
the new feedback scheme implemented
in our model is inspired; the similarity with results from model \sagb~is
not surprising since they adopt a parameter $\beta=1.25$.

\cite{Xie17} argue that central galaxies with underestimated
SFRs are responsible for the decreasing trend of the sSFR 
for low stellar masses.
In fact, from Fig.~\ref{fig:sSFRmstarz}, we can clearly see that
both central and satellite galaxies contribute to the
decrease of sSFR with decreasing stellar mass.
The responsible of this trend is the 
additional modulation with the virial velocity introduced in the estimation of reheated and ejected mass (eqs.~\ref{eq:feedfire} and \ref{eq:ejecfire}), following the broken power law suggested by \citet{muratov15}.
The exponent $\alpha_\text{F}=-3.2$ adopted for low-mass galaxies ($V_{\rm vir} < 60\,{\rm km}\,{\rm s}^{-1}$) makes the SN feedback too strong for them, which then
have too low SFR. We have verified this by running
a variant of model \textsc{sag$_{\beta1.3}$} in which the slope $\alpha_{\rm F}$ is fixed to -1 (the value corresponding to $V_{\rm vir} > 60\,{\rm km}\,{\rm s}^{-1}$), for any virial velocity.
In this case, the main sequence of central galaxies become flatter at low stellar masses, while satellites present a mild trend in which sSFR increases monotonically with decreasing stellar mass.

The differences in normalization
and shape of the main sequence of central and satellite galaxies 
are not explained by the particular treatment of 
environmental effects introduced in \sag. Avoiding the action of RPS on the hot
and cold gas phases only produces a higher number of active satellite galaxies,
without changing the mean values of the sSFR of star-forming ones.
In order to understand the origin of these differences,
we examine the relationship between stellar mass and DM halo mass.
We consider the mass of the main host haloes for central galaxies 
and the subhaloes masses for satellite galaxies, 
excluding orphans. 
Fig.~\ref{fig:SHMratio} shows such relations,
representing galaxies for the calibrated model \sag~with a coloured contour map
on top of which
the mean values of the stellar mass for each bin of halo mass 
(depicted by squares) are superimposed.
Mean values of model \sagb~are also added (represented by triangles), 
showing that the SHM relation is not appreciably affected by
a change in the parameter $\beta$.
Mean values are compared with
the fitting function for the stellar-to-halo mass relation at $z=0$
derived by \citet{Moster10}; such parametrization was estimated 
by requiring that the observed SMF obtained
from SDSS DR3 \citep{Panter04} is reproduced when  
populating with galaxies the haloes and subhaloes in an {\em N}-body simulation.
Note that this comparison is not fair for satellite galaxies,
but it is done with the purpose of highlighting the differences
in the  relation for centrals and satellites. 
The general trend is very well reproduced by model central galaxies.
For halo mass $M_{\rm vir}\lesssim 10^{12}\,{\rm M}_{\odot}$, differences
between centrals and satellites become more evident.  
For a given stellar mass, central galaxies inhabit more massive haloes, while
satellites prefer less massive ones, being the dispersion of the
relation (vertical error bars) larger for smaller stellar masses.
Systematic differences in the halo virial mass between main host 
and satellite haloes translate into differences
in the hot halo mass 
and associated
cooling rates, which impact on the SF activity leaving
different imprints in the main sequence of central and satellite galaxies. 

\begin{figure*}
  \includegraphics[width=2.0\columnwidth]{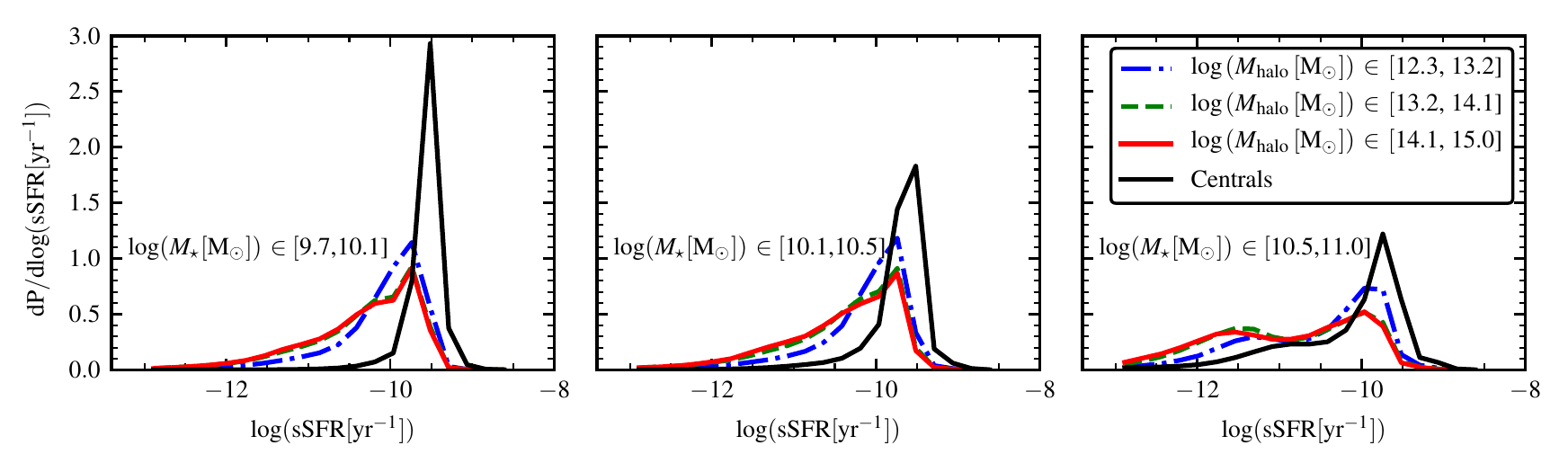}
  \caption{Specific star formation rate distributions for galaxies 
with stellar masses in different mass bins at $z=0$: 
$\log(M_{\star} [{\rm M}_{\odot}])\in[9.7, 10.1]$ (left panel),
$\log(M_{\star} [{\rm M}_{\odot}])\in[10.1, 10.5]$ (middle panel) and
$\log(M_{\star} [{\rm M}_{\odot}])\in[10.5, 11.0]$ (right panel).
Coloured thin lines of different style represent satellite galaxies 
residing in main host DM haloes within different mass ranges,
as indicated in the legend. 
Central galaxies are identified with black thick solid lines; all central
galaxies are considered, without any restriction regarding 
their main host DM haloes.
A bimodal distribution
emerges only for the most massive galaxies considered, with a break at
${\rm sSFR}=10^{-10.7}\,{\rm yr}^{-1}$, a value
in good agreement with the one adopted by \citet{Brown17} and
slightly higher than the one identified from the observational
catalogue constructed by W12.
}
  \label{fig:sSFRhisto-mstar-mhalo}
\end{figure*}

\begin{figure}
  \centering
  \includegraphics[width=0.99\columnwidth]{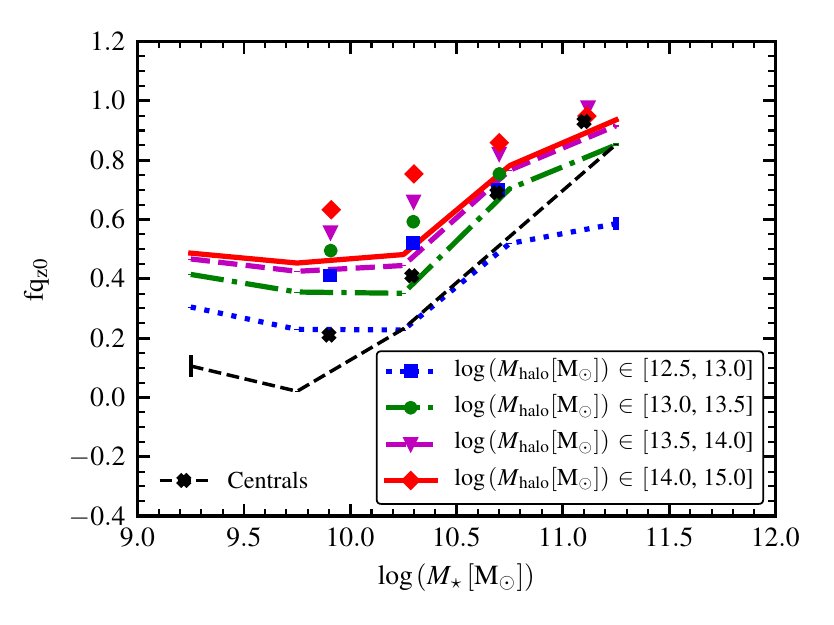}
  \includegraphics[width=0.99\columnwidth]{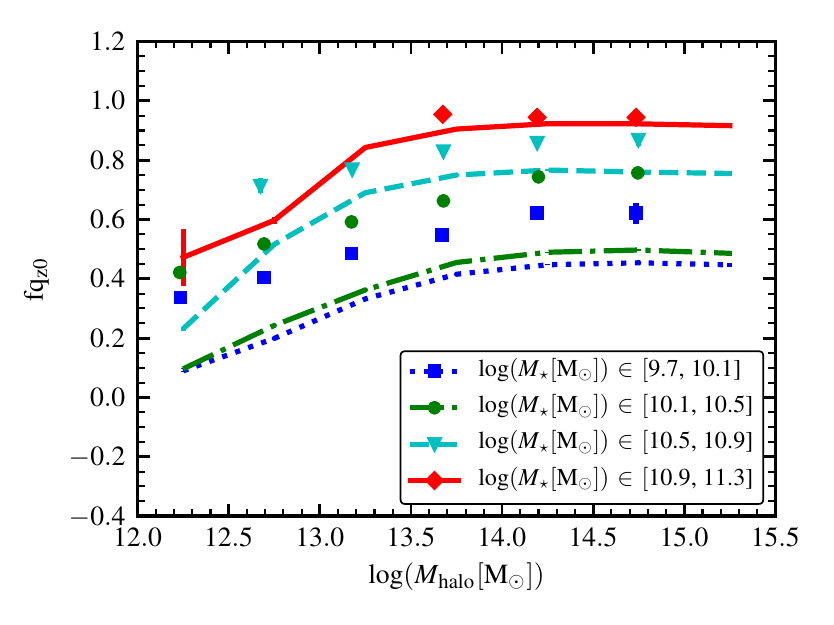}
  \caption{Fraction of quenched galaxies 
(${\rm sSFR} < 10^{-10.7}\,{\rm yr}^{-1}$) in \sag~model at $z=0$ as
a function of stellar mass (top panel) and main host halo mass (bottom panel).  
Satellite galaxies are binned according to their main host 
halo or stellar mass, and are represented by coloured lines of different style, 
as indicated in the legend. Central galaxies, 
represented by a thin black dashed line, are only included in the top panel
without making any distinction to the mass of the halo they reside in.
Error bars show the $68$ per cent bayesian confidence interval estimated
following
\citet{Cameron11}; they are hardly visible for the satellite population.
These fractions are compared to
those obtained by W12 (different symbols associated to different
line styles).
The calibrated \sag~model underpredicts the fraction of quenched galaxies.
Only the most massive galaxies, those in the stellar mass range
$10.9 < {\text {\rm log} (M_{\star} [{\rm M}_{\odot}])} < 11.3$
within DM host haloes with masses  $M_{\rm halo}\gtrsim 10^{14}\,{\rm M}_{\odot}$
achieve a quenched fraction similar to the observed one.
This lack of general agreement is related to  
the underpredicted SFRD at high redshifts,
as discussed in Section~\ref{sec:SAG-pred-sfr}.
The value $\beta=1.99$
that regulates the redshift dependence of the reheated and
ejected mass in \sag~model is too high and produces
a shift of the SF activity to lower redshifts not leaving enough
time for quenching.
}
  \label{fig:fq_mstar_mhalo-erebos}
\end{figure}

\begin{figure}
  \centering
  \includegraphics{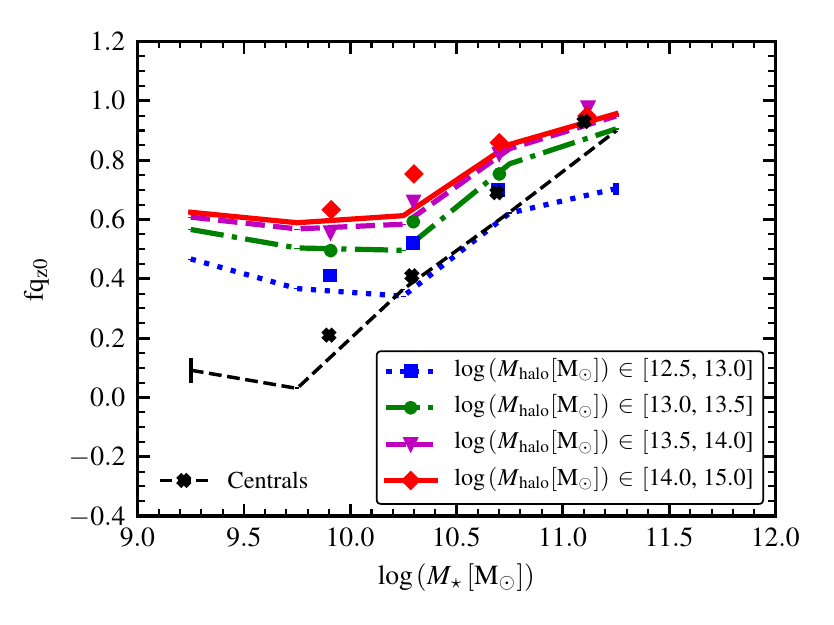}
  \includegraphics{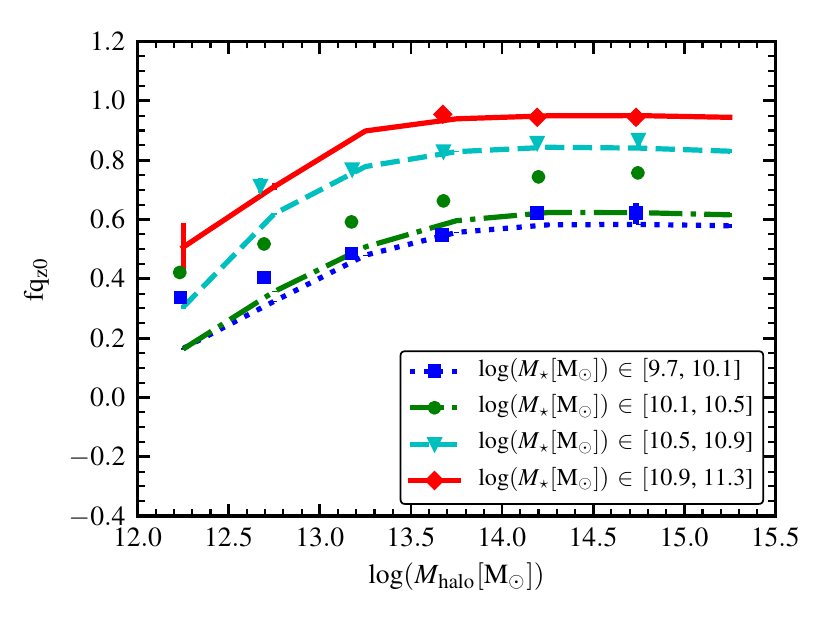}
  \caption{Same as Fig.~\ref{fig:fq_mstar_mhalo-erebos} 
for galaxies generated with 
model \sagb.
The general agreement with observations is rather good except for the
underprediction of the quenched fraction for galaxies within 
the mass range $\log (M_{\star} [{\rm M}_{\odot}]) \in [10.1, 10.5]$.
}
  \label{fig:fq_mstar_mhalo}
\end{figure}

\subsection[]{Specific star formation rate: distribution}
\label{sec:SAG-pred-ssfr-distr}

We analyse the distribution of sSFR
for central and satellite galaxies within different stellar mass ranges,
discriminating the latter 
according to the mass of their main host 
DM haloes\footnote{From hereafter, 
all the sampling of satellite galaxies is done according to the masses of
their main host haloes, even if it is simply referred as \textsl{halo}.}.
We select galaxies within the same stellar and halo mass ranges as 
W12, who construct galaxy group catalogues
using a  group-finding algorithm based on the one presented
by \citet{Yang05}.
W12 derive stellar masses and sSFR
for galaxies
within groups and clusters from the spectroscopic information
provided by the SDSS Data Release 7; the median redshift of the sample is
$z = 0.045$. 
Left, middle and right panels of
Fig.~\ref{fig:sSFRhisto-mstar-mhalo} show, respectively, the sSFR 
distributions\footnote{Probability density function at the bin, normalized 
such that the integral over the range is equal to unity.} 
of galaxies with stellar masses within the ranges 
$\log(M_{\star} [{\rm M}_{\odot}]) \in [9.7,10.1]$,
$\log(M_{\star} [{\rm M}_{\odot}]) \in [10.1,10.5]$ and
$\log(M_{\star} [{\rm M}_{\odot}]) \in [10.5,11.0]$.
Satellite galaxies are discriminated according to the
mass of their main host haloes.
In all cases, both central and satellite galaxies have a well
distinguished peak at high values of sSFR 
(${\rm sSFR} > 10^{-10}\,{\rm yr}^{-1}$).
A bimodal distribution
emerges only for the most massive 
galaxies considered.
The break of the bimodality takes place at 
${\rm sSFR}=10^{-10.7}\,{\rm yr}^{-1}$, a value 
in good agreement with the one adopted by \citet{Brown17} and
slightly higher than the one identified from the observational
catalogue constructed by W12.
While the sSFR distribution of galaxies in their observational sample 
exhibits a clear bimodality 
at all stellar masses, regardless of their classification in centrals or
satellites (see their fig. 1), 
the authors clarify that the strong sharpness of the peak near
${\rm sSFR}=10^{-12}\,{\rm yr}^{-1}$ is a result of an artifact in spectral reductions.
They also emphasize that the distributions exhibit
a tail to much lower sSFR, like the sSFR distributions
of our model galaxies do, which are in better agreement with the behaviour
shown by \citet{Salim07}, as it was described in 
Section~\ref{sec:SAG-pred-ssfr-mainseq}.
Morever, \citet{Kauffmann14} concludes
that such a distribution is expected for galaxies characterized by a bursty mode of star formation with times of formation between $1$ and $10\,{\rm Gyrs}$. This is consistent with the results we obtain from a detailed
analysis of quenching time-scales of satellites \citep[][submitted]{Cora18b}.
Consistent with the discussion also presented in Section~\ref{sec:SAG-pred-ssfr-mainseq}
when analysing the main sequence of central and satellite galaxies at $z=0$,
the high sSFR peak of the distributions of satellite galaxies
is slightly shifted towards lower sSFR values than for centrals, while
such a trend is not evident in the observational sample.

Despite these differences between our results and those of
W12, there are several similarities in relevant aspects of 
the distributions.
As in the observational sample, more massive central galaxies 
have larger probability of having low sSFR at the expense of a reduction
in the peak height at high sSFR. However, satellites only show such a trend
for those cases where the bimodality becomes evident in the model.
Both in the simulated and observed sample, there are
more satellites with low sSFR than centrals while the opposite occurs for
high sSFR, regardless of the stellar mass bin considered.
Besides, the number of satellite galaxies with low(high) sSFR 
increases(decreases) systematically as we consider more massive
DM haloes.
These general results remain valid for galaxies generated by model \sagb.
A quantitative comparison with observational results is attained through
the estimation of the fraction of quenched galaxies for different stellar
mass and halo mass ranges. 

\subsection[]{Quenched fractions of central and satellite galaxies from the model SAG}
\label{sec:SAG-pred-fq}

From a given population of model galaxies 
selected according to their stellar mass and/or main host halo mass, 
we estimate
the quenched fraction by considering those galaxies 
that are passive at a given time of interest, that is, those that satisfy
the criterion 
${\rm sSFR} < 10^{-10.7}\,{\rm yr}^{-1}$,
as emerges from the distribution of sSFR analysed in the previous section.
We compare model results  
with those obtained by W12.
Although these authors consider a galaxy as passive when its 
${\rm sSFR} < 10^{-11}\,{\rm yr}^{-1}$, we  
prefer to estimate the quenched fraction
taking into account the separation between
active and passive galaxies that emerge from the model at the expense of 
accepting a shift in the sSFR cut with respect to the one 
proposed by W12. 
Stellar mass and main host halo mass dependence of the quenched 
fraction at $z=0$ ($fq_{\rm z0}$) are shown
in the left and right panels
of Fig.~\ref{fig:fq_mstar_mhalo-erebos}, respectively. 
In both cases, the same mass bins as those chosen by W12 
are used.   
Model results are identified with lines.
Error bars show the $68$ per cent bayesian confidence interval estimated
following 
\citet{Cameron11}; their method is applied to estimate errors of fractions
in all subsequent plots.
The overwhelmingly numerous galaxy population obtained with the MDPL2 simulation makes a very good statistics in these results, so most of these error bars are too small to be distinguished.
For the stellar mass dependence, we present the quenched fraction
of both central (thin dashed line) and satellite (different
line styles) galaxies;  
the latter are grouped according to the halo mass they reside in.
For the halo mass dependence, only satellites within different
mass ranges are considered.  
Symbols depict the results of W12 (see their fig. 3(a) and 3(b)).
From this comparison, we can see that the 
model underpredicts the fraction of quenched galaxies.
Only the most massive galaxies, those in the stellar mass range
$\log(M_\star [\textrm{M}_\odot]) \in [10.9, 11.3]$,
residing in haloes with masses  
$M_\textrm{halo}\gtrsim 10^{14}\,\textrm{M}_\odot$
are characterized by a quenched fraction similar to the observed one.

This drawback of the model is the result of the high
value of the parameter $\beta$ achieved during the calibration
process when trying to satisfy the constraint imposed by the
SMF at $z=2$, giving raise to an underprediction of the SFRD at
high redshifts, as discussed in Section~\ref{sec:SAG-pred-sfr}. 
Thus, the star formation activity is shifted to later epochs having less time
to be quenched.
We have also shown that fixing $\beta$ in the lower value suggested 
by \citet{muratov15}
allows to obtain a SFRD in better agreement with observations at high 
redshift, 
at the expense of having an excess in the number density of low-mass galaxies
at $z=2$. 
Results on the quenched fractions are in favour of choosing a model
that prioritises the right 
evolution of the SFRD. 
In the following, we show that quenched fractions are considerably 
improved for galaxies generated from model \sagb.

\subsection[]{Quenched fractions of central and satellite galaxies from the model SAG$_{\beta 1.3}$}
\label{sec:fq-ms-mh-r}

Fig.~\ref{fig:fq_mstar_mhalo} shows the stellar and main host halo mass
dependence of the quenched fraction of galaxies generated by model \sagb.
The fractions of quenched galaxies at $z=0$
are larger than those obtained with the calibrated model \sag~characterized by the parameter
$\beta=1.99$.
The general agreement with values inferred by W12 is rather good, reproducing
the observed trends
for both central and satellite galaxies. This fact highlights the importance of an adequate efficiency of SN feedback at high redshifts in determining the passive fraction of galaxies.
Despite this general good agreement, some differences in a certain range of 
stellar mass for satellites still remain.
Namely, the quenched fractions of satellite galaxies with stellar masses
$\log (M_{\star} [{\rm M}_{\odot}]) \in [10.1, 10.5]$ are clearly
under-predicted for
any halo mass considered. 
This mass range is responsible of the excess of the SFRD at $z=0$
(see Fig.~\ref{fig:SFRDvsz}).
However, 
satellites do not contribute significantly to the SFRD excess. 
This under-prediction is caused by the lack of satellite galaxies
in this stellar mass range as a result of over-merging, as it is
evident from the depletion in the conditional stellar mass function
of satellites at $\log (M_{\star} [h^{-2}\,{\rm M}_{\odot}]) \approx 10$ (see Fig.~\ref{fig:CSMFz0}).

As noted by W12, both central and satellite galaxies
are more likely to be quenched as their stellar masses increase,
being this dependence stronger for the former and mainly 
produced by self-regulating processes such as AGN and SN feedback,
i.e. mass quenching 
\citep[e.g.][]{Peng10,Henriques17}.
Instead, satellites have a milder dependence of the
quenched fraction on stellar mass, and its increment is more
gradual for satellites hosted by more massive haloes, as a result
of the larger effect of environmental processes on satellites
with lower stellar masses.  
Hence, for a given stellar mass, the fraction of passive satellites
increases with increasing halo mass, being the dependence
on halo mass much milder for higher mass galaxies. 
This is more clearly shown in 
the right panel of Fig.~\ref{fig:fq_mstar_mhalo}, where
we can also appreciate that, 
for a given halo mass, more massive satellites
are more likely to be quenched.

\begin{figure}
  \centering
  \includegraphics{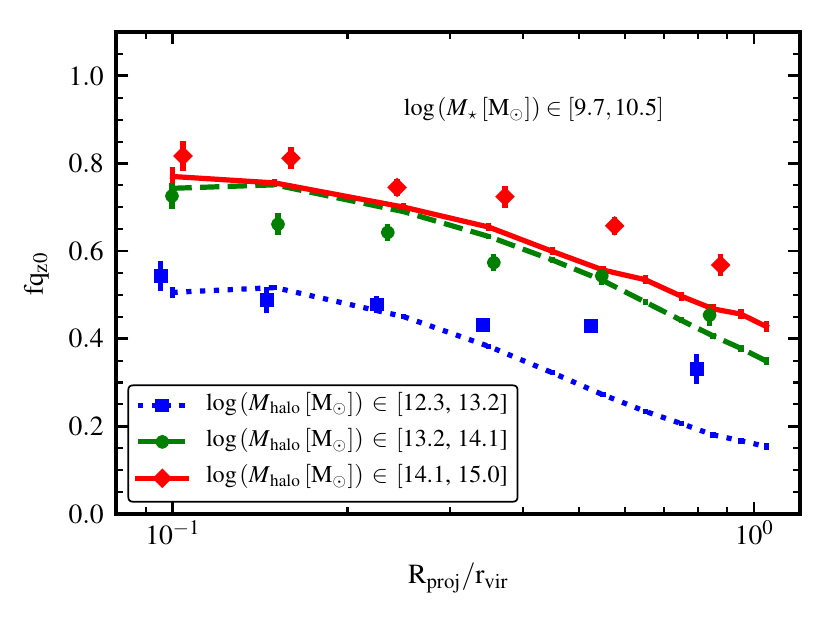}
  \includegraphics{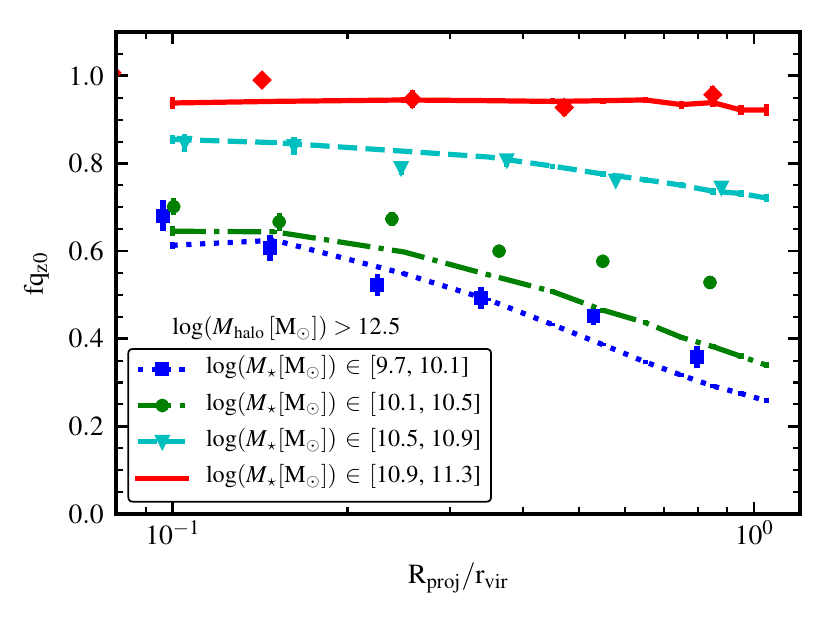}
   \caption{Fraction of quenched satellite galaxies
(${\rm sSFR} < 10^{-10.7}\,{\rm yr}^{-1}$) in model \sagb~at $z=0$ as
a function of the projected halo-centric distance normalized with
the virial radius of the main host halo. 
Satellites are binned according to their
main host halo mass (top panel) and stellar mass (bottom panel).
Model results are shown by different lines;
error bars show the $68$ per cent Bayesian confidence interval estimated
following \citet{Cameron11}.
Only satellites within the stellar mass range
$ \log (M_{\star} [{\rm M}_{\odot}]) \in [9.7, 10.5]$
are considered in the top panel and with main host halo mass 
$\log (M_{\rm halo} [{\rm M}_{\odot}]) > 12.5$ in the bottom one.
Such selection is done following W12 for comparison purposes.
Their data are represented by different symbols according to
the different stellar or halo mass ranges.
}
  \label{fig:fq_r}
\end{figure}

Fig.~\ref{fig:fq_r} shows how the fraction of $z=0$ quenched satellite galaxies 
in model \sagb~varies with the projected halo-centric distance, 
normalized by the
virial radius of the main host DM halo in which the galaxies reside.
Satellites are grouped according to their main host halo mass (top panel) 
or stellar
mass (bottom panel), and their quenched fractions are depicted
by different line styles. In the first case, 
only galaxies within the stellar mass
range $\log (M_{\star} [{\rm M}_{\odot}]) \in [9.7, 10.5]$ are considered.
In the second case, all galaxies within haloes with masses 
$M_{\rm halo} > 3.16 \times 10^{12}\,{\rm M}_{\odot}$ are taken into account.
These selections  
allow to make a direct comparison with the quenched
fractions of the sample analysed
by W12, represented by different
symbols according to the stellar or halo mass considered. 
In general terms,
the radial gradient of the observed quenched fraction is well recovered
by the model for both the stellar mass and halo mass selected sets.

When the halo mass is varied (top panel), the radial profiles are a bit steeper
than observed.  
The model gives a good match of $fq_{\rm z0}$ 
for $R_{\rm proj}/r_{\rm vir}\lesssim 0.2$
and $R_{\rm proj}/r_{\rm vir}\gtrsim 0.5$ for galaxies
within haloes of mass 
$\log (M_{\rm halo} [{\rm M}_{\odot}]) \in [12.3, 13.2]$ 
and
$\log (M_{\rm halo} [{\rm M}_{\odot}]) \in [13.2, 14.1]$, respectively.
Quenched fractions for the highest halo mass bin
($\log (M_{\rm halo} [{\rm M}_{\odot}]) \in [14.1, 15.0]$)
are below W12 results for all halo-centric distances, and 
similar to those corresponding to the immediately smaller halo mass 
range.
This result is consistent with the rather flat behaviour of the 
predicted quenched 
fractions as a function of halo mass for masses
$M_{\rm halo} \gtrsim 4 \times 10^{13}\,{\rm M}_{\odot}$,
shown in the bottom panel of  
Fig.~\ref{fig:fq_mstar_mhalo}.

The discrepancies found 
when the halo mass is varied 
are reflected in the mismatch between model
predictions and values inferred from observations 
for the two lowest stellar mass bins
considered in the bottom panel of Fig.~\ref{fig:fq_r}.
As we can see,
predicted values of $fq_{\rm z0}$ are comprised within the range delimited
by the corresponding observed quenched fractions, 
with galaxies within the stellar mass range
$\log (M_{\star} [{\rm M}_{\odot}]) \in [10.1, 10.5]$
only achieving a good agreement near the cluster centre;
this highlights once again the inability of the model 
in making adequate predictions
for this particular mass range 
because of over-merger,
as already noted.
This might have affected the fraction of quenched low-mass satellites,
since most of surviving orphans are those on less eccentric orbits 
that keep far from the central galaxy 
avoiding the merging region and preventing orphans from experiencing 
medium to strong-level RP.

\begin{figure*}
  \centering
   \includegraphics{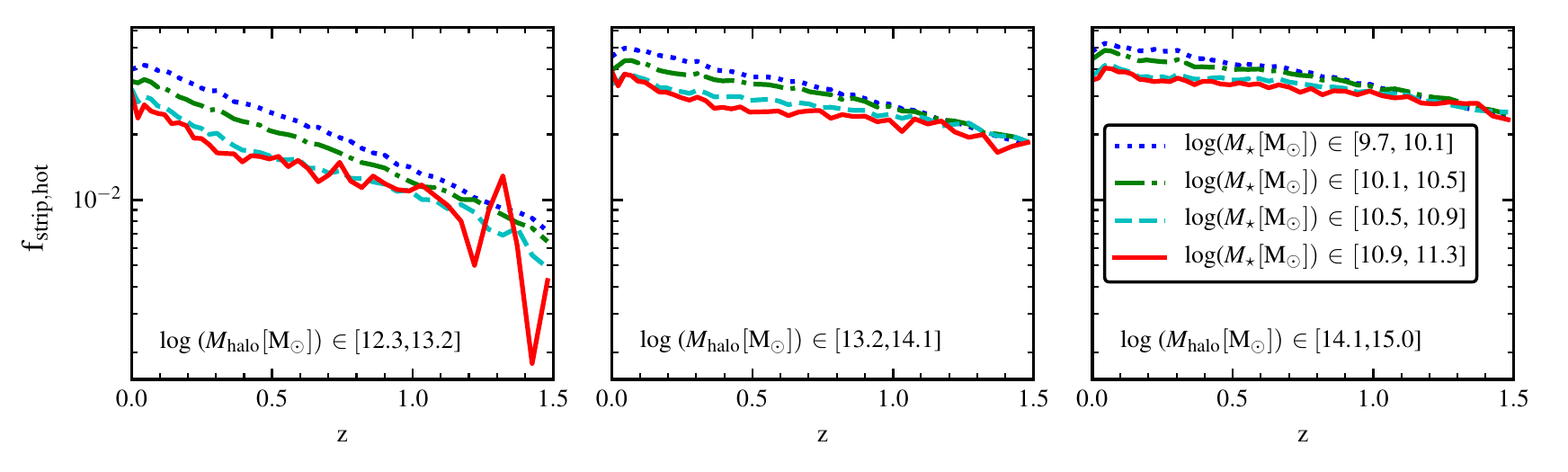}
  \caption{Mean values of the
fraction of hot gas mass
stripped by RP at each snapshot of the simulation
as a function of redshift for model \sagb.
All satellites (passive and active) at $z=0$ are included. 
They are grouped according to their local  
stellar mass (different line styles)
and main host halo mass (different panels), as indicated in the legends. 
For any galaxy mass and 
and halo mass, 
the fraction of hot gas mass stripped by RP
increases with decreasing redshift.
}
  \label{fig:stripfrac_hot}
\end{figure*}

\begin{figure}
  \centering
   \includegraphics{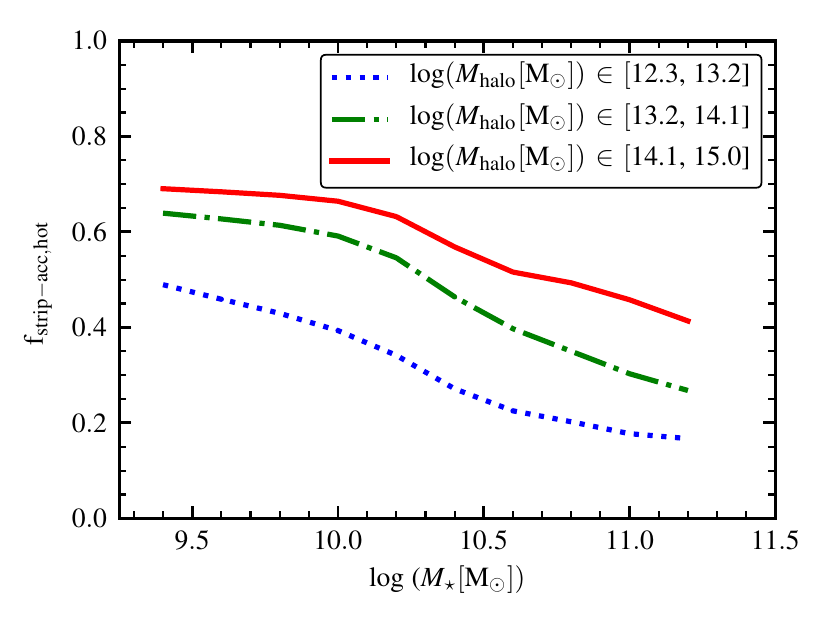}
  \caption{Mean values of the
fraction of accumulated stripped hot gas by RP as a function of stellar mass for model \sagb.
All satellites (passive and active) at $z=0$ are included. 
They are grouped according to their local  
main host halo mass (different line styles), as indicated in the legend. 
}
  \label{fig:stripfrac_hot_acc}
\end{figure}

\begin{figure}
  \centering
  \includegraphics{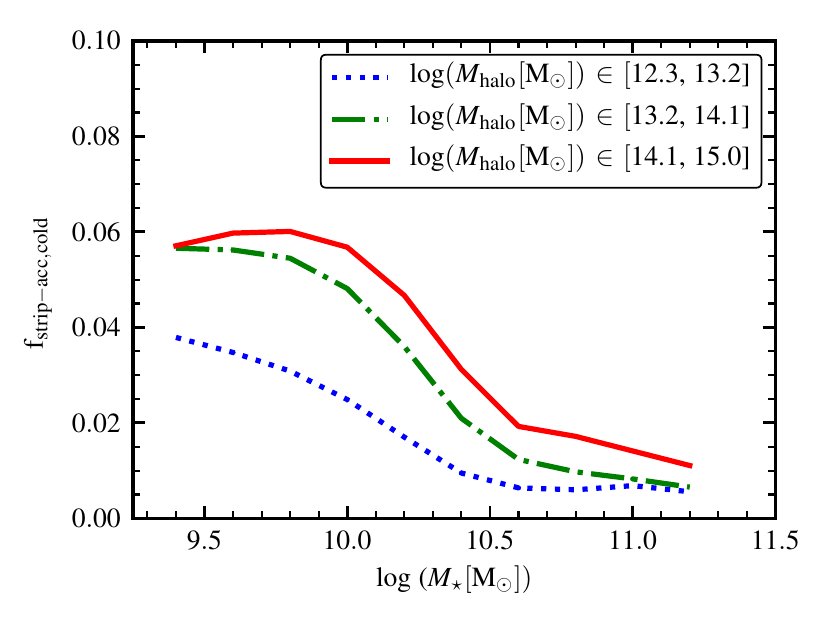}
    \caption{Mean values of the fraction of accumulated stripped cold gas by RP as a function of stellar mass for satellites in model \sagb~which interstellar medium 
is not longer shielded by the hot gas halo.   
They are grouped according to the mass of their local
main host haloes (different line styles), as indicated in the legend.}
  \label{fig:stripfrac_cold}
\end{figure}

On the other hand,
the predicted quenched fractions of
satellites within the two more massive stellar mass bins
($\log (M_{\star} [{\rm M}_{\odot}]) > 10.5$)
follow the observed radial profiles very well.
Since SF in more massive galaxies 
is mainly suppressed by mass quenching processes at both
low and high redshifts \citep{Lin14, Kawinwanichakij17},
both the radial and halo mass dependence of the fractions of quenched massive galaxies   
predicted by model 
\sagb~support the modelling
of those mass quenching processes that are particularly relevant 
for galaxies with stellar masses 
$\log (M_{\star} [{\rm M}_{\odot}]) > 10.5$, such as  
AGN feedback and disc instabilities.  
While the former reduces the amount of cooled gas,
the latter produces starbursts leading to a rapid exhaustion
of the cold gas reservoir.
These self-regulating physical processes
explain the SF quenching in massive central galaxies
but are still active as quenching mechanisms
even when galaxies become satellites, as demonstrated by \citet{Peng12},
playing a dominant role with respect to environmental effects.
This does not mean than environmental processes are irrelevant for high-mass galaxies. Indeed, these galaxies suffer the effects of RPS but they are smaller than those experienced by low-mass satellites, as shown in the next Section.

\section[]{Role of environment: Stripped mass and atomic gas content}
\label{sec:ts-rps}

In order to understand the role of environment on 
satellite galaxies of different mass, 
we show in Fig.~\ref{fig:stripfrac_hot} 
mean values of the fraction of hot gas mass 
stripped by RP, $f_{\rm strip,hot}$, at each snapshot of the simulation
as a function of redshift, for model \sagb. This quantity is estimated considering
the ratio between the stripped mass and the mass of hot gas just before
being stripped;
the latter is given by the sum of the stripped mass and the
mass of hot gas at the redshift considered.
Satellites are grouped according to their current stellar mass
(different lines) and the virial mass of their current main host 
DM haloes (different panels).
As expected, RPS is more efficient in removing hot gas from galaxies
with smaller stellar masses 
as demonstrated from both the analysis of observational data
\citep[e.g.][]{Fillingham16} and hydrodynamical simulations
\citep{Bahe15}. 
For any galaxy mass and 
and halo mass, 
the fraction of hot gas mass stripped by RP
increases with decreasing redshift. 
This is explained  
by the mass growth of the host DM halo and the consequent  
higher densities in the ICM, and high redshift infalling
galaxies that have more time to approach the cluster core, 
thus suffering stronger RP 
(\citealt{Jaffe15, Jaffe16}, Vega-Mart\'inez et al., in prep.).
Local values of the removed fraction at each stripping episode
are quite similar for any halo mass,
being of the order of $\approx 0.03-0.05$, with lower values corresponding
to lower mass haloes.
The dependence of this fraction with halo
mass becomes more evident at higher redshifts; 
the effect of RP is more pronounced in clusters
of $M_{\rm vir} \approx 10^{15}\,{\rm M}_\odot$ since $z=1.5$ than in less massive ones,
consistent with the distribution of RP values presented by
\citet[][see their fig. 5]{tecce10}. 

Fig.~\ref{fig:stripfrac_hot_acc} shows the mean values of the cumulative stripped hot gas fraction 
as a function of stellar mass of satellite galaxies binned according to
the mass of the main host halo they inhabit. This fraction is estimated as the ratio between the accumulated stripped hot gas mass since the first
RP stripping event suffered by the satellite and the sum of its current hot gas mass and the accumulated
stripped hot gas.
Again, we see that the mass stripped by RP is larger for less massive satellites residing in more massive halos. For satellites within low-mass haloes 
($\log (M_{\rm halo} [{\rm M}_{\odot}]) \in [12.3, 13.2]$), the cumulative stripped hot gas fractions
decrease from $\approx 50$ per cent for $M_{\star} \approx 3 \times 10^{9}\,{\rm M}_\odot$ to $\approx 20$ per cent for $M_{\star} \approx 10^{11}\,{\rm M}_\odot$. For high-mass haloes
($\log (M_{\rm halo} [{\rm M}_{\odot}]) \in [14.1, 15.]$), these fractions increase to $\approx 70$ and $\approx 40$ per cent, respectively.

In our model, both RP and TS act on the hot gas halo, but in general the former process gives rise to a smaller stripping radius than the latter, indicating that RPS dominates TS.
Thus, TS
is considered as a secondary effect
\citep{mccarthy2008, font2008, Bahe15}.
This is also consistent with inferences from observational results,
like the efficient SF quenching detected in 
Virgo cluster
\citep{Boselli16}. 

We demonstrate 
that low-mass satellites are the galaxies mainly affected by the
way in which environmental processes regulate the content of the hot gas 
reservoir
by making a test in which
satellites are allowed to keep their hot gas halo at infall
but neither RPS nor TS are activated, so that the reduction of hot gas
is a result of ejection and/or gas cooling. 
Only low-mass satellites 
($\log (M_{\star} [{\rm M}_{\odot}]) \in [9.5, 10.5]$) 
experience a reduction in the quenched fraction which is as high as
$\approx 0.15$ for galaxies 
residing within the most 
massive haloes 
($\log (M_{\rm halo} [{\rm M}_{\odot}]) \in [14.1, 15.]$).
However, the quenched fractions of more massive satellites remain
unchanged.
The lack of significant change in the fraction of quenched massive satellites 
when their hot gas reservoir suffers all type of possible effects
(strangulation, gradual removal, suppression of environmental processes)
is a strong evidence 
that environmental processes do not play a significant role
on the SF activity in massive satellites 
although these galaxies do suffer them as shown 
in Figs.~\ref{fig:stripfrac_hot} and \ref{fig:stripfrac_hot_acc}.
This was already deduced from 
the good match between model results and observations for the radial
distribution of the fraction of quenched massive galaxies 
(see Fig.~\ref{fig:fq_r}).

\begin{figure}
  \centering
  \includegraphics{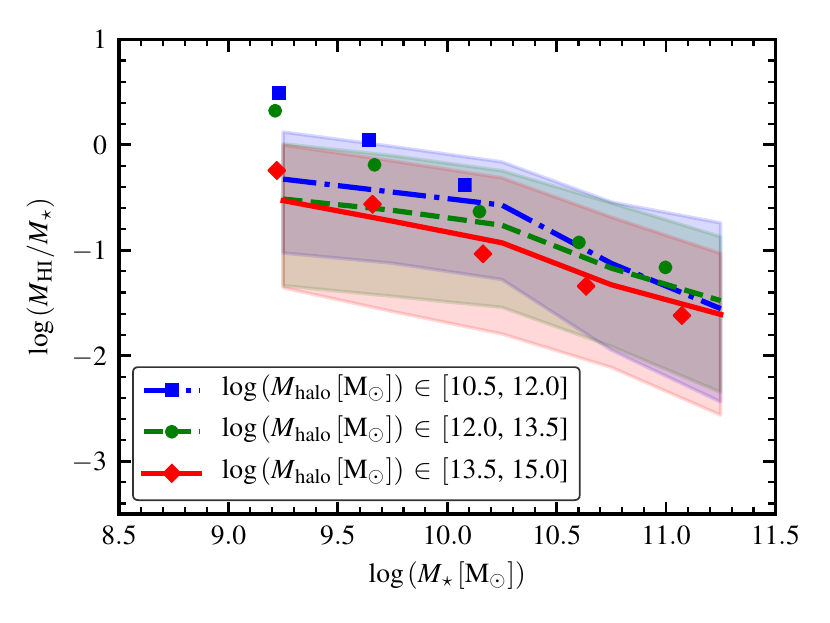}
  \caption{ 
Atomic hydrogen gas fraction as a function of stellar mass for satellite
galaxies in model \sagb~within different ranges of main host halo mass,
as indicated in the legend.
Different lines 
and associated shaded areas represent the corresponding median values
and $10$ and $90$ percentiles, respectively.
Median values are compared with the
average atomic gas fraction presented by 
\citet{Brown17}, who consider
HI data from ALFALFA survey (different symbols).
The match between model results and observations is rather good for
satellites with stellar mass larger than $10^{10}\,{\rm M}_{\odot}$
residing in haloes more massive than
$\approx 10^{12}\, {\rm M}_{\odot}$,
consistent with the good behaviour of the radial dependence of
the fraction of quenched massive satellites.
}
  \label{fig:mHI-fraction-halo-bins}
\end{figure}

The cold gas reservoir of satellite galaxies can also be reduced by the 
action of RPS, 
as inferred from images of highly asymmetric HI distribution
well within the stellar disc in cluster spirals 
\citep[e.g.][]{Abramson11,Kenney15, Bellhouse17}.
Fig.~\ref{fig:stripfrac_cold} shows the mean values of the cumulative stripped cold gas fraction as a function of stellar mass of satellite galaxies in model \sagb~which interstellar medium 
is not longer shielded by the hot gas halo. They are  grouped according to the mass of their main host haloes. 
The dependences of these
fractions with stellar mass
and halo mass are quite similar to those
characterising the stripped hot gas 
(Fig.~\ref{fig:stripfrac_hot}), although the values
of the fractions of cumulative stripped cold gas are one order of magnitude smaller 
than for the hot gas. 

The small fraction of cold gas removed by RPS
has no impact on the fractions of quenched galaxies,
which remain unchanged if RPS of cold gas is suppressed.
However, this slight reduction of the cold gas content allows to obtain
a better agreement with the observed 
relation of the atomic hydrogen (HI) gas content
as a function of stellar mass 
than in the case where RPS of the cold disc is deactivated.
This good agreement is shown in Fig.~\ref{fig:mHI-fraction-halo-bins}
for satellites
within main host haloes of different mass in model \sagb. 
HI mass is estimated
assuming that total cold gas is the sum of the contribution
of molecular (H2) and atomic hydrogen content and $30$ per cent helium
\citep[][eq. 2]{boselli14}. The molecular hydrogen content
is estimated adopting the scaling relation of 
molecular gas-to-stellar mass ratio \citep[][table 3]{boselli14}, in which
the coefficients correspond to molecular hydrogen masses 
estimated from CO intensities using the H-band luminosity-dependent
CO-to-H2 conversion factor. 
We compare against measurements
by \citet{Brown17} from a spectral stacking technique
applied to a multi-wavelength sample of satellite galaxies
selected 
from the SDSS, 
with HI data from ALFALFA survey,
with stellar masses larger than $10^{9}\,{\rm M}_{\odot}$ (different
symbols correspond to average values for different bins of halo mass). 
The match between model results and observations is good for
satellites with stellar mass $M_{\star}\gtrsim 10^{10}\,{\rm M}_{\odot}$
residing in haloes more massive than 
$\approx 10^{12}\, {\rm M}_{\odot}$,
consistent with the good behaviour of the radial dependence of
the quenched fraction for this galaxy population, giving additional
support to the modelling of mass quenching processes, as discussed in  
section~\ref{sec:fq-ms-mh-r}.
For smaller stellar masses, 
the mean HI-to-stellar mass fraction is underestimated by 
\sagb~regardless of the host halo mass.
This is not related to the modelling of RPS 
since this
general trend persists even when the action of RPS on the cold gas is 
deactivated,
in which case there is a systematic increase of this fraction
by $\approx 0.25$ dex for all stellar masses.
Hence, the low-mass content of atomic gas might be explained by
levels of star formation higher than expected in low-mass
galaxies as becomes evident from the excess of the cosmic SFRD at $z=0$ (see
Fig.~\ref{fig:SFRDvsz}). 

Results from \sagb~regarding the levels of current SF and atomic gas content
call for a more refined treatment
of quiescent SF, like  
distinguishing the neutral and molecular phases
of the cold gas as \citep[e.g. ][]{Lagos11a}.
Molecular gas is not significantly affected
by RPS as arises both from observations
\citep[e.g][]{Abramson11} and hydrodynamical simulations \citep{Tonnesen09};
this cold gas phase could be highly perturbed by high-speed tidal interactions
\citep{Scott15}.

\section{Comparison with other SAMs}
\label{sec:SAMcomp}

In a recent work, \citet{Henriques17} study 
environmental and mass quenching 
using the model described in \citet{henriques_mcmc_2015}, 
which is an updated version of the Munich semi-analytic model 
{\textsc {l-galaxies}}
\citep{guo11}.
Since we adopt the prescription
used by \citet{Henriques17} to model AGN feedback, the discrepancies between
the predicted fractions of quenched galaxies from {\textsc {l-galaxies}} and
those inferred from observations seem to originate in 
the modelling of other baryonic (e.g. SN feedback, SF, recycling) and environmental
processes, which affect the complex baryon cycle.

The new features implemented in \sag, including the new SN feedback scheme and 
treatment of environmental processes, 
differ considerably 
with respect to the corresponding implementation in {\textsc {l-galaxies}}. 
Regarding gas disruption processes, the most relevant difference is that
\sag~allows gradual removal of hot gas
reservoir by TS and RPS in both types of satellite galaxies (those
that keep their DM subhaloes and orphans), while
in {\textsc {l-galaxies}}  
orphan galaxies
lose immediately their hot gas reservoir by TS, as 
this process is considered responsible of the disruption of their DM 
subhaloes (in general, hot gas is assumed to be tidally stripped 
at the same rate as dark matter). 
Their assumption might be too strong, not taking into account
the fact that the mass resolution of the simulation prevents subhalo detection
in some cases.
Therefore, the assumption adopted in \textsc{l-galaxies}
prevents orphans from the gradual removal of hot gas through RPS.
The action of RPS is only allowed 
in satellites that keep their substructures.
As described in \citet{guo11}, the stripping radius is estimated
from the balance of self-gravity and ram pressure.  
After the stripping event, the remaining hot gas expands out to the 
subhalo~\rvir~recorded at infall.
\citet{guo11} argue that feedback could provide the required energy 
to redistribute the gas in this manner, which would require outflow velocities
comparable to the circular velocity of the host group. This assumption 
on the distribution of the remaining hot gas results in an increased efficiency
of~RPS compared to our procedure, since there is always hot gas 
beyond the current stripping radius.
Besides, the halo becomes increasingly more diffuse which leads to reduced cooling
rates. Therefore, their model results in
an increased fraction of passive low-mass satellites
\citep{guo11,henriques13,Hirschmann14}.

With the aim of reducing the excess
of passive satellites predicted by \textsc{l-galaxies},
\citet{henriques_mcmc_2015} define a minimum threshold halo mass 
($10^{14}\,{\rm M}_{\odot}$) for RP to be effective as 
an stripping mechanism of the hot gas 
halo of satellites that keep their substructure, 
below which the action of RP is suppressed.
However, despite of this additional condition, discrepancies between predicted and
observed fractions of quenched galaxies still remain.
Their condition for the action of RP
not only differs from the one we adopt (equation~\ref{eq:rpshot}),
which action is self-regulated without any restriction,
but it also considers that the orbital
velocity of the satellite is approximated by the virial
circular velocity of the main halo. This aspect is improved in our
implementation by taking the information of the actual orbits of satellites, 
including those of orphans. 
Furthermore, model \textsc{l-galaxies} 
does not include RPS of the cold gas disc.

\citet{GonzalezPerez14} present a 
variant of \textsc{galform} semi-analytic model with 
a molecular-based SF law that also assumes 
gradual stripping of the hot gas by the action of RP
adopting the same parametrization 
for the gravitational restoring as in \sag~
(see equation~\ref{eq:rpshot}) but with the value $\alpha_{\rm RP}=2$
suggested by \citet{mccarthy2008}. The main difference with
our implementation is that RP is set to its maximum value for the whole
orbit of the satellite galaxy since 
it is  
calculated at its pericentre. Therefore, the stripped hot gas is overestimated
leading to lower fractions of atomic gas than those inferred from observational
data \citep[see fig. 7b of][]{Brown17}. 
This disagreement would become even larger if RPS of cold gas were included
in their model.

The impact of both cold gas and hot gas stripping 
on 
the HI fractions of local galaxies
is investigated by \citet{Stevens17} using an updated version of the  
\darksage~semi-analytic model,
which evolves the one-dimensional structure of
galactic discs in annuli of fixed specific angular momentum.
Quiescent SF depends on the molecular gas fraction of each annulus;
they consider
both metallicity- and pressure-based prescriptions for
determining the ratio of H2.
The action of RP on both hot gas and cold gas phases
is regulated by conditions similar to those applied in our model \sag, that is,
those given by \citet{mccarthy2008} and \citet{gg72}, 
respectively. An additional important similarity to note is that 
\darksage~also considers
that the hot gas halo protects the cold gas disc from the
action of RP. However, the condition adopted to allow RPS 
differs from ours, since 
RP is able to remove cold gas from the disc 
when the total
baryonic mass of the galaxy (cold gas and stars) exceeds the
mass of hot gas halo.
\citet{Stevens17} show that 
the satellite HI fractions
are underpredicted by their full 
model even when it has been calibrated using the
observed HI fraction of galaxies as a constraint with the greatest weight.
These fractions
become closer to the observational data when cold gas stripping is suppressed.
Among the possible explanations of these results offered by the authors,
there is one directly related to the limitations of SAMs, namely,
the fact that the hot gas of satellite
galaxies is not replenished by accretion from the ICM.

The effect of RPS of cold gas is also taken into account by \citet{luo16} in a branch of \textsc{l-galaxies} \citep{Fu13} which also
assumes galaxy discs divided into multiple rings, and a SFR dependent on the local surface density of the 
molecular gas. This model is unable to reproduce the observed trends of quenched fractions of satellites and centrals. The number of low-mass passive galaxies are over-predicted despite gradual stripping of the hot gas is considered. The action of RPS of the cold gas increase the discrepancies with observational data. However, it
is difficult to quantify the effect of this environmental process because the passive fraction of central galaxies is under-predicted for any stellar mass hinting to problems not related with environment.

Even with the aforementioned drawback affecting the hot gas halo of satellites
and a simple modelling of galaxy discs
and quiescent star formation, our model \sagb~is able to achieve better
agreement with observed atomic gas content and quiescent fractions
than other current SAMs.

\section{Summary and conclusions}
\label{sec:conclu}

The updated version of our semi-analytic
model of galaxy formation \sag~has been used to
generate one of the galaxy catalogues of the
\textsc{MultiDark Galaxies} project \citep{knebe17},
which is based on the Planck cosmology $1\,h^{-1} \,{\rm Gpc}$
\textsc{MultiDark} simulation MDPL2. The model has been calibrated using
the PSO technique \citep{ruiz2015} combined with a particular set of
constraints and the observational
data defined in \citet{Knebe18}. This catalogue is publicly available
in the \textsc{CosmoSim} database.
We also consider a second galaxy population generated by a slightly different
version of the model, characterized by a change in one of the
parameters involved in the modelling of SN feedback (model \sagb).
We analyse the properties of the galaxies in these two catalogues,
both of star forming and quiescent galaxies, splitting the population
in centrals and satellites, and also sampling them according to the mass
of the main host haloes they inhabit.
The detailed analysis performed allowed us to evaluate the impact of
the improved treatment of environmental effects implemented in the model, 
and the modification introduced in the SN feedback scheme, given by
the addition of an explicit
redshift dependence in the estimation of the reheated and ejected mass.

We implement
the gradual removal of hot gas in satellites from RPS and TS,
allowing also the action of these processes on the cold gas disc
under certain conditions.
This is an improvement of our model
with respect to previous studies that ignore the RPS of the cold gas
\citep{guo11, kimm2011, GonzalezPerez14, Henriques17}.
The advantage of our implementation of RPS
with respect to previous works \citep[e.g. ][]{Henriques17, Stevens17}
resides in the use of a
fitting formulae for RP
experienced by galaxies in haloes of different mass as a function of
halo-centric distance and redshift
instead of analytic estimations.
The effect
of TS is considered as an additional mechanism contributing to the
removal of the hot and cold gas. It can also affect the stellar components.
Another important improvement with respect to other
SAMs is the integration of orphan satellites
(Vega-Mart\'inez et al., in preparation).

In the following, we summarize the main results of this work
obtained from the analysis of the two galaxy catalogues generated with
the calibrated model \sag~and model \sagb:

\begin{itemize}
\item SN feedback plays an important role in
the star formation history of galaxies.
In order to avoid the excess of the faint end of the SMF
at $z=2$ and to give rise to downsizing in the
stellar mass assembly, it is necessary to reduce the availability of cold
gas for star formation at high redshifts.
Thus, both the reheating of the cold gas phase and the ejection of part
of the hot gas reservoir, with the possibility of being reincorporated later,
must be stronger at earlier epochs.
This has been demonstrated by the inclusion of an explicit redshift dependence
of both the reheated and ejected mass in the new feedback scheme
implemented in \sag. Results are sensitive to the value of the power-law slope of such
dependence, denoted by the parameter $\beta$ in our model.
When $\beta=1.99$, as emerged from the calibration process in which the
free parameters are restricted by imposing the SMF at $z=2$ as constraint,
the predicted SFRD becomes too low with respect to observational data
at high redshifts. When fixing that parameter in the value
suggested by the fit of \citet{muratov15} obtained from the FIRE hydrodynamical
simulations ($\beta=1.3$),
the trend of SFRD is reconciled with the observed
one (Fig.~\ref{fig:SFRDvsz}) at the expense of an excess in the faint end
of the SMF at $z=2$ (Fig.~\ref{fig:SAGconstraints_SMF}).

\item
The gradual removal of hot halo gas through a
robust treatment of environmental effects, as the one implemented in our model,
becomes crucial to produce main sequences that
evolve consistently with observational data for both
central and satellite galaxies (Fig~\ref{fig:sSFRmstarz}).
Cooling flows keep replenishing the cold gas reservoir of satellites
after first infall during gradual starvation, delaying the beginning of the quenching of SF.
The differences in normalization and shape of the main
sequence of central and satellite galaxies that still remain
arise because, for a given stellar mass,
central
galaxies inhabit more massive haloes than satellites
(Fig.~\ref{fig:SHMratio}), which impact on the mass of the hot halo
and associated cooling rates.
The fact that galaxies start to form stars earlier in model \sagb~is
reflected in
the lower normalization of the main sequence of star-forming galaxies
achieved by this model with
respect to model \sag.
In both models, the main sequences 
of central and satellite galaxies are characterized by a decreasing
trend of sSFR with decreasing
stellar mass for $M_{\star} \lesssim 10^{10}\,{\rm M}_{\odot}$. The strong dependence with the virial velocity for low-mass galaxies introduced in the estimation of the reheated and ejected mass is responsible for this effect.

\item
The higher efficiency of SN feedback at high 
redshifts, required to recover the downsizing in stellar mass
assembly, must be characterized by a
mild redshift dependence of the reheated and ejected mass,
as in model \sagb, in order to achieve the
expected behaviour of the fractions of quenched galaxies
(those with
$sSFR < 10^{10.7}\,{\rm yr}^{-1}$)
as a function of stellar mass, halo mass and the halo-centric distances
(Figs.~\ref{fig:fq_mstar_mhalo} and~\ref{fig:fq_r}).
The right fractions
of quenched massive galaxies
($\log(M_{\star}[{\rm M}_{\odot}]) > 10.5$)
as a function of both halo mass
and halo-centric distance
support the modelling of those
self-regulating physical processes
that are particularly relevant for massive galaxies
(AGN feedback; disk instabilities), which are still active as quenching
mechanisms even when galaxies become satellites,
dominating over environmental quenching;
the effect exerted by the latter is milder in high-mass galaxies than in low-mass ones.

\item 
RPS plays a dominant role among the environmental
processes considered in our model and contributes to regulate
adequately the mass of the hot gas halo and the cold
gas disc (when it is not longer shielded by the hot halo),
being more efficient in removing
gas from galaxies with smaller stellar masses, as expected
(Figs.~\ref{fig:stripfrac_hot}, ~\ref{fig:stripfrac_hot_acc} and~\ref{fig:stripfrac_cold}).
For any galaxy mass, the fraction of
hot and cold gas mass stripped by RP increases with decreasing redshift,
as a result of higher densities achieved by the ICM,
associated to the mass growth of the main host DM haloes, and
the fact
that high redshift infalling galaxies have more time to approach
the cluster core where they suffer strong RP.
The total stripped mass 
is higher 
in more massive clusters
($M_{\rm vir} \approx 10^{15}\,{\rm M}_\odot$)
than in less massive ones, consistent with the distribution of RP values
inherent to groups and clusters \citep{tecce10}.
The cumulative stripped hot gas fraction increases 
from $\approx 20$ to $\approx 50$ per cent for satellites with $M_{\star} \approx 3 \times 10^{9}\,{\rm M}_\odot$ and $M_{\star} \approx 10^{11}\,{\rm M}_\odot$, respectively, residing in low-mass haloes  
($\log (M_{\rm halo} [{\rm M}_{\odot}]) \in [12.3, 13.2]$). For high-mass haloes
($\log (M_{\rm halo} [{\rm M}_{\odot}]) \in [14,1 15.]$), these fractions increase to $\approx 40$ and $\approx 70$ per cent, respectively.

\item The cold gas disc is less affected by RP than the hot gas,
with 
cumulative stripped cold gas fractions
being one order of magnitude smaller for the former than for the latter.
Thus, RPS on the cold gas does not affect the fraction of quenched galaxies
but it definitely contributes to reach
the right atomic hydrogen (HI) gas content
for satellites with different stellar mass,
especially for more massive ones
($M_{\star} \gtrsim 10^{10}\,{\rm M}_{\odot}$)
residing in haloes with masses
$M_{\rm vir} \gtrsim 10^{12}\, {\rm M}_{\odot}$
(Fig.~\ref{fig:mHI-fraction-halo-bins}).
Distinguishing the neutral and
molecular phases of the cold gas could help to
reproduce the observed levels of HI gas content of low-mass satellites.

\end{itemize}

Our results highlight the impact of specific aspects of mass and environmental quenching on galaxy evolution.
The stronger effect of SN feedback at higher
redshifts and the physics of environmental processes
captured by the RP fitting formulae
combined with the orbital evolution of orphan galaxies
make our model \sagb~able to generate a galaxy population   
with fractions of gas and
levels of SF that allow to achieve 
better
agreement with observed atomic gas content and quiescent fractions
than other current SAMs.
These attainments of our model
are reached even though it does not include a
detailed treatment of galaxy discs
as in {\textsc {dark sage}} \citep{Stevens17}, or a molecular-based SF law as {\textsc {galform}}
\citep{GonzalezPerez14}. 
Differences still found for particular ranges of stellar and halo mass
give hints for further improvement of the model.

\section*{Acknowledgements}
The authors gratefully acknowledge the Gauss Centre for Supercomputing e.V. 
(www.gauss-centre.eu) and the Partnership for Advanced Supercomputing in Europe 
(PRACE, www.prace-ri.eu) for funding the \textsc{MultiDark} simulation project 
by providing computing time on the GCS Supercomputer SuperMUC at Leibniz 
Supercomputing Centre (LRZ, www.lrz.de). The MDPL2 simulation has been performed 
under grant pr87yi. Our collaboration has been supported by the DFG grant GO 563/24-1.
This work was done in part using the Geryon computer at the
Center for Astro-Engineering UC, part of the BASAL PFB-06, which received
additional funding from QUIMAL 130008 and Fondequip AIC-57 for upgrades.
We thank the referee for useful comments and suggestions that have contributed to improve this work.
We acknowledge Tom\'as Tecce for his valuable contribution in 
initial stages of the development of the current version of \sag~and Mario Abadi for useful discussion.
SAC acknowledges funding from {\it Consejo Nacional de Investigaciones
Cient\'{\i}ficas y T\'ecnicas} (CONICET, PIP-0387), {\it Agencia Nacional
de Promoci\'on Cient\'ifica y Tecnol\'ogica} (ANPCyT, PICT-2013-0317), and {\it Universidad Nacional de La Plata} (G11-124),
Argentina.
CVM, TH, FC and IDG acknowledge CONICET, Argentina, for their supporting fellowships.
ANR acknowledges funding from ANPCyT (PICT-2014-2862) and from {\it Secretar\'ia de Ciencia y Tecnolog\'ia
de la Universidad Nacional de C\'ordoba} (PID 30720450100484).
AO acknowledges  support  from  project  AYA2015-66211-C2-2  of  the
Spanish {\it Ministerio de Economia, Industria y Competitividad}.
AMMA acknowledges support from CONICYT-PCHA/Doctorado Nacional 2011-21110870, BASAL PFB-06 and FONDECYT grant 3160776.
GY acknowledges  financial  support  from  the {\it Ministerio de Econom\'ia y 
Competitividad} and the {\it Fondo Europeo de Desarrollo Regional} 
(MINECO/FEDER, UE) in Spain through grant AYA2015-63810-P.

%%%%%%%%%%%%%%%%%%%%%%%%%%%%%%%%%%%%%%%%%%%%%%%%%%


\bibliographystyle{mnras}
\bibliography{references}

\begin{thebibliography}{}
\makeatletter
\relax
\def\mn@urlcharsother{\let\do\@makeother \do\$\do\&\do\#\do\^\do\_\do\%\do\~}
\def\mn@doi{\begingroup\mn@urlcharsother \@ifnextchar [ {\mn@doi@}
  {\mn@doi@[]}}
\def\mn@doi@[#1]#2{\def\@tempa{#1}\ifx\@tempa\@empty \href
  {http://dx.doi.org/#2} {doi:#2}\else \href {http://dx.doi.org/#2} {#1}\fi
  \endgroup}
\def\mn@eprint#1#2{\mn@eprint@#1:#2::\@nil}
\def\mn@eprint@arXiv#1{\href {http://arxiv.org/abs/#1} {{\tt arXiv:#1}}}
\def\mn@eprint@dblp#1{\href {http://dblp.uni-trier.de/rec/bibtex/#1.xml}
  {dblp:#1}}
\def\mn@eprint@#1:#2:#3:#4\@nil{\def\@tempa {#1}\def\@tempb {#2}\def\@tempc
  {#3}\ifx \@tempc \@empty \let \@tempc \@tempb \let \@tempb \@tempa \fi \ifx
  \@tempb \@empty \def\@tempb {arXiv}\fi \@ifundefined
  {mn@eprint@\@tempb}{\@tempb:\@tempc}{\expandafter \expandafter \csname
  mn@eprint@\@tempb\endcsname \expandafter{\@tempc}}}

\bibitem[\protect\citeauthoryear{{Abadi}, {Moore}  \& {Bower}}{{Abadi}
  et~al.}{1999}]{Abadi99}
{Abadi} M.~G.,  {Moore} B.,   {Bower} R.~G.,  1999, \mn@doi [\mnras]
  {10.1046/j.1365-8711.1999.02715.x}, \href
  {http://adsabs.harvard.edu/abs/1999MNRAS.308..947A} {308, 947}

\bibitem[\protect\citeauthoryear{{Abramson}, {Kenney}, {Crowl}, {Chung}, {van
  Gorkom}, {Vollmer}  \& {Schiminovich}}{{Abramson} et~al.}{2011}]{Abramson11}
{Abramson} A.,  {Kenney} J.~D.~P.,  {Crowl} H.~H.,  {Chung} A.,  {van Gorkom}
  J.~H.,  {Vollmer} B.,   {Schiminovich} D.,  2011, \mn@doi [\aj]
  {10.1088/0004-6256/141/5/164}, \href
  {http://adsabs.harvard.edu/abs/2011AJ....141..164A} {141, 164}

\bibitem[\protect\citeauthoryear{{Bah{\'e}} \& {McCarthy}}{{Bah{\'e}} \&
  {McCarthy}}{2015}]{Bahe15}
{Bah{\'e}} Y.~M.,  {McCarthy} I.~G.,  2015, \mn@doi [\mnras]
  {10.1093/mnras/stu2293}, \href
  {http://adsabs.harvard.edu/abs/2015MNRAS.447..969B} {447, 969}

\bibitem[\protect\citeauthoryear{{Bait}, {Barway}  \& {Wadadekar}}{{Bait}
  et~al.}{2017}]{Bait17}
{Bait} O.,  {Barway} S.,   {Wadadekar} Y.,  2017, \mn@doi [\mnras]
  {10.1093/mnras/stx1688}, \href
  {http://adsabs.harvard.edu/abs/2017MNRAS.471.2687B} {471, 2687}

\bibitem[\protect\citeauthoryear{{Baldry}, {Glazebrook}, {Brinkmann},
  {Ivezi{\'c}}, {Lupton}, {Nichol}  \& {Szalay}}{{Baldry}
  et~al.}{2004}]{Baldry04}
{Baldry} I.~K.,  {Glazebrook} K.,  {Brinkmann} J.,  {Ivezi{\'c}} {\v Z}.,
  {Lupton} R.~H.,  {Nichol} R.~C.,   {Szalay} A.~S.,  2004, \mn@doi [\apj]
  {10.1086/380092}, \href {http://adsabs.harvard.edu/abs/2004ApJ...600..681B}
  {600, 681}

\bibitem[\protect\citeauthoryear{{Baldry}, {Balogh}, {Bower}, {Glazebrook},
  {Nichol}, {Bamford}  \& {Budavari}}{{Baldry} et~al.}{2006}]{baldry2006}
{Baldry} I.~K.,  {Balogh} M.~L.,  {Bower} R.~G.,  {Glazebrook} K.,  {Nichol}
  R.~C.,  {Bamford} S.~P.,   {Budavari} T.,  2006, \mn@doi [\mnras]
  {10.1111/j.1365-2966.2006.11081.x}, \href
  {http://adsabs.harvard.edu/abs/2006MNRAS.373..469B} {373, 469}

\bibitem[\protect\citeauthoryear{{Baldry}, {Glazebrook}  \& {Driver}}{{Baldry}
  et~al.}{2008}]{baldry08}
{Baldry} I.~K.,  {Glazebrook} K.,   {Driver} S.~P.,  2008, \mn@doi [\mnras]
  {10.1111/j.1365-2966.2008.13348.x}, \href
  {http://adsabs.harvard.edu/abs/2008MNRAS.388..945B} {388, 945}

\bibitem[\protect\citeauthoryear{{Baldry} et~al.,}{{Baldry}
  et~al.}{2012}]{baldry_smf_2012}
{Baldry} I.~K.,  et~al., 2012, \mn@doi [\mnras]
  {10.1111/j.1365-2966.2012.20340.x}, \href
  {http://adsabs.harvard.edu/abs/2012MNRAS.421..621B} {421, 621}

\bibitem[\protect\citeauthoryear{{Balogh}, {Navarro}  \& {Morris}}{{Balogh}
  et~al.}{2000}]{balogh2000}
{Balogh} M.~L.,  {Navarro} J.~F.,   {Morris} S.~L.,  2000, \mn@doi [\apj]
  {10.1086/309323}, \href {http://adsabs.harvard.edu/abs/2000ApJ...540..113B}
  {540, 113}

\bibitem[\protect\citeauthoryear{{Bamford} et~al.,}{{Bamford}
  et~al.}{2009}]{Bamford09}
{Bamford} S.~P.,  et~al., 2009, \mn@doi [\mnras]
  {10.1111/j.1365-2966.2008.14252.x}, \href
  {http://adsabs.harvard.edu/abs/2009MNRAS.393.1324B} {393, 1324}

\bibitem[\protect\citeauthoryear{{Behroozi}, {Wechsler}  \& {Wu}}{{Behroozi}
  et~al.}{2013a}]{Behroozi_rockstar}
{Behroozi} P.~S.,  {Wechsler} R.~H.,   {Wu} H.-Y.,  2013a, \mn@doi [\apj]
  {10.1088/0004-637X/762/2/109}, \href
  {http://cdsads.u-strasbg.fr/abs/2013ApJ...762..109B} {762, 109}

\bibitem[\protect\citeauthoryear{{Behroozi}, {Wechsler}, {Wu}, {Busha},
  {Klypin}  \& {Primack}}{{Behroozi} et~al.}{2013b}]{Behroozi_ctrees}
{Behroozi} P.~S.,  {Wechsler} R.~H.,  {Wu} H.-Y.,  {Busha} M.~T.,  {Klypin}
  A.~A.,   {Primack} J.~R.,  2013b, \mn@doi [\apj]
  {10.1088/0004-637X/763/1/18}, \href
  {http://adsabs.harvard.edu/abs/2013ApJ...763...18B} {763, 18}

\bibitem[\protect\citeauthoryear{{Behroozi}, {Wechsler}  \&
  {Conroy}}{{Behroozi} et~al.}{2013c}]{behroozi13c}
{Behroozi} P.~S.,  {Wechsler} R.~H.,   {Conroy} C.,  2013c, \mn@doi [\apj]
  {10.1088/0004-637X/770/1/57}, \href
  {http://adsabs.harvard.edu/abs/2013ApJ...770...57B} {770, 57}

\bibitem[\protect\citeauthoryear{{Bekki}}{{Bekki}}{2009}]{bekki2009}
{Bekki} K.,  2009, \mn@doi [\mnras] {10.1111/j.1365-2966.2009.15431.x}, \href
  {http://adsabs.harvard.edu/abs/2009MNRAS.399.2221B} {399, 2221}

\bibitem[\protect\citeauthoryear{{Bekki}}{{Bekki}}{2014}]{Bekki14}
{Bekki} K.,  2014, \mn@doi [\mnras] {10.1093/mnras/stt2216}, \href
  {http://adsabs.harvard.edu/abs/2014MNRAS.438..444B} {438, 444}

\bibitem[\protect\citeauthoryear{{Bellhouse} et~al.,}{{Bellhouse}
  et~al.}{2017}]{Bellhouse17}
{Bellhouse} C.,  et~al., 2017, \mn@doi [\apj] {10.3847/1538-4357/aa7875}, \href
  {http://adsabs.harvard.edu/abs/2017ApJ...844...49B} {844, 49}

\bibitem[\protect\citeauthoryear{{Benson}}{{Benson}}{2012}]{Benson12}
{Benson} A.~J.,  2012, \mn@doi [\na] {10.1016/j.newast.2011.07.004}, \href
  {http://adsabs.harvard.edu/abs/2012NewA...17..175B} {17, 175}

\bibitem[\protect\citeauthoryear{{Bernardi}, {Meert}, {Sheth}, {Fischer},
  {Huertas-Company}, {Maraston}, {Shankar}  \& {Vikram}}{{Bernardi}
  et~al.}{2017}]{Bernardi17}
{Bernardi} M.,  {Meert} A.,  {Sheth} R.~K.,  {Fischer} J.-L.,
  {Huertas-Company} M.,  {Maraston} C.,  {Shankar} F.,   {Vikram} V.,  2017,
  \mn@doi [\mnras] {10.1093/mnras/stx176}, \href
  {http://adsabs.harvard.edu/abs/2017MNRAS.tmp..182B} {}

\bibitem[\protect\citeauthoryear{{Boselli} \& {Gavazzi}}{{Boselli} \&
  {Gavazzi}}{2006}]{BoselliGavazzi06}
{Boselli} A.,  {Gavazzi} G.,  2006, \mn@doi [\pasp] {10.1086/500691}, \href
  {http://adsabs.harvard.edu/abs/2006PASP..118..517B} {118, 517}

\bibitem[\protect\citeauthoryear{{Boselli}, {Cortese}, {Boquien}, {Boissier},
  {Catinella}, {Lagos}  \& {Saintonge}}{{Boselli} et~al.}{2014}]{boselli14}
{Boselli} A.,  {Cortese} L.,  {Boquien} M.,  {Boissier} S.,  {Catinella} B.,
  {Lagos} C.,   {Saintonge} A.,  2014, \mn@doi [\aap]
  {10.1051/0004-6361/201322312}, \href
  {http://adsabs.harvard.edu/abs/2014A%26A...564A..66B} {564, A66}

\bibitem[\protect\citeauthoryear{{Boselli} et~al.,}{{Boselli}
  et~al.}{2016}]{Boselli16}
{Boselli} A.,  et~al., 2016, \mn@doi [\aap] {10.1051/0004-6361/201629221},
  \href {http://adsabs.harvard.edu/abs/2016A%26A...596A..11B} {596, A11}

\bibitem[\protect\citeauthoryear{{Brown} et~al.,}{{Brown}
  et~al.}{2017}]{Brown17}
{Brown} T.,  et~al., 2017, \mn@doi [\mnras] {10.1093/mnras/stw2991}, \href
  {http://adsabs.harvard.edu/abs/2017MNRAS.466.1275B} {466, 1275}

\bibitem[\protect\citeauthoryear{{Br{\"u}ggen} \& {De Lucia}}{{Br{\"u}ggen} \&
  {De Lucia}}{2008}]{bdl2008}
{Br{\"u}ggen} M.,  {De Lucia} G.,  2008, \mn@doi [\mnras]
  {10.1111/j.1365-2966.2007.12670.x}, \href
  {http://adsabs.harvard.edu/abs/2008MNRAS.383.1336B} {383, 1336}

\bibitem[\protect\citeauthoryear{{Cameron}}{{Cameron}}{2011}]{Cameron11}
{Cameron} E.,  2011, \mn@doi [\pasa] {10.1071/AS10046}, \href
  {http://adsabs.harvard.edu/abs/2011PASA...28..128C} {28, 128}

\bibitem[\protect\citeauthoryear{{Chabrier}}{{Chabrier}}{2003}]{Chabrier03}
{Chabrier} G.,  2003, \mn@doi [\pasp] {10.1086/376392}, \href
  {http://adsabs.harvard.edu/abs/2003PASP..115..763C} {115, 763}

\bibitem[\protect\citeauthoryear{{Chang}, {Macci{\`o}}  \& {Kang}}{{Chang}
  et~al.}{2013}]{Chang13}
{Chang} J.,  {Macci{\`o}} A.~V.,   {Kang} X.,  2013, \mn@doi [\mnras]
  {10.1093/mnras/stt434}, \href
  {http://adsabs.harvard.edu/abs/2013MNRAS.431.3533C} {431, 3533}

\bibitem[\protect\citeauthoryear{{Chen} et~al.,}{{Chen} et~al.}{2012}]{Chen12}
{Chen} Y.-M.,  et~al., 2012, \mn@doi [\mnras]
  {10.1111/j.1365-2966.2011.20306.x}, \href
  {http://adsabs.harvard.edu/abs/2012MNRAS.421..314C} {421, 314}

\bibitem[\protect\citeauthoryear{{Coenda}, {Mart{\'{\i}}nez}  \&
  {Muriel}}{{Coenda} et~al.}{2018}]{Coenda18}
{Coenda} V.,  {Mart{\'{\i}}nez} H.~J.,   {Muriel} H.,  2018, \mn@doi [\mnras]
  {10.1093/mnras/stx2707}, \href
  {http://adsabs.harvard.edu/abs/2018MNRAS.473.5617C} {473, 5617}

\bibitem[\protect\citeauthoryear{{Cole}, {Lacey}, {Baugh}  \& {Frenk}}{{Cole}
  et~al.}{2000}]{cole2000}
{Cole} S.,  {Lacey} C.~G.,  {Baugh} C.~M.,   {Frenk} C.~S.,  2000, \mn@doi
  [\mnras] {10.1046/j.1365-8711.2000.03879.x}, \href
  {http://adsabs.harvard.edu/abs/2000MNRAS.319..168C} {319, 168}

\bibitem[\protect\citeauthoryear{{Conselice}}{{Conselice}}{2006}]{conselice2006}
{Conselice} C.~J.,  2006, \mn@doi [\mnras] {10.1111/j.1365-2966.2006.11114.x},
  \href {http://adsabs.harvard.edu/abs/2006MNRAS.373.1389C} {373, 1389}

\bibitem[\protect\citeauthoryear{{Cora}}{{Cora}}{2006}]{cora2006}
{Cora} S.~A.,  2006, \mn@doi [\mnras] {10.1111/j.1365-2966.2006.10271.x}, \href
  {http://adsabs.harvard.edu/abs/2006MNRAS.368.1540C} {368, 1540}

\bibitem[\protect\citeauthoryear{{Cora}, {Hough}, {Vega-Mart{\'{\i}}nez}  \&
  {Orsi}}{{Cora} et~al.}{2018}]{Cora18b}
{Cora} S.~A.,  {Hough} T.,  {Vega-Mart{\'{\i}}nez} C.~A.,   {Orsi} {\'A}.,
  2018, preprint, \href {http://adsabs.harvard.edu/abs/2018arXiv180103884C} {}
  (\mn@eprint {arXiv} {1801.03884})

\bibitem[\protect\citeauthoryear{{Croton} et~al.,}{{Croton}
  et~al.}{2016}]{Croton16}
{Croton} D.~J.,  et~al., 2016, \mn@doi [\apjs] {10.3847/0067-0049/222/2/22},
  \href {http://adsabs.harvard.edu/abs/2016ApJS..222...22C} {222, 22}

\bibitem[\protect\citeauthoryear{{Daddi} et~al.,}{{Daddi}
  et~al.}{2007}]{Daddi07}
{Daddi} E.,  et~al., 2007, \mn@doi [\apj] {10.1086/521818}, \href
  {http://adsabs.harvard.edu/abs/2007ApJ...670..156D} {670, 156}

\bibitem[\protect\citeauthoryear{{Darvish}, {Mobasher}, {Martin}, {Sobral},
  {Scoville}, {Stroe}, {Hemmati}  \& {Kartaltepe}}{{Darvish}
  et~al.}{2017}]{Darvish17}
{Darvish} B.,  {Mobasher} B.,  {Martin} D.~C.,  {Sobral} D.,  {Scoville} N.,
  {Stroe} A.,  {Hemmati} S.,   {Kartaltepe} J.,  2017, \mn@doi [\apj]
  {10.3847/1538-4357/837/1/16}, \href
  {http://adsabs.harvard.edu/abs/2017ApJ...837...16D} {837, 16}

\bibitem[\protect\citeauthoryear{{Dom{\'{\i}}nguez S{\'a}nchez}
  et~al.,}{{Dom{\'{\i}}nguez S{\'a}nchez} et~al.}{2011}]{dominguezsanchez11}
{Dom{\'{\i}}nguez S{\'a}nchez} H.,  et~al., 2011, \mn@doi [\mnras]
  {10.1111/j.1365-2966.2011.19263.x}, \href
  {http://adsabs.harvard.edu/abs/2011MNRAS.417..900D} {417, 900}

\bibitem[\protect\citeauthoryear{{Fillingham}, {Cooper}, {Pace},
  {Boylan-Kolchin}, {Bullock}, {Garrison-Kimmel}  \& {Wheeler}}{{Fillingham}
  et~al.}{2016}]{Fillingham16}
{Fillingham} S.~P.,  {Cooper} M.~C.,  {Pace} A.~B.,  {Boylan-Kolchin} M.,
  {Bullock} J.~S.,  {Garrison-Kimmel} S.,   {Wheeler} C.,  2016, \mn@doi
  [\mnras] {10.1093/mnras/stw2131}, \href
  {http://adsabs.harvard.edu/abs/2016MNRAS.463.1916F} {463, 1916}

\bibitem[\protect\citeauthoryear{{Font}, {Bower}, {McCarthy}  \& et al.}{{Font}
  et~al.}{2008}]{font2008}
{Font} A.~S.,  {Bower} R.~G.,  {McCarthy} I.~G.,   et al. 2008, \mn@doi
  [\mnras] {10.1111/j.1365-2966.2008.13698.x}, \href
  {http://adsabs.harvard.edu/abs/2008MNRAS.389.1619F} {389, 1619}

\bibitem[\protect\citeauthoryear{{Foster}, {Ji}, {Smith}  \&
  {Brickhouse}}{{Foster} et~al.}{2012}]{foster2012}
{Foster} A.~R.,  {Ji} L.,  {Smith} R.~K.,   {Brickhouse} N.~S.,  2012, \mn@doi
  [\apj] {10.1088/0004-637X/756/2/128}, \href
  {http://adsabs.harvard.edu/abs/2012ApJ...756..128F} {756, 128}

\bibitem[\protect\citeauthoryear{{Fu} et~al.,}{{Fu} et~al.}{2013}]{Fu13}
{Fu} J.,  et~al., 2013, \mn@doi [\mnras] {10.1093/mnras/stt1117}, \href
  {http://adsabs.harvard.edu/abs/2013MNRAS.434.1531F} {434, 1531}

\bibitem[\protect\citeauthoryear{{Gan}, {Kang}, {van den Bosch}  \&
  {Hou}}{{Gan} et~al.}{2010}]{gan2010}
{Gan} J.,  {Kang} X.,  {van den Bosch} F.~C.,   {Hou} J.,  2010, \mn@doi
  [\mnras] {10.1111/j.1365-2966.2010.17266.x}, \href
  {http://adsabs.harvard.edu/abs/2010MNRAS.408.2201G} {408, 2201}

\bibitem[\protect\citeauthoryear{{Gargiulo} et~al.,}{{Gargiulo}
  et~al.}{2015}]{Gargiulo15}
{Gargiulo} I.~D.,  et~al., 2015, \mn@doi [\mnras] {10.1093/mnras/stu2272},
  \href {http://adsabs.harvard.edu/abs/2015MNRAS.446.3820G} {446, 3820}

\bibitem[\protect\citeauthoryear{{Gonzalez-Perez}, {Lacey}, {Baugh}, {Lagos},
  {Helly}, {Campbell}  \& {Mitchell}}{{Gonzalez-Perez}
  et~al.}{2014}]{GonzalezPerez14}
{Gonzalez-Perez} V.,  {Lacey} C.~G.,  {Baugh} C.~M.,  {Lagos} C.~D.~P.,
  {Helly} J.,  {Campbell} D.~J.~R.,   {Mitchell} P.~D.,  2014, \mn@doi [\mnras]
  {10.1093/mnras/stt2410}, \href
  {http://adsabs.harvard.edu/abs/2014MNRAS.439..264G} {439, 264}

\bibitem[\protect\citeauthoryear{{Greggio} \& {Renzini}}{{Greggio} \&
  {Renzini}}{1983}]{gr83}
{Greggio} L.,  {Renzini} A.,  1983, \aap, \href
  {http://adsabs.harvard.edu/abs/1983A%26A...118..217G} {118, 217}

\bibitem[\protect\citeauthoryear{{Gruppioni} et~al.,}{{Gruppioni}
  et~al.}{2015}]{gruppioni15}
{Gruppioni} C.,  et~al., 2015, \mn@doi [\mnras] {10.1093/mnras/stv1204}, \href
  {http://adsabs.harvard.edu/abs/2015MNRAS.451.3419G} {451, 3419}

\bibitem[\protect\citeauthoryear{{Gunn} \& {Gott}}{{Gunn} \&
  {Gott}}{1972}]{gg72}
{Gunn} J.~E.,  {Gott} J.~R.~I.,  1972, \apj, \href
  {http://adsabs.harvard.edu/abs/1972ApJ...176....1G} {176, 1}

\bibitem[\protect\citeauthoryear{{Guo}, {White}, {Boylan-Kolchin}  \& et
  al.}{{Guo} et~al.}{2011}]{guo11}
{Guo} Q.,  {White} S.,  {Boylan-Kolchin} M.,   et al. 2011, \mn@doi [\mnras]
  {10.1111/j.1365-2966.2010.18114.x}, \href
  {http://adsabs.harvard.edu/abs/2011MNRAS.413..101G} {413, 101}

\bibitem[\protect\citeauthoryear{{Guo} et~al.,}{{Guo} et~al.}{2016}]{guo16}
{Guo} Q.,  et~al., 2016, \mn@doi [\mnras] {10.1093/mnras/stw1525}, \href
  {http://adsabs.harvard.edu/abs/2016MNRAS.461.3457G} {461, 3457}

\bibitem[\protect\citeauthoryear{{Henriques}, {White}, {Thomas}, {Angulo},
  {Guo}, {Lemson}  \& {Springel}}{{Henriques} et~al.}{2013}]{henriques13}
{Henriques} B.~M.~B.,  {White} S.~D.~M.,  {Thomas} P.~A.,  {Angulo} R.~E.,
  {Guo} Q.,  {Lemson} G.,   {Springel} V.,  2013, \mn@doi [\mnras]
  {10.1093/mnras/stt415}, \href
  {http://adsabs.harvard.edu/abs/2013MNRAS.431.3373H} {431, 3373}

\bibitem[\protect\citeauthoryear{{Henriques}, {White}, {Thomas}, {Angulo},
  {Guo}, {Lemson}, {Springel}  \& {Overzier}}{{Henriques}
  et~al.}{2015}]{henriques_mcmc_2015}
{Henriques} B.~M.~B.,  {White} S.~D.~M.,  {Thomas} P.~A.,  {Angulo} R.,  {Guo}
  Q.,  {Lemson} G.,  {Springel} V.,   {Overzier} R.,  2015, \mn@doi [\mnras]
  {10.1093/mnras/stv705}, \href
  {http://adsabs.harvard.edu/abs/2015MNRAS.451.2663H} {451, 2663}

\bibitem[\protect\citeauthoryear{{Henriques}, {White}, {Thomas}, {Angulo},
  {Guo}, {Lemson}  \& {Wang}}{{Henriques} et~al.}{2017}]{Henriques17}
{Henriques} B.~M.~B.,  {White} S.~D.~M.,  {Thomas} P.~A.,  {Angulo} R.~E.,
  {Guo} Q.,  {Lemson} G.,   {Wang} W.,  2017, \mn@doi [\mnras]
  {10.1093/mnras/stx1010}, \href
  {http://adsabs.harvard.edu/abs/2017MNRAS.469.2626H} {469, 2626}

\bibitem[\protect\citeauthoryear{{Hernquist}}{{Hernquist}}{1990}]{hernquist1990}
{Hernquist} L.,  1990, \mn@doi [\apj] {10.1086/168845}, \href
  {http://adsabs.harvard.edu/abs/1990ApJ...356..359H} {356, 359}

\bibitem[\protect\citeauthoryear{{Hirschmann}, {De Lucia}, {Wilman},
  {Weinmann}, {Iovino}, {Cucciati}, {Zibetti}  \& {Villalobos}}{{Hirschmann}
  et~al.}{2014}]{Hirschmann14}
{Hirschmann} M.,  {De Lucia} G.,  {Wilman} D.,  {Weinmann} S.,  {Iovino} A.,
  {Cucciati} O.,  {Zibetti} S.,   {Villalobos} {\'A}.,  2014, \mn@doi [\mnras]
  {10.1093/mnras/stu1609}, \href
  {http://adsabs.harvard.edu/abs/2014MNRAS.444.2938H} {444, 2938}

\bibitem[\protect\citeauthoryear{{Hirschmann}, {De Lucia}  \&
  {Fontanot}}{{Hirschmann} et~al.}{2016}]{hirschmann16}
{Hirschmann} M.,  {De Lucia} G.,   {Fontanot} F.,  2016, \mn@doi [\mnras]
  {10.1093/mnras/stw1318}, \href
  {http://cdsads.u-strasbg.fr/abs/2016MNRAS.461.1760H} {461, 1760}

\bibitem[\protect\citeauthoryear{{Ilbert} et~al.,}{{Ilbert}
  et~al.}{2013}]{ilbert13}
{Ilbert} O.,  et~al., 2013, \mn@doi [\aap] {10.1051/0004-6361/201321100}, \href
  {http://adsabs.harvard.edu/abs/2013A%26A...556A..55I} {556, A55}

\bibitem[\protect\citeauthoryear{{Jaff{\'e}}, {Poggianti}, {Verheijen},
  {Deshev}  \& {van Gorkom}}{{Jaff{\'e}} et~al.}{2013}]{Jaffe13}
{Jaff{\'e}} Y.~L.,  {Poggianti} B.~M.,  {Verheijen} M.~A.~W.,  {Deshev} B.~Z.,
   {van Gorkom} J.~H.,  2013, \mn@doi [\mnras] {10.1093/mnras/stt250}, \href
  {http://adsabs.harvard.edu/abs/2013MNRAS.431.2111J} {431, 2111}

\bibitem[\protect\citeauthoryear{{Jaff{\'e}}, {Smith}, {Candlish}, {Poggianti},
  {Sheen}  \& {Verheijen}}{{Jaff{\'e}} et~al.}{2015}]{Jaffe15}
{Jaff{\'e}} Y.~L.,  {Smith} R.,  {Candlish} G.~N.,  {Poggianti} B.~M.,  {Sheen}
  Y.-K.,   {Verheijen} M.~A.~W.,  2015, \mn@doi [\mnras]
  {10.1093/mnras/stv100}, \href
  {http://adsabs.harvard.edu/abs/2015MNRAS.448.1715J} {448, 1715}

\bibitem[\protect\citeauthoryear{{Jaff{\'e}} et~al.,}{{Jaff{\'e}}
  et~al.}{2016}]{Jaffe16}
{Jaff{\'e}} Y.~L.,  et~al., 2016, \mn@doi [\mnras] {10.1093/mnras/stw984},
  \href {http://adsabs.harvard.edu/abs/2016MNRAS.461.1202J} {461, 1202}

\bibitem[\protect\citeauthoryear{{Jeltema}, {Binder}  \& {Mulchaey}}{{Jeltema}
  et~al.}{2008}]{jeltema2008}
{Jeltema} T.~E.,  {Binder} B.,   {Mulchaey} J.~S.,  2008, \mn@doi [\apj]
  {10.1086/587508}, \href {http://adsabs.harvard.edu/abs/2008ApJ...679.1162J}
  {679, 1162}

\bibitem[\protect\citeauthoryear{{Jian} et~al.,}{{Jian} et~al.}{2017}]{Jian17}
{Jian} H.-Y.,  et~al., 2017, \mn@doi [\apj] {10.3847/1538-4357/aa7de2}, \href
  {http://adsabs.harvard.edu/abs/2017ApJ...845...74J} {845, 74}

\bibitem[\protect\citeauthoryear{{Jiang}, {Jing}, {Faltenbacher}, {Lin}  \&
  {Li}}{{Jiang} et~al.}{2008}]{jiang2008}
{Jiang} C.~Y.,  {Jing} Y.~P.,  {Faltenbacher} A.,  {Lin} W.~P.,   {Li} C.,
  2008, \mn@doi [\apj] {10.1086/526412}, \href
  {http://adsabs.harvard.edu/abs/2008ApJ...675.1095J} {675, 1095}

\bibitem[\protect\citeauthoryear{{Jim{\'e}nez}, {Cora}, {Bassino}, {Tecce}  \&
  {Smith Castelli}}{{Jim{\'e}nez} et~al.}{2011}]{Jimenez11}
{Jim{\'e}nez} N.,  {Cora} S.~A.,  {Bassino} L.~P.,  {Tecce} T.~E.,   {Smith
  Castelli} A.~V.,  2011, \mn@doi [\mnras] {10.1111/j.1365-2966.2011.19328.x},
  \href {http://adsabs.harvard.edu/abs/2011MNRAS.417..785J} {417, 785}

\bibitem[\protect\citeauthoryear{{Kang}}{{Kang}}{2014}]{Kang14}
{Kang} X.,  2014, \mn@doi [\mnras] {10.1093/mnras/stt2132}, \href
  {http://adsabs.harvard.edu/abs/2014MNRAS.437.3385K} {437, 3385}

\bibitem[\protect\citeauthoryear{{Kang} \& {van den Bosch}}{{Kang} \& {van den
  Bosch}}{2008}]{Kang08}
{Kang} X.,  {van den Bosch} F.~C.,  2008, \mn@doi [\apjl] {10.1086/587620},
  \href {http://adsabs.harvard.edu/abs/2008ApJ...676L.101K} {676, L101}

\bibitem[\protect\citeauthoryear{{Kannan}, {Macci{\`o}}, {Fontanot}, {Moster},
  {Karman}  \& {Somerville}}{{Kannan} et~al.}{2015}]{Kannan15}
{Kannan} R.,  {Macci{\`o}} A.~V.,  {Fontanot} F.,  {Moster} B.~P.,  {Karman}
  W.,   {Somerville} R.~S.,  2015, \mn@doi [\mnras] {10.1093/mnras/stv1633},
  \href {http://adsabs.harvard.edu/abs/2015MNRAS.452.4347K} {452, 4347}

\bibitem[\protect\citeauthoryear{{Kauffmann}}{{Kauffmann}}{2014}]{Kauffmann14}
{Kauffmann} G.,  2014, \mn@doi [\mnras] {10.1093/mnras/stu752}, \href
  {http://adsabs.harvard.edu/abs/2014MNRAS.441.2717K} {441, 2717}

\bibitem[\protect\citeauthoryear{{Kawinwanichakij} et~al.,}{{Kawinwanichakij}
  et~al.}{2017}]{Kawinwanichakij17}
{Kawinwanichakij} L.,  et~al., 2017, preprint, \href
  {http://adsabs.harvard.edu/abs/2017arXiv170603780K} {} (\mn@eprint {arXiv}
  {1706.03780})

\bibitem[\protect\citeauthoryear{{Kazantzidis}, {{\L}okas}, {Callegari},
  {Mayer}  \& {Moustakas}}{{Kazantzidis} et~al.}{2011}]{Kazantzidis11}
{Kazantzidis} S.,  {{\L}okas} E.~L.,  {Callegari} S.,  {Mayer} L.,
  {Moustakas} L.~A.,  2011, \mn@doi [\apj] {10.1088/0004-637X/726/2/98}, \href
  {http://adsabs.harvard.edu/abs/2011ApJ...726...98K} {726, 98}

\bibitem[\protect\citeauthoryear{{Kenney}, {Abramson}  \&
  {Bravo-Alfaro}}{{Kenney} et~al.}{2015}]{Kenney15}
{Kenney} J.~D.~P.,  {Abramson} A.,   {Bravo-Alfaro} H.,  2015, \mn@doi [\aj]
  {10.1088/0004-6256/150/2/59}, \href
  {http://adsabs.harvard.edu/abs/2015AJ....150...59K} {150, 59}

\bibitem[\protect\citeauthoryear{{Kimm}, {Somerville}, {Yi}  \& et al.}{{Kimm}
  et~al.}{2009}]{kimm2009}
{Kimm} T.,  {Somerville} R.~S.,  {Yi} S.~K.,   et al. 2009, \mn@doi [\mnras]
  {10.1111/j.1365-2966.2009.14414.x}, \href
  {http://adsabs.harvard.edu/abs/2009MNRAS.394.1131K} {394, 1131}

\bibitem[\protect\citeauthoryear{{Kimm}, {Yi}  \& {Khochfar}}{{Kimm}
  et~al.}{2011}]{kimm2011}
{Kimm} T.,  {Yi} S.~K.,   {Khochfar} S.,  2011, \mn@doi [\apj]
  {10.1088/0004-637X/729/1/11}, \href
  {http://adsabs.harvard.edu/abs/2011ApJ...729...11K} {729, 11}

\bibitem[\protect\citeauthoryear{{Klypin}, {Yepes}, {Gottl{\"o}ber}, {Prada}
  \& {He{\ss}}}{{Klypin} et~al.}{2016}]{Klypin16}
{Klypin} A.,  {Yepes} G.,  {Gottl{\"o}ber} S.,  {Prada} F.,   {He{\ss}} S.,
  2016, \mn@doi [\mnras] {10.1093/mnras/stw248}, \href
  {http://cdsads.u-strasbg.fr/abs/2016MNRAS.457.4340K} {457, 4340}

\bibitem[\protect\citeauthoryear{{Knebe} et~al.,}{{Knebe}
  et~al.}{2015}]{knebe15}
{Knebe} A.,  et~al., 2015, \mn@doi [\mnras] {10.1093/mnras/stv1149}, \href
  {http://adsabs.harvard.edu/abs/2015MNRAS.451.4029K} {451, 4029}

\bibitem[\protect\citeauthoryear{{Knebe} et~al.,}{{Knebe}
  et~al.}{2017b}]{Knebe18}
{Knebe} A.,  et~al., 2017b, preprint, \href
  {http://adsabs.harvard.edu/abs/2017arXiv171206420K} {} (\mn@eprint {arXiv}
  {1712.06420})

\bibitem[\protect\citeauthoryear{{Knebe} et~al.,}{{Knebe}
  et~al.}{2017a}]{knebe17}
{Knebe} A.,  et~al., 2017a, preprint, \href
  {http://adsabs.harvard.edu/abs/2017arXiv171008150K} {} (\mn@eprint {arXiv}
  {1710.08150})

\bibitem[\protect\citeauthoryear{{Kormendy} \& {Ho}}{{Kormendy} \&
  {Ho}}{2013}]{kormendy_bhb_2013}
{Kormendy} J.,  {Ho} L.~C.,  2013, \mn@doi [\araa]
  {10.1146/annurev-astro-082708-101811}, \href
  {http://adsabs.harvard.edu/abs/2013ARA%26A..51..511K} {51, 511}

\bibitem[\protect\citeauthoryear{{Kova{\v c}} et~al.,}{{Kova{\v c}}
  et~al.}{2014}]{Kovac14}
{Kova{\v c}} K.,  et~al., 2014, \mn@doi [\mnras] {10.1093/mnras/stt2241}, \href
  {http://adsabs.harvard.edu/abs/2014MNRAS.438..717K} {438, 717}

\bibitem[\protect\citeauthoryear{{Lagos}, {Cora}  \& {Padilla}}{{Lagos}
  et~al.}{2008}]{lcp08}
{Lagos} C.~D.~P.,  {Cora} S.~A.,   {Padilla} N.~D.,  2008, \mn@doi [\mnras]
  {10.1111/j.1365-2966.2008.13456.x}, \href
  {http://adsabs.harvard.edu/abs/2008MNRAS.388..587L} {388, 587}

\bibitem[\protect\citeauthoryear{{Lagos}, {Lacey}, {Baugh}, {Bower}  \&
  {Benson}}{{Lagos} et~al.}{2011}]{Lagos11a}
{Lagos} C.~D.~P.,  {Lacey} C.~G.,  {Baugh} C.~M.,  {Bower} R.~G.,   {Benson}
  A.~J.,  2011, \mn@doi [\mnras] {10.1111/j.1365-2966.2011.19160.x}, \href
  {http://adsabs.harvard.edu/abs/2011MNRAS.416.1566L} {416, 1566}

\bibitem[\protect\citeauthoryear{{Lanzoni}, {Guiderdoni}, {Mamon}, {Devriendt}
  \& {Hatton}}{{Lanzoni} et~al.}{2005}]{lanzoni2005}
{Lanzoni} B.,  {Guiderdoni} B.,  {Mamon} G.~A.,  {Devriendt} J.,   {Hatton} S.,
   2005, \mn@doi [\mnras] {10.1111/j.1365-2966.2005.09252.x}, \href
  {http://adsabs.harvard.edu/abs/2005MNRAS.361..369L} {361, 369}

\bibitem[\protect\citeauthoryear{{Larson}, {Tinsley}  \& {Caldwell}}{{Larson}
  et~al.}{1980}]{larson80}
{Larson} R.~B.,  {Tinsley} B.~M.,   {Caldwell} C.~N.,  1980, \mn@doi [\apj]
  {10.1086/157917}, \href {http://adsabs.harvard.edu/abs/1980ApJ...237..692L}
  {237, 692}

\bibitem[\protect\citeauthoryear{{Li} \& {White}}{{Li} \&
  {White}}{2009}]{li_smf_2009}
{Li} C.,  {White} S.~D.~M.,  2009, \mn@doi [\mnras]
  {10.1111/j.1365-2966.2009.15268.x}, \href
  {http://adsabs.harvard.edu/abs/2009MNRAS.398.2177L} {398, 2177}

\bibitem[\protect\citeauthoryear{{Lia}, {Portinari}  \& {Carraro}}{{Lia}
  et~al.}{2002}]{Lia2002}
{Lia} C.,  {Portinari} L.,   {Carraro} G.,  2002, \mn@doi [MNRAS]
  {10.1046/j.1365-8711.2002.05118.x}, \href
  {http://adsabs.harvard.edu/abs/2002MNRAS.330..821L} {330, 821}

\bibitem[\protect\citeauthoryear{{Lin} et~al.,}{{Lin} et~al.}{2014}]{Lin14}
{Lin} L.,  et~al., 2014, \mn@doi [\apj] {10.1088/0004-637X/782/1/33}, \href
  {http://adsabs.harvard.edu/abs/2014ApJ...782...33L} {782, 33}

\bibitem[\protect\citeauthoryear{{Luo}, {Kang}, {Kauffmann}  \& {Fu}}{{Luo}
  et~al.}{2016}]{luo16}
{Luo} Y.,  {Kang} X.,  {Kauffmann} G.,   {Fu} J.,  2016, \mn@doi [\mnras]
  {10.1093/mnras/stw268}, \href
  {http://cdsads.u-strasbg.fr/abs/2016MNRAS.458..366L} {458, 366}

\bibitem[\protect\citeauthoryear{{Madau} \& {Dickinson}}{{Madau} \&
  {Dickinson}}{2014}]{Madau2014}
{Madau} P.,  {Dickinson} M.,  2014, \mn@doi [\araa]
  {10.1146/annurev-astro-081811-125615}, \href
  {http://adsabs.harvard.edu/abs/2014ARA%26A..52..415M} {52, 415}

\bibitem[\protect\citeauthoryear{{McCarthy}, {Frenk}, {Font}, {Lacey}, {Bower},
  {Mitchell}, {Balogh}  \& {Theuns}}{{McCarthy} et~al.}{2008}]{mccarthy2008}
{McCarthy} I.~G.,  {Frenk} C.~S.,  {Font} A.~S.,  {Lacey} C.~G.,  {Bower}
  R.~G.,  {Mitchell} N.~L.,  {Balogh} M.~L.,   {Theuns} T.,  2008, \mn@doi
  [\mnras] {10.1111/j.1365-2966.2007.12577.x}, \href
  {http://adsabs.harvard.edu/abs/2008MNRAS.383..593M} {383, 593}

\bibitem[\protect\citeauthoryear{{McConnell} \& {Ma}}{{McConnell} \&
  {Ma}}{2013}]{mcconnell_bhb_2013}
{McConnell} N.~J.,  {Ma} C.-P.,  2013, \mn@doi [\apj]
  {10.1088/0004-637X/764/2/184}, \href
  {http://adsabs.harvard.edu/abs/2013ApJ...764..184M} {764, 184}

\bibitem[\protect\citeauthoryear{{Mendel}, {Simard}, {Palmer}, {Ellison}  \&
  {Patton}}{{Mendel} et~al.}{2014}]{Mendel14}
{Mendel} J.~T.,  {Simard} L.,  {Palmer} M.,  {Ellison} S.~L.,   {Patton} D.~R.,
   2014, \mn@doi [\apjs] {10.1088/0067-0049/210/1/3}, \href
  {http://adsabs.harvard.edu/abs/2014ApJS..210....3M} {210, 3}

\bibitem[\protect\citeauthoryear{{Merritt}}{{Merritt}}{1983}]{Merritt83}
{Merritt} D.,  1983, \mn@doi [\apj] {10.1086/160571}, \href
  {http://adsabs.harvard.edu/abs/1983ApJ...264...24M} {264, 24}

\bibitem[\protect\citeauthoryear{{Merritt}, {van Dokkum}, {Abraham}  \&
  {Zhang}}{{Merritt} et~al.}{2016}]{Merritt16}
{Merritt} A.,  {van Dokkum} P.,  {Abraham} R.,   {Zhang} J.,  2016, \mn@doi
  [\apj] {10.3847/0004-637X/830/2/62}, \href
  {http://adsabs.harvard.edu/abs/2016ApJ...830...62M} {830, 62}

\bibitem[\protect\citeauthoryear{{Mo}, {Mao}  \& {White}}{{Mo}
  et~al.}{1998}]{mmw98}
{Mo} H.~J.,  {Mao} S.,   {White} S.~D.~M.,  1998, \mnras, \href
  {http://adsabs.harvard.edu/abs/1998MNRAS.295..319M} {295, 319}

\bibitem[\protect\citeauthoryear{{Moster}, {Somerville}, {Maulbetsch}, {van den
  Bosch}, {Macci{\`o}}, {Naab}  \& {Oser}}{{Moster} et~al.}{2010}]{Moster10}
{Moster} B.~P.,  {Somerville} R.~S.,  {Maulbetsch} C.,  {van den Bosch} F.~C.,
  {Macci{\`o}} A.~V.,  {Naab} T.,   {Oser} L.,  2010, \mn@doi [\apj]
  {10.1088/0004-637X/710/2/903}, \href
  {http://adsabs.harvard.edu/abs/2010ApJ...710..903M} {710, 903}

\bibitem[\protect\citeauthoryear{{Mu{\~n}oz Arancibia}, {Navarrete}, {Padilla},
  {Cora}, {Gawiser}, {Kurczynski}  \& {Ruiz}}{{Mu{\~n}oz Arancibia}
  et~al.}{2015}]{munnozarancibia2015}
{Mu{\~n}oz Arancibia} A.~M.,  {Navarrete} F.~P.,  {Padilla} N.~D.,  {Cora}
  S.~A.,  {Gawiser} E.,  {Kurczynski} P.,   {Ruiz} A.~N.,  2015, \mn@doi
  [\mnras] {10.1093/mnras/stu2237}, \href
  {http://adsabs.harvard.edu/abs/2015MNRAS.446.2291M} {446, 2291}

\bibitem[\protect\citeauthoryear{{Mulchaey} \& {Jeltema}}{{Mulchaey} \&
  {Jeltema}}{2010}]{mj2010}
{Mulchaey} J.~S.,  {Jeltema} T.~E.,  2010, \mn@doi [\apjl]
  {10.1088/2041-8205/715/1/L1}, \href
  {http://adsabs.harvard.edu/abs/2010ApJ...715L...1M} {715, L1}

\bibitem[\protect\citeauthoryear{{Muratov}, {Kere{\v s}},
  {Faucher-Gigu{\`e}re}, {Hopkins}, {Quataert}  \& {Murray}}{{Muratov}
  et~al.}{2015}]{muratov15}
{Muratov} A.~L.,  {Kere{\v s}} D.,  {Faucher-Gigu{\`e}re} C.-A.,  {Hopkins}
  P.~F.,  {Quataert} E.,   {Murray} N.,  2015, \mn@doi [\mnras]
  {10.1093/mnras/stv2126}, \href
  {http://cdsads.u-strasbg.fr/abs/2015MNRAS.454.2691M} {454, 2691}

\bibitem[\protect\citeauthoryear{{Muzzin} et~al.,}{{Muzzin}
  et~al.}{2012}]{Muzzin12}
{Muzzin} A.,  et~al., 2012, \mn@doi [\apj] {10.1088/0004-637X/746/2/188}, \href
  {http://adsabs.harvard.edu/abs/2012ApJ...746..188M} {746, 188}

\bibitem[\protect\citeauthoryear{{Muzzin} et~al.,}{{Muzzin}
  et~al.}{2013}]{muzzin13}
{Muzzin} A.,  et~al., 2013, \mn@doi [\apj] {10.1088/0004-637X/777/1/18}, \href
  {http://adsabs.harvard.edu/abs/2013ApJ...777...18M} {777, 18}

\bibitem[\protect\citeauthoryear{{Okamoto} \& {Nagashima}}{{Okamoto} \&
  {Nagashima}}{2003}]{on2003}
{Okamoto} T.,  {Nagashima} M.,  2003, \mn@doi [\apj] {10.1086/368251}, \href
  {http://adsabs.harvard.edu/abs/2003ApJ...587..500O} {587, 500}

\bibitem[\protect\citeauthoryear{{Orsi}, {Padilla}, {Groves}, {Cora}, {Tecce},
  {Gargiulo}  \& {Ruiz}}{{Orsi} et~al.}{2014}]{orsi14}
{Orsi} {\'A}.,  {Padilla} N.,  {Groves} B.,  {Cora} S.,  {Tecce} T.,
  {Gargiulo} I.,   {Ruiz} A.,  2014, \mn@doi [\mnras] {10.1093/mnras/stu1203},
  \href {http://adsabs.harvard.edu/abs/2014MNRAS.443..799O} {443, 799}

\bibitem[\protect\citeauthoryear{{Padovani} \& {Matteucci}}{{Padovani} \&
  {Matteucci}}{1993}]{pm93}
{Padovani} P.,  {Matteucci} F.,  1993, \mn@doi [\apj] {10.1086/173212}, \href
  {http://adsabs.harvard.edu/abs/1993ApJ...416...26P} {416, 26}

\bibitem[\protect\citeauthoryear{{Panter}, {Heavens}  \& {Jimenez}}{{Panter}
  et~al.}{2004}]{Panter04}
{Panter} B.,  {Heavens} A.~F.,   {Jimenez} R.,  2004, \mn@doi [\mnras]
  {10.1111/j.1365-2966.2004.08355.x}, \href
  {http://adsabs.harvard.edu/abs/2004MNRAS.355..764P} {355, 764}

\bibitem[\protect\citeauthoryear{{Peng} et~al.,}{{Peng} et~al.}{2010}]{Peng10}
{Peng} Y.-j.,  et~al., 2010, \mn@doi [\apj] {10.1088/0004-637X/721/1/193},
  \href {http://adsabs.harvard.edu/abs/2010ApJ...721..193P} {721, 193}

\bibitem[\protect\citeauthoryear{{Peng}, {Lilly}, {Renzini}  \&
  {Carollo}}{{Peng} et~al.}{2012}]{Peng12}
{Peng} Y.-j.,  {Lilly} S.~J.,  {Renzini} A.,   {Carollo} M.,  2012, \mn@doi
  [\apj] {10.1088/0004-637X/757/1/4}, \href
  {http://adsabs.harvard.edu/abs/2012ApJ...757....4P} {757, 4}

\bibitem[\protect\citeauthoryear{{Planck Collaboration} et~al.,}{{Planck
  Collaboration} et~al.}{2014}]{Planck2013}
{Planck Collaboration} et~al., 2014, \mn@doi [\aap]
  {10.1051/0004-6361/201321591}, \href
  {http://adsabs.harvard.edu/abs/2014A%26A...571A..16P} {571, A16}

\bibitem[\protect\citeauthoryear{{Poggianti} et~al.,}{{Poggianti}
  et~al.}{2017}]{Poggianti17}
{Poggianti} B.~M.,  et~al., 2017, \mn@doi [\apj] {10.3847/1538-4357/aa78ed},
  \href {http://adsabs.harvard.edu/abs/2017ApJ...844...48P} {844, 48}

\bibitem[\protect\citeauthoryear{{Powell}, {Urry}, {Cardamone}, {Simmons},
  {Schawinski}, {Young}  \& {Kawakatsu}}{{Powell} et~al.}{2017}]{Powell17}
{Powell} M.~C.,  {Urry} C.~M.,  {Cardamone} C.~N.,  {Simmons} B.~D.,
  {Schawinski} K.,  {Young} S.,   {Kawakatsu} M.,  2017, \mn@doi [\apj]
  {10.3847/1538-4357/835/1/22}, \href
  {http://adsabs.harvard.edu/abs/2017ApJ...835...22P} {835, 22}

\bibitem[\protect\citeauthoryear{{Quilis}, {Planelles}  \&
  {Ricciardelli}}{{Quilis} et~al.}{2017}]{Quilis17}
{Quilis} V.,  {Planelles} S.,   {Ricciardelli} E.,  2017, \mn@doi [\mnras]
  {10.1093/mnras/stx770}, \href
  {http://adsabs.harvard.edu/abs/2017MNRAS.469...80Q} {469, 80}

\bibitem[\protect\citeauthoryear{{Rodrigues}, {Vernon}  \& {Bower}}{{Rodrigues}
  et~al.}{2017}]{Rodrigues17}
{Rodrigues} L.~F.~S.,  {Vernon} I.,   {Bower} R.~G.,  2017, \mn@doi [\mnras]
  {10.1093/mnras/stw3269}, \href
  {http://adsabs.harvard.edu/abs/2017MNRAS.466.2418R} {466, 2418}

\bibitem[\protect\citeauthoryear{{Roediger} \& {Br{\"u}ggen}}{{Roediger} \&
  {Br{\"u}ggen}}{2006}]{RoedigerBrueggen06}
{Roediger} E.,  {Br{\"u}ggen} M.,  2006, \mn@doi [\mnras]
  {10.1111/j.1365-2966.2006.10335.x}, \href
  {http://adsabs.harvard.edu/abs/2006MNRAS.369..567R} {369, 567}

\bibitem[\protect\citeauthoryear{{Roediger} \& {Br{\"u}ggen}}{{Roediger} \&
  {Br{\"u}ggen}}{2007}]{RoedigerBrueggen07}
{Roediger} E.,  {Br{\"u}ggen} M.,  2007, \mn@doi [\mnras]
  {10.1111/j.1365-2966.2007.12241.x}, \href
  {http://adsabs.harvard.edu/abs/2007MNRAS.380.1399R} {380, 1399}

\bibitem[\protect\citeauthoryear{{Ruggiero} \& {Lima Neto}}{{Ruggiero} \& {Lima
  Neto}}{2017}]{Ruggiero17}
{Ruggiero} R.,  {Lima Neto} G.~B.,  2017, \mn@doi [\mnras]
  {10.1093/mnras/stx744}, \href
  {http://adsabs.harvard.edu/abs/2017MNRAS.468.4107R} {468, 4107}

\bibitem[\protect\citeauthoryear{{Ruiz} et~al.,}{{Ruiz}
  et~al.}{2015}]{ruiz2015}
{Ruiz} A.~N.,  et~al., 2015, \mn@doi [\apj] {10.1088/0004-637X/801/2/139},
  \href {http://adsabs.harvard.edu/abs/2015ApJ...801..139R} {801, 139}

\bibitem[\protect\citeauthoryear{{Saintonge} et~al.,}{{Saintonge}
  et~al.}{2016}]{Saintonge16}
{Saintonge} A.,  et~al., 2016, \mn@doi [\mnras] {10.1093/mnras/stw1715}, \href
  {http://adsabs.harvard.edu/abs/2016MNRAS.462.1749S} {462, 1749}

\bibitem[\protect\citeauthoryear{{Salim} et~al.,}{{Salim}
  et~al.}{2007}]{Salim07}
{Salim} S.,  et~al., 2007, \mn@doi [\apjs] {10.1086/519218}, \href
  {http://adsabs.harvard.edu/abs/2007ApJS..173..267S} {173, 267}

\bibitem[\protect\citeauthoryear{{Schawinski} et~al.,}{{Schawinski}
  et~al.}{2014}]{Schawinski14}
{Schawinski} K.,  et~al., 2014, \mn@doi [\mnras] {10.1093/mnras/stu327}, \href
  {http://adsabs.harvard.edu/abs/2014MNRAS.440..889S} {440, 889}

\bibitem[\protect\citeauthoryear{{Scott}, {Usero}, {Brinks}, {Bravo-Alfaro},
  {Cortese}, {Boselli}  \& {Argudo-Fern{\'a}ndez}}{{Scott}
  et~al.}{2015}]{Scott15}
{Scott} T.~C.,  {Usero} A.,  {Brinks} E.,  {Bravo-Alfaro} H.,  {Cortese} L.,
  {Boselli} A.,   {Argudo-Fern{\'a}ndez} M.,  2015, \mn@doi [\mnras]
  {10.1093/mnras/stv1592}, \href
  {http://adsabs.harvard.edu/abs/2015MNRAS.453..328S} {453, 328}

\bibitem[\protect\citeauthoryear{{Shankar} et~al.,}{{Shankar}
  et~al.}{2015}]{Shankar15}
{Shankar} F.,  et~al., 2015, \mn@doi [\apj] {10.1088/0004-637X/802/2/73}, \href
  {http://adsabs.harvard.edu/abs/2015ApJ...802...73S} {802, 73}

\bibitem[\protect\citeauthoryear{{Smethurst}, {Lintott}, {Bamford}, {Hart},
  {Kruk}, {Masters}, {Nichol}  \& {Simmons}}{{Smethurst}
  et~al.}{2017}]{Smethurst17}
{Smethurst} R.~J.,  {Lintott} C.~J.,  {Bamford} S.~P.,  {Hart} R.~E.,  {Kruk}
  S.~J.,  {Masters} K.~L.,  {Nichol} R.~C.,   {Simmons} B.~D.,  2017, \mn@doi
  [\mnras] {10.1093/mnras/stx973}, \href
  {http://adsabs.harvard.edu/abs/2017MNRAS.469.3670S} {469, 3670}

\bibitem[\protect\citeauthoryear{{Springel}, {White}, {Tormen}  \&
  {Kauffmann}}{{Springel} et~al.}{2001}]{springel2001}
{Springel} V.,  {White} S.~D.~M.,  {Tormen} G.,   {Kauffmann} G.,  2001,
  \mn@doi [\mnras] {10.1046/j.1365-8711.2001.04912.x}, \href
  {http://adsabs.harvard.edu/abs/2001MNRAS.328..726S} {328, 726}

\bibitem[\protect\citeauthoryear{{Steinhauser}, {Schindler}  \&
  {Springel}}{{Steinhauser} et~al.}{2016}]{Steinhauser16}
{Steinhauser} D.,  {Schindler} S.,   {Springel} V.,  2016, \mn@doi [\aap]
  {10.1051/0004-6361/201527705}, \href
  {http://adsabs.harvard.edu/abs/2016A%26A...591A..51S} {591, A51}

\bibitem[\protect\citeauthoryear{{Stevens} \& {Brown}}{{Stevens} \&
  {Brown}}{2017}]{Stevens17}
{Stevens} A.~R.~H.,  {Brown} T.,  2017, \mn@doi [\mnras]
  {10.1093/mnras/stx1596}, \href
  {http://adsabs.harvard.edu/abs/2017MNRAS.471..447S} {471, 447}

\bibitem[\protect\citeauthoryear{{Sun}, {Jones}, {Forman}, {Vikhlinin},
  {Donahue}  \& {Voit}}{{Sun} et~al.}{2007}]{sun2007}
{Sun} M.,  {Jones} C.,  {Forman} W.,  {Vikhlinin} A.,  {Donahue} M.,   {Voit}
  M.,  2007, \mn@doi [\apj] {10.1086/510895}, \href
  {http://adsabs.harvard.edu/abs/2007ApJ...657..197S} {657, 197}

\bibitem[\protect\citeauthoryear{{Tecce}, {Cora}, {Tissera}, {Abadi}  \&
  {Lagos}}{{Tecce} et~al.}{2010}]{tecce10}
{Tecce} T.~E.,  {Cora} S.~A.,  {Tissera} P.~B.,  {Abadi} M.~G.,   {Lagos}
  C.~D.~P.,  2010, \mn@doi [\mnras] {10.1111/j.1365-2966.2010.17262.x}, \href
  {http://adsabs.harvard.edu/abs/2010MNRAS.408.2008T} {408, 2008}

\bibitem[\protect\citeauthoryear{{Tecce}, {Cora}  \& {Tissera}}{{Tecce}
  et~al.}{2011}]{tecce11}
{Tecce} T.~E.,  {Cora} S.~A.,   {Tissera} P.~B.,  2011, \mn@doi [\mnras]
  {10.1111/j.1365-2966.2011.19267.x}, \href
  {http://adsabs.harvard.edu/abs/2011MNRAS.416.3170T} {416, 3170}

\bibitem[\protect\citeauthoryear{{Tomczak} et~al.,}{{Tomczak}
  et~al.}{2014}]{tomczak14}
{Tomczak} A.~R.,  et~al., 2014, \mn@doi [\apj] {10.1088/0004-637X/783/2/85},
  \href {http://adsabs.harvard.edu/abs/2014ApJ...783...85T} {783, 85}

\bibitem[\protect\citeauthoryear{{Tonnesen} \& {Bryan}}{{Tonnesen} \&
  {Bryan}}{2008}]{Tonnesen08}
{Tonnesen} S.,  {Bryan} G.~L.,  2008, \mn@doi [\apjl] {10.1086/592066}, \href
  {http://adsabs.harvard.edu/abs/2008ApJ...684L...9T} {684, L9}

\bibitem[\protect\citeauthoryear{{Tonnesen} \& {Bryan}}{{Tonnesen} \&
  {Bryan}}{2009}]{Tonnesen09}
{Tonnesen} S.,  {Bryan} G.~L.,  2009, \mn@doi [\apj]
  {10.1088/0004-637X/694/2/789}, \href
  {http://adsabs.harvard.edu/abs/2009ApJ...694..789T} {694, 789}

\bibitem[\protect\citeauthoryear{{Tonnesen}, {Bryan}  \& {van
  Gorkom}}{{Tonnesen} et~al.}{2007}]{Tonnesen07}
{Tonnesen} S.,  {Bryan} G.~L.,   {van Gorkom} J.~H.,  2007, \mn@doi [\apj]
  {10.1086/523034}, \href {http://adsabs.harvard.edu/abs/2007ApJ...671.1434T}
  {671, 1434}

\bibitem[\protect\citeauthoryear{{Villalobos}, {De Lucia}  \&
  {Murante}}{{Villalobos} et~al.}{2014}]{Villalobos14}
{Villalobos} {\'A}.,  {De Lucia} G.,   {Murante} G.,  2014, \mn@doi [\mnras]
  {10.1093/mnras/stu1278}, \href
  {http://adsabs.harvard.edu/abs/2014MNRAS.444..313V} {444, 313}

\bibitem[\protect\citeauthoryear{{Wagner}, {McDonald}  \& {Courteau}}{{Wagner}
  et~al.}{2017}]{Wagner17}
{Wagner} C.~R.,  {McDonald} M.,   {Courteau} S.,  2017, preprint, \href
  {http://adsabs.harvard.edu/abs/2017arXiv171102675W} {} (\mn@eprint {arXiv}
  {1711.02675})

\bibitem[\protect\citeauthoryear{{Wang} et~al.,}{{Wang} et~al.}{2017}]{Wang17}
{Wang} H.,  et~al., 2017, preprint, \href
  {http://adsabs.harvard.edu/abs/2017arXiv170709002W} {} (\mn@eprint {arXiv}
  {1707.09002})

\bibitem[\protect\citeauthoryear{{Weinmann}, {van den Bosch}, {Yang}  \&
  {Mo}}{{Weinmann} et~al.}{2006}]{weinmann2006}
{Weinmann} S.~M.,  {van den Bosch} F.~C.,  {Yang} X.,   {Mo} H.~J.,  2006,
  \mn@doi [\mnras] {10.1111/j.1365-2966.2005.09865.x}, \href
  {http://adsabs.harvard.edu/abs/2006MNRAS.366....2W} {366, 2}

\bibitem[\protect\citeauthoryear{{Weinmann}, {Kauffmann}, {von der Linden}  \&
  {De Lucia}}{{Weinmann} et~al.}{2010}]{weinmann2010}
{Weinmann} S.~M.,  {Kauffmann} G.,  {von der Linden} A.,   {De Lucia} G.,
  2010, \mn@doi [\mnras] {10.1111/j.1365-2966.2010.16855.x}, \href
  {http://adsabs.harvard.edu/abs/2010MNRAS.406.2249W} {406, 2249}

\bibitem[\protect\citeauthoryear{{Weinmann}, {Pasquali}, {Oppenheimer},
  {Finlator}, {Mendel}, {Crain}  \& {Macci{\`o}}}{{Weinmann}
  et~al.}{2012}]{Weinmann12}
{Weinmann} S.~M.,  {Pasquali} A.,  {Oppenheimer} B.~D.,  {Finlator} K.,
  {Mendel} J.~T.,  {Crain} R.~A.,   {Macci{\`o}} A.~V.,  2012, \mn@doi [\mnras]
  {10.1111/j.1365-2966.2012.21931.x}, \href
  {http://adsabs.harvard.edu/abs/2012MNRAS.426.2797W} {426, 2797}

\bibitem[\protect\citeauthoryear{{Wetzel}, {Tinker}  \& {Conroy}}{{Wetzel}
  et~al.}{2012}]{Wetzel12}
{Wetzel} A.~R.,  {Tinker} J.~L.,   {Conroy} C.,  2012, \mn@doi [\mnras]
  {10.1111/j.1365-2966.2012.21188.x}, \href
  {http://adsabs.harvard.edu/abs/2012MNRAS.424..232W} {424, 232}

\bibitem[\protect\citeauthoryear{{Wilkins}, {Trentham}  \& {Hopkins}}{{Wilkins}
  et~al.}{2008}]{Wilkins2008}
{Wilkins} S.~M.,  {Trentham} N.,   {Hopkins} A.~M.,  2008, \mn@doi [\mnras]
  {10.1111/j.1365-2966.2008.12885.x}, \href
  {http://adsabs.harvard.edu/abs/2008MNRAS.385..687W} {385, 687}

\bibitem[\protect\citeauthoryear{{Xie}, {De Lucia}, {Hirschmann}, {Fontanot}
  \& {Zoldan}}{{Xie} et~al.}{2017}]{Xie17}
{Xie} L.,  {De Lucia} G.,  {Hirschmann} M.,  {Fontanot} F.,   {Zoldan} A.,
  2017, \mn@doi [\mnras] {10.1093/mnras/stx889}, \href
  {http://adsabs.harvard.edu/abs/2017MNRAS.469..968X} {469, 968}

\bibitem[\protect\citeauthoryear{{Yang}, {Mo}, {van den Bosch}  \&
  {Jing}}{{Yang} et~al.}{2005}]{Yang05}
{Yang} X.,  {Mo} H.~J.,  {van den Bosch} F.~C.,   {Jing} Y.~P.,  2005, \mn@doi
  [\mnras] {10.1111/j.1365-2966.2005.08560.x}, \href
  {http://adsabs.harvard.edu/abs/2005MNRAS.356.1293Y} {356, 1293}

\bibitem[\protect\citeauthoryear{{Yang}, {Mo}  \& {van den Bosch}}{{Yang}
  et~al.}{2009}]{Yang09b}
{Yang} X.,  {Mo} H.~J.,   {van den Bosch} F.~C.,  2009, \mn@doi [\apj]
  {10.1088/0004-637X/695/2/900}, \href
  {http://adsabs.harvard.edu/abs/2009ApJ...695..900Y} {695, 900}

\bibitem[\protect\citeauthoryear{{Yozin} \& {Bekki}}{{Yozin} \&
  {Bekki}}{2015}]{YozinBekki15}
{Yozin} C.,  {Bekki} K.,  2015, \mn@doi [\mnras] {10.1093/mnras/stv1593}, \href
  {http://adsabs.harvard.edu/abs/2015MNRAS.453...14Y} {453, 14}

\bibitem[\protect\citeauthoryear{{Zentner}, {Berlind}, {Bullock}, {Kravtsov}
  \& {Wechsler}}{{Zentner} et~al.}{2005}]{Zentner05}
{Zentner} A.~R.,  {Berlind} A.~A.,  {Bullock} J.~S.,  {Kravtsov} A.~V.,
  {Wechsler} R.~H.,  2005, \mn@doi [\apj] {10.1086/428898}, \href
  {http://adsabs.harvard.edu/abs/2005ApJ...624..505Z} {624, 505}

\bibitem[\protect\citeauthoryear{{van der Wel}, {Rix}, {Holden}, {Bell}  \&
  {Robaina}}{{van der Wel} et~al.}{2009}]{vanderWel09}
{van der Wel} A.,  {Rix} H.-W.,  {Holden} B.~P.,  {Bell} E.~F.,   {Robaina}
  A.~R.,  2009, \mn@doi [\apjl] {10.1088/0004-637X/706/1/L120}, \href
  {http://adsabs.harvard.edu/abs/2009ApJ...706L.120V} {706, L120}

\makeatother
\end{thebibliography}

%%%%%%%%%%%%%%%%%%%%%%%%%%%%%%%%%%%%%%%%%%%%%%%%%%

%%%%%%%%%%%%%%%%% APPENDICES %%%%%%%%%%%%%%%%%%%%%

\appendix

\section{Ram pressure stripping radius of the hot gas}
\label{ap:rpscalc}
%\normalsize{
Assuming that the DM and hot gas in a satellite distribute following an
isothermal sphere density profile, the total mass of the
satellite from equation~\eqref{eq:msat} can be written as
\begin{equation}\label{eq:msatsis}
  M_\text{sat}(r_\text{sat}) = M_\text{gx} + \left(
  \frac{M_\text{hot}}{r_\text{hot}} + \frac{M_\text{DM}}{r_\text{DM}}
  \right) r_\text{sat}.
\end{equation}
where we have defined $M_\text{gx} \equiv M_* + M_\text{cold}$ and
$r_\text{sat}$ is  
the current radius of the satellite.
We assume that
\rdm~has already been determined by the effect of TS. Using
equation~\eqref{eq:msatsis} in the RPS condition for a spherically symmetrical
distribution of gas from \citeauthor{mccarthy2008} (2008, equation~\ref{eq:rpshot}
in this paper) and rearranging terms a bit, we can obtain
\begin{equation}
  P_\text{ram} = \frac{\alpha_\text{RP} M_\text{hot}}{4\pi
  r_\text{hot} r_\text{sat}^3} \left[ G M_\text{gx} + \left(
  \frac{G M_\text{hot}}{r_\text{hot}} + \frac{G M_\text{DM}}{r_\text{DM}}
  \right) r_\text{sat} \right].
\end{equation}
The equality actually holds when $r_\text{sat}$ equals the gas stripping
radius 
$r_{\rm s}$, which represents the radius $r_{\rm s,hot}^{\rm RPS}$ defined in Section~\ref{sec:hotgasstrip}.
Noting that terms of the form $G M / r$ have units of velocity
squared, we can define
\begin{equation}
  V_h^2 \equiv \frac{G M_\text{hot}}{r_\text{hot}} + \frac{G
  M_\text{DM}}{r_\text{DM}} \equiv V_\text{hot}^2 + V_\text{DM}^2
\end{equation}
and also
\begin{equation}
  V_\text{gx}^2 \equiv G M_\text{gx} / r_\text{DM}.
\end{equation}
We then have
\begin{equation}
  P_\text{ram}  r_{\rm s}^3 = \frac{\alpha_\text{RP} M_\text{hot}}{4\pi r_\text{hot}}
  \left[ V_\text{gx}^2 r_\text{DM} + V_h^2 r_{\rm s} \right].
\end{equation}
Rearranging terms once again, we obtain that the stripping radius~$r_{\rm s}$
must
satisfy the following condition:
\begin{equation}\label{eq:rstriphot}
   r_{\rm s}^3 - \frac{\alpha_\text{RP} M_\text{hot}}{4\pi r_\text{hot} P_\text{ram}}
  \left[ V_\text{gx}^2 r_\text{DM} + V_h^2 r_{\rm s} \right] = 0.
\end{equation}
This is a third-order polynomial $a r_{\rm s}^3 + b r_{\rm s}^2 + c r_{\rm s} + d$, with
$a$~=~1, $b$~=~0 and both $c$ and $d$ are always negative. A polynomial of this
form has only one real and positive root, which corresponds to the RP stripping
radius; equation~\eqref{eq:rstriphot} must be solved numerically to obtain~
$r_{\rm s}$.

%%%%%%%%%%%%%%%%%%%%%%%%%%%%%%%%%%%%%%%%%%%%%%%%%%

% Don't change these lines
\bsp	% typesetting comment
\label{lastpage}
\end{document}